# Unsupervised and Supervised Structure Learning for Protein Contact Prediction

by

Siqi Sun

A thesis submitted in partial fulfillment of the requirements for

The degree of

Doctor of Philosophy in Computer Science

at the

TOYOTA TECHNOLOGICAL INSTITUTE AT CHICAGO

June 2019

Thesis Committee:

Jinbo Xu (Thesis Advisor)

Gregory Shakhnarovich

Dong Xu

# Unsupervised and Supervised Structure Learning for Protein Contact Prediction

a thesis presented

by

Siqi Sun

In partial fulfillment of the requirements for the degree of
Doctor of Philosophy in Computer Science
Toyota Technological Institute at Chicago
June 2019

-Thesis Committee-

Jinbo Xu

| Committee member | Signature | Date |

Gregory Shakhnarovich

| Committee member | Signature | Date |

Dong Xu

| Committee member | Signature | Date |

# Unsupervised and Supervised Structure Learning for Protein Contact Prediction

by

Siqi Sun

## Abstract

Protein contacts provide key information for the understanding of protein structure and function, and therefore contact prediction from sequences is an important problem. Recent research shows that some correctly predicted long-range contacts could help topology-level structure modeling. Thus, contact prediction and contact-assisted protein folding also proves the importance of this problem. In this thesis, I will briefly introduce the extant related work, then show how to establish the contact prediction through unsupervised graphical models with topology constraints. Further, I will explain how to use the supervised deep learning methods to further boost the accuracy of contact prediction. Finally, I will propose a scoring system called diversity score to measure the novelty of contact predictions, as well as an algorithm that predicts contacts with respect to the new scoring system.

Thesis Supervisor: Jinbo Xu
Title: Professor

# Acknowledgements

First, I want to thank my advisor, Jinbo Xu; without his sincere help, I would never have been able to finish my PhD. I also want to thank him for his encouragement and patience in our conversations, his deep insights about bioinformatics and protein always amazed me. The most important thing I learned from him is to find an interesting and impactful direction, then do your best to push the limits of the field. I am grateful to my thesis committee members, Gregory Shakhnarovich and Dong Xu, for being kind enough to agree to be in my committee.

I would also like to thank my friends from TTIC and University of Chicago, Yuancheng Zhu, Zhengrong Xing, Sheng Wang, Hai Wang, Qingming Tang, Payman Yadollahpour, Hao Tang, Jianzhu Ma, Zhiyong Wang, Somaye Hashemifar, Feng Zhao, Behnam Neyshabur, Liwen Zhang, Tong Lu, to name a few, for many conversations related or not related to my study.

I want to thank my friends from Planet 9, Ang Min, Jian Xu, Yun Li, Jialei Wang, and Lifu Tu. It was a great experience that will always have a place in my memory.

Finally, I would like to thank my parents, Qingxin Sun and Guifei Chen, and my wife Mengwen Zhang for their unconditional support over the past few years; without their help and understanding, I would never have been able to finish my PhD.

# Contents

# List of Figures

# List of Tables

# Chapter 1
# Introduction

Protein residue-residue contact prediction is used to predict whether two positions in a protein sequence are spatially proximal to each other in the 3D structure. In this thesis, we define a contact between two residues if the Euclidian distance between their $C_\beta$ atom less than 8Å [4]. Recently, some research shows that correctly predicted contact prediction plays an important role in protein folding, especially for long-range contacts that are between sequentially distant residues [1]. Therefore, designing an accurate and reliable contact prediction algorithm is a very important task in computational biology. However, it is also a very challenging task; even the prediction quality of current state-of-the-art predictors is not sufficient for accurate contact-assisted protein folding [2，3], especially for those without many sequence homologs. This motivated us to develop more accurate methods for this problem.

Currently, there are two types of contact prediction methods: evolutional coupling analysis (ECA) and supervised machine learning methods. ECA predicts contact by identifying co-evolution pairs since a pair of co-evolved residues is often found to be spatially close in the 3D structure. An initial method of ECA calculated mutual information between a pair of residues to detect contacts from multiple sequence alignment (MSA) [77, 20]. Though the results seemed promising, its accuracy was still low because interactions can also happen when more than two positions show substitute patterns, leading to many false positive predictions. Years later, a maximum-entropy approach was developed by replacing mutual information with a pairwise graphical model [7-9, 76], and applied successfully to distinguish the direct coupling from indirect ones, thus improving the prediction quality substantially. To ease the computation of the graphical model, an iterative message passing approach like [8] or mean field approximation approach like EVfold [9] was employed.

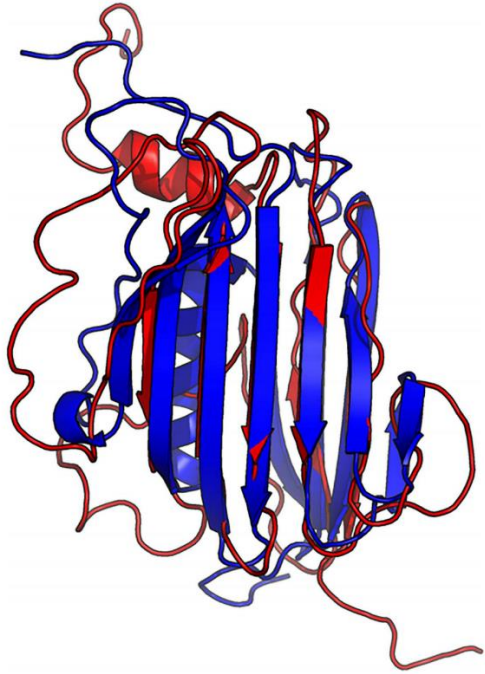

FIGURE 1 SUPERIMPOSITION BETWEEN A CONTACT ASSISTED FOLDING (RED) AND ITS NATIVE STRUCTURE (BLUE) FOR A CAMEO PROTEIN (PDB ID 2NC8 AND CHAIN A). FIGURE IS FROM [58].

In 2011, Jones et al. proposed the use of sparse covariance inverse to detect co-evolved pairs by assuming that MSA follows Gaussian graphical models, and further improved performance. Instead of using message passing or mean field to approximate likelihood function, a more recently proposed method utilizes a pseudo-likelihood maximization approach for Potts model [12, 13], and has state-of-the-art results for contact prediction in the direct coupling analysis (DCA) category. Gremlin[11] is another work based on optimizing towards pseudo-likelihood, but it includes more prior information, such as secondary structure. Note that there are also other methods that don't follow this Markov random field line of work. For example, Burger et al. used Bayesian network to disentangle direct co-evolution residues from indirect ones [78]. Representative tools of recent DCA method for contact prediction include EV fold [9], PSICOV [10], GREMLIN [11] and plmDCA [12, 13]. See [14] for a more detailed review of EC analysis.

Unlike supervised machine learning models, ECA is an unsupervised approach and therefore does not require any labeled training data. Rather, recently developed EC methods take all positions from MSA into consideration to predict contacts between two residues. Thus, it has rich global information and better accuracy for long-range contact prediction. However, ECA's prediction accuracy for many proteins is still low, even when only the top L/10 (L is the sequence length) predicted contacts are evaluated. Moreover, to be able to detect the co-evolution patterns between residues, it requires a larger number of homologs.

On the other hand, supervised machine learning methods predict contacts by using various types of information, including the derived features from MSAs, such as predicted secondary structures, predicted solvent accessibility, and even the results of ECA methods. Example methods are SVMSEQ [15], PconsC2 [16], MetaPSICOV [17], coinDCA-NN [18], CMAPpro [73] and PhyCMAP [19]. Existing supervised machine learning methods typically perform better than ECA by a large margin in terms of contact prediction accuracy due to their supervised approach. However, the predictions are still quite limited for accurate contact-assisted protein folding because of the model's shallow architecture. For example, coinDCA-NN and MetaPSICOV use a neural network with only two layers; PconsC2 uses a neural network with only five layers. CMAPpro applies a deep learning model with many more layers, but its performance saturates at approximately ten layers.

To understand why the number of layers matters, let us go through the basic processes of these approaches. To make predictions between two residues i and j, supervised approaches typically extract features from a fixed window around both i and j. Sometimes they also include the features around (i+j)/2. However, if i and j are far apart on the sequence level, it is very difficult for shallow architectures to model the relations between them because the fixed windows are not large enough. Additionally, for these supervised machine approaches, the prediction of contact at (i, j) is independent of that of their neighbors—e.g., (i-1, j), (i+1, j) and (i, j-1)— because of the independent computation of loss function. Intuitively, those distances are supposed to be highly correlated. This motivates us to develop a better contact prediction method with more layers and a loss function that takes the whole contact map into consideration, especially for proteins without many sequence homologs.

In this thesis, I will investigate both approaches and try to improve the results from theoretical and practical points of view. More specifically,

   (a) We first use a nonparametric Bayesian model to incorporate a cluster topology constraint into the Gaussian graphical model, which improves the plain Gaussian graphical model, i.e., PSICOV. As mentioned earlier, the unsupervised structure learning approach is important because it carries rich global information about MSAs and could be used later as a potential input feature for supervised approaches.

   (b) Next, we relax the Gaussian assumption and assume the data can be of almost any distribution. By using a novel score-matching approach, we bypass the computation of partition function, which is typically intractable. Another way to avoid computing the partition function is to use pseudo-likelihood to approximate true likelihood (e.g., plmDCA). We will compare those two approaches in synthetic data settings and real data settings.

   (c) Finally, we present a very deep residual neural network for contact prediction. Additionally, we show that our proposed model can capture extremely complex sequence-contact relationships and high-order contact correlations due to its deep architecture. We also describe the detailed model architecture and the training procedure with selected hyperparameters. Moreover, our research leads to another intriguing question of whether we can learn contact prediction end-to-end, as preferred in the deep learning community for its high performance. We investigate this problem through a detailed ablation study and evaluate the features' importance.

# Chapter 2

# Background and Existing Methods

Denote the target sequence as $R = r_1 r_2 \ldots r_L$, where L is the sequence length, $r_i \in S'$ is an amino acid, and $S' = \{A,C,D,E,F,G,H,I,K,L,M,N,P,Q,R,S,T,V,W,Y\}$ is the set of all 20 amino acids. To get the multiple sequence alignment, we typically run PSI-BLAST to search the non-redundant protein sequence database for its sequence homologs, then build its multiple sequence alignment, and sequence profile and other features. Denote the multiple sequence alignment as $X_{N \times L}$, where N is the number of homologs, L is sequence length, and each $X_{ij}$ is a categorical variable that can take values from $S'$ or a gap (-). For simplicity, we denote $S = S' \cup \{-\}$. Please see figure 2 for an example of multiple sequence alignment.

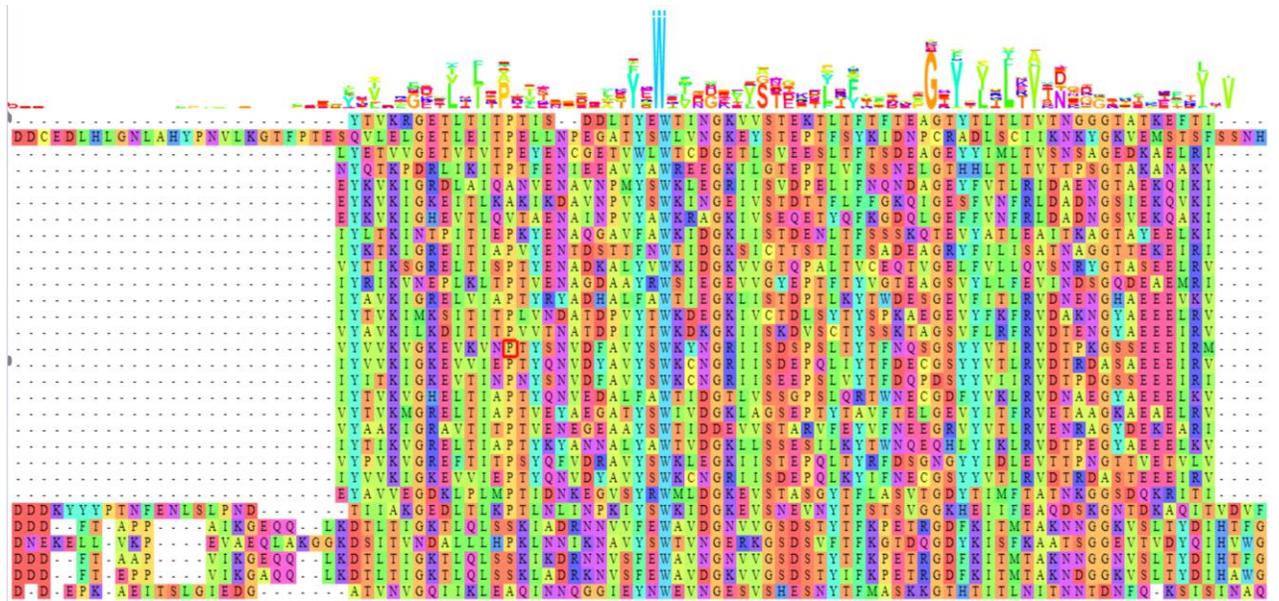

FIGURE 2 AN EXAMPLE OF MULTIPLE SEQUENCE ALIGNMENT FOR T0845-D1 FROM CASP 11. THE VIEW IS GENERATED BASED ON CODE FROM HTTPS://GITHUB.COM/WILZBACH/MSA.

A pair of residues that co-evolve in tandem, thus preserving relative physiochemical properties, is more likely to form contacts. The most common method for detecting correlated mutations in multiple sequence alignment is to compute the mutual information between any two sites and then to select pairs based on the rank of their mutual information. To compute the mutual information between position i and position j,

$$MI_{ij} = \sum_{ab} f(X_{.i}X_{.j} = ab) \log \frac{f(X_{.i}X_{.j} = ab)}{f(X_{.i} = a)f(X_{.j} = b)}$$

, where $X_{.i}$ denotes the i-th column for multiple sequence alignment X, $f(X_{.i} = a)$ is the observed relative frequency of amino acid type a at column i, and $f(X_{.i}X_{.j} = ab)$ is the observed relative frequency of amino acid pair ab at columns i, j. To simplify the notation, we will use $f_i(a) = f(X_{.i} = a)$ and $f_{ij}(a,b) = f(X_{.i}X_{.j} = ab)$. Then, the mutual information can be simplified as

$$MI_{ij} = \sum_{ab} f_{ij}(ab) \log \frac{f_{ij}(ab)}{f_i(a)f_j(b)}$$

The result can be further improved by utilizing normalization to take into account bias; please see [20] for more details. Despite doing a reasonable job for the prediction of contact maps, such a method cannot reduce the chaining within the contact map.

## 2.1 Unsupervised Structure Learning Algorithm for Contact Prediction

Instead of modeling the local statistical information, such as $MI_{ij}$, researchers start to use a global statistical approach to model the distribution of MSAs. Denote $x = (x_1, x_2, ..., x_L)$ as L random variables, where each $x_i$ ($i \in \{1, 2, .., L\}$) represents the MSA's distribution at position i. Then the joint probability of MSA is assumed to be in a Markov Random Field (MRF), or P(x) as follows

$$P(x) = \frac{1}{Z} \exp(\sum_{i<j} \theta_{ij}(x_i, x_j) + \sum_i b_i(x_i)),$$

where Z is the partition function such that P(x) is a probability density distribution. The MRF has a property such that $x_i$ and $x_j$ are conditionally independent if and only if

$\theta_{ij}(x_i, x_j) = 0$. Therefore, the conditional dependency's zero pattern ($\theta_{ij}(x_i, x_j)$ is 0 or not) can be thought as degree of covariation between residue from position i and j due to direct effects of i and j.

The objective therefore is to maximize the log-likelihood given the MSAs by estimating $\theta_{ij}$ and $b_i$. However, the gradient of objective function is typically intractable because the computation of partition function Z involves summation over an exponential number of terms with respect to MSA's length L.

To solve this problem, several approximation approaches have been proposed. Lapedes et al. [78] first proposed a resource demanding Monte Carlo method. Years later, Weigt et al. developed an algorithm based on message passing to compute the partition function [8]. Unfortunately, these methods are still very computationally intensive. To overcome this problem, more computationally efficient approaches have been proposed recently, and we will discuss three successful ones, i.e., mean-field approximation [9], Gaussian graphical model approximation [10] and pseudo-likelihood approximation [11, 12, 13].

### 2.1.1 Mean-Field Approximation

We assume that each $x_i$ is sampled from S with 21 categories, thus the parameter $\theta_{ij}$ is a 21×21 matrix, and $\theta_{ij}(S_i, S_j)$ could be used to measure the interaction strength between positions i and j, taking residues $S_i$ $and$ $S_j$; a higher value of $\theta_{ij}$ indicates that two positions (i and j) are more likely to form a physical contact. Then, Marks et al. assumes the single and pairwise residue marginal probability defined by the exponential family above is coherent with the empirical single and pairwise frequency counts from MSA; more specifically,

$$P_i(x_i = A_i) = \sum_{x_l, l \neq i} P(A_1, A_2, .., A_L) = f_i(A_i)$$

$$P_i(x_i = A_i, x_j = A_j) = \sum_{x_l, l \neq i,j} P(A_1, A_2, .., A_L) = f_{ij}(A_i, A_j).$$

With these constraints, the model is then optimized by maximizing the entropy using Lagrange multipliers. Further, define

$$Z(\alpha) = \sum_{A_i, i=1,\dots,L} \exp\left(\alpha \sum_{i<j} \theta_{ij}(A_i, A_j) + \sum_i h_i(A_i)\right)$$

and its Legendre transform as

$$G(\alpha) = \log Z(\alpha) - \sum_i \sum_{A_i} h_i(A_i) P_i(A_i).$$

Note that when α=1, Z(α) reduces to partition function. Then G(α) can be approximated by its first order Taylor series expansion:

$$G(\alpha) \approx G(0) + \frac{\partial G(\alpha)}{\partial \alpha}\bigg|_{\alpha=0} \alpha.$$

In this approximation, we can further obtain

$$(C^{-1})_{ij}(A_i, A_j)|_{\alpha=0} = -\theta_{ij}(A_i, A_j),$$

where $C_{ij}(A_i, A_j) = f_{ij}(A_i, A_j) - f_i(A_i)f_j(A_j)$ is the empirical correlation matrix. Therefore, $\theta_{ij}(A_i, A_j)$ can be estimated by computing the inverse of the empirical correlation matrix. Finally, the interaction strength between two positions i and j is defined as the relative entropy between $P_{ij}$ and the independent position distribution. Please refer to [9] for more details about how to calculate interaction strength given estimated parameters.

### 2.1.2 Gaussian Graph Model Approximation

Instead of approximating the partition function, PSICOV [10] attempts to correct the above effects by using Gaussian Graphical Models (GGMs) [22] because GGMs have close-formed partition function. For GGMs, given L random variable, $x = (x_1, x_2, \dots, x_L)$ that follow multiple variable Gaussian distribution, it is known that the graph structure of the model is encoded in the sparsity pattern of a precision matrix $\Omega_{L \times L}$ (inverse of covariance matrix $\Sigma_{L \times L}$), i.e., $x_i$ and $x_j$ are conditionally independent if and only if $\Omega_{ij} = 0$. In a contact map setting, we can represent each $X_{ij}$ in multiple sequence alignment using a 21-dimensional one hot vector. Assuming the new multiple sequence alignment $X_{N\times(L\times 21)}$ follows multivariate Gaussian distribution, we can estimate the sparse precision matrix by graphical lasso [22] and treat the resulting graph structure as the predicted contact map.

More specifically, each entry in the sample covariance matrix $S$ can be estimated as follows:

$$S_{ij}^{ab} = \frac{1}{n}\sum_{k=1}^{N}(x_i^{ak} - \bar{x}_i^a)(x_j^{bk} - \bar{x}_j^b)$$

, where $x_i^{ak}$ is a binary variable that indicates the presence or absence of amino acid a at position i and in sequence k. Note that the dimension of $S$ is $21L \times 21L$ because there are 21L variables in the new multiple sequence alignment. The objective function based on Gaussian Graphical Model assumption is

$$L(\Omega) = \sum_{ij=1}^{d} S_{ij}\Omega_{ij} - \log \det \Omega + \rho \sum_{ij=1}^{d} |\Omega_{ij}|$$

, where the first two terms are the negative log-likelihood and the last term is the $l_1$ penalty such that the estimated precision matrix ($\widehat{\Omega}$) is sparse. To translate the estimated $\Omega$ back to contact map, the final processing step is to compute the $l_1$ norm for the 20×20 submatrix of $\Omega$ corresponding to all of the amino acid pairs ab in any two columns, i.e., $Score_{ij} = \sum_{ab \in S'} |\widehat{\Omega}_{ij}|$, where the contribution of gap is ignored. Similar to the mutual information case, the final contact is predicted by the rank of pairs based on $Score_{ij}$. There are several more post-processing steps; please refer to [10] for more details.

### 2.1.3 Pseudo-likelihood Approximation

The main problem of PSICOV is that the assumption of the model is not accurate enough because observed data is binary rather than continuous. Ekeberg1 et al. proposed using the Potts model (named plmDCA) to replace the GGM because the Potts model assumes that each random variable follows categorical distribution[1], which fits the data more reasonably than Gaussian [13], since each position in a multiple sequence alignment can take 21 discrete values. As we noted earlier, the model is difficult to optimize because it is hard to compute the partition function for the corresponding graphical model. Instead of using mean-field approximation, a pseudo-likelihood approach is used to approximate the likelihood by using

$$P(x_1, x_2, \ldots, x_L) = P(x_1|x_{-1})P(x_2|x_{-2}) \ldots P(x_L|x_{-L}),$$

where $P(x_i|x_{-i})$ is further modeled by a multi-output logistic regression and $x_{-i}$ indicates all other variables except for $x_i$. Much more accurate than GGM, the algorithm is also very easy to compute in parallel. A faster version (called CCMpred) of the algorithm was later proposed and implemented on GPU [12], and is now more widely used. Unlike PSICOV, where the $l_2$ norm penalty is used, a sequence reweighting is used in plmDCA. Additionally, a different interaction scoring approach is used. For more details about the post processing method, please see [13].

---

[1] For Bernoulli distribution, the model is reduced to Ising model

## 2.2 Supervised Structure Learning Algorithm for Contact Prediction

In the Protein Data Bank (PDB) [23], there are thousands of proteins with known contact maps. They can therefore be used to train a machine learning model, such as CMAPpro [4], SVMSEQ [15], PconsC2 [16], MetaPSICOV [17], coinDCA [18], and PhyCMAP [19]. Typically, supervised methods outperform unsupervised approaches, but the performance of supervised methods is still limited due to shallow architecture. Here we introduce and focus on only MetaPSICOV because it has been the best performing algorithm to date, and we will use it extensively to make comparisons.

### 2.2.1 MetaPSICOV

MetaPSICOV was trained on highly resolved protein chains with 624 proteins and tested on the original PSICOV test set. To make a fair comparison, the authors removed proteins that overlap with test set from the training set. To prepare for training data, for any two amino acids i and j, features were extracted from within a fixed window size around i, j, and the mid-point (i+j)/2. The features can be divided into several categories: (1) column features, such as amino acid composition, predicted secondary structures, and predicted solvent accessibility; (2) coevolution features, such as mutual information, PSICOV score, and CCMpred score; (3) sequence separation features, such as |i-j|<5 and |i-j|<17; and (4) global sequence features, such as log sequence length and log effective number of sequences.

Overall, 672 features are used in the first stage classifier. Then, those features are fed into a two-layer fully connected neural network with 55 hidden neurons. In the training procedure, 10% of the original training data was also used as a validation set to select which epoch of model to use.

The results on the test set prove that MetaPSICOV outperforms both coevolution methods, such as PSICOV and CCMpred, and PconsC on the same benchmark set [17].

## 2.3   Introduction to Dataset and Metrics

In this thesis, we used a much larger training set containing 6,767 proteins from a subset of the protein data bank created in February 2015, in which any two proteins share less than 25% sequence identity. From among those 6,767 proteins, we randomly selected 400 as a validation set to select hyperparameters, such as epoch, step size, and number of layers. For inclusion in this training set, the proteins satisfied all of the following conditions: (i) has a sequence length between 26 and 700; (ii) has a resolution better than 2.5Å; (iii) has no domains made up of multiple protein chains; (iv) has DSSP information; and (v) has no inconsistency between its PDB, DSSP, and ASTRAL sequences. To remove redundancy with the test sets, we further excluded any training proteins sharing >25% sequence identity or having BLAST E-value < 0.1 with any test proteins.

We used three publicly available benchmark datasets as our test set, including 108 proteins from CASP 11, 76 hard proteins from CAMEO released in 2015, and 396 membrane proteins. All test membrane proteins have a length of no more than 400 residues, and any two membrane proteins share less than 40% sequence identity. For the

CASP test proteins, we used the official domain definitions, but we did not parse a CAMEO or membrane protein into domains.

### 2.3.1 Evaluation Metric

Contact can be divided into three groups based on sequence level distance. More specifically, denote i, j as an index of two amino acids that form a contact. If:

(1)     $6 \leq |i - j| < 12$, it is called a short-range contact;

(2)     $12 \leq |i - j| < 24$, it is called a medium-range contact;

(3)     $|i - j| \geq 24$, it is called a long-range contact.

Because longer-range contact can provide more information for protein folding, long-range contact prediction is more important and informative than medium-range contact, and medium-range contact is more important and informative than short-range contact. Generally, researchers are not interested in contacts with a sequence distance of less than 6. Consequently, in this thesis we only consider medium- and long-range contacts.

Denoting the sequence length as L, each prediction algorithm will predict top L/10, L/5, L/2 and L pairs that are most likely to form a medium- or long-range contact, and then compute the accuracy for each pair. Therefore, the resulting metric for each algorithm will be an 8-dimensional vector (10 dimensions if 2L prediction is included), rather than scalar. Additionally, note that possibly even ground truth cannot achieve 100% accuracy, especially for medium-range contact and top L prediction, simply because the number of contacts is not enough to fill all the prediction slots. For more details, please see Table 1 for the upper bounds of all training and testing set prediction accuracy. Note that for medium-range top L prediction, the upper bounds are all far less than 1 because there are far fewer medium-range predictions.

| Dataset | Medium-range upper bounds | | | | Long-range upper bounds | | | |
|---|---|---|---|---|---|---|---|---|
| | L/10 | L/5 | L/2 | L | L/10 | L/5 | L/2 | L |
| PDB | 0.99 | 0.98 | 0.93 | 0.79 | 0.99 | 0.99 | 0.98 | 0.96 |
| CASP11 | 1.00 | 0.99 | 0.78 | 0.44 | 0.98 | 0.98 | 0.97 | 0.90 |
| CAMEO | 0.97 | 0.95 | 0.86 | 0.73 | 0.96 | 0.95 | 0.92 | 0.85 |
| MEMS | 0.99 | 0.97 | 0.91 | 0.81 | 0.99 | 0.99 | 0.99 | 0.96 |

TABLE 1 UPPER BOUND OF EACH DATASET FOR ALL EVALUATION METRICS. NOTE THAT THE UPPER BOUND FOR MEDIUM RANGE CONTACT IS MUCH SMALLER THAN 1 SIMPLY BECAUSE THERE ARE NOT ENOUGH NUMBER OF CONTACTS IN THIS RANGE.

### 2.3.2 Effective number of sequences (Meff)

Given the target and the multiple sequence alignment of all of its homologs, the effective number of sequences [19], $M_{eff}$, is computed as

$$M_{eff} = \Sigma_i \frac{1}{\Sigma_j S_{ij}},$$

where i and j go over all the sequence homologs and $S_{ij}$ is a binary similarity value between two proteins. Following Coinfold [3], we computed the similarity of two sequence homologs using their hamming distance. That is, $S_{ij} = 1$ if the normalized hamming distance is less than 0.3; 0 otherwise. This measures the number of non-redundant sequences in a multiple sequence alignment. The smaller the number, the harder it is to make predictions. The distribution of $\log M_{eff}$ for the test datasets can be seen in Figure 3. Interestingly, all of them except for the membrane protein dataset have a significant number of sequences that have nearly no homologs, which demonstrates the difficulty of this problem.

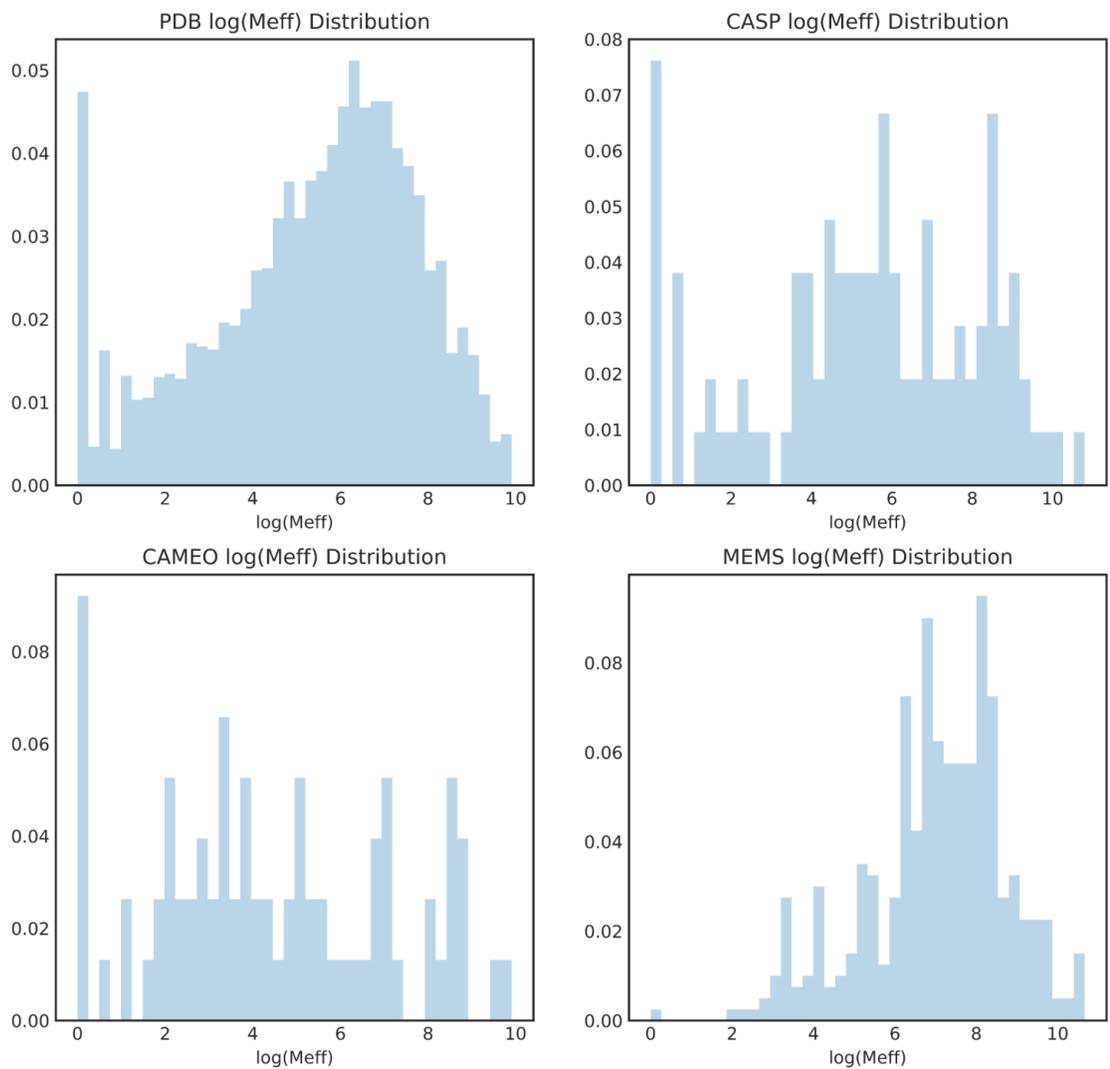

FIGURE 3 LOG(MEFF) HISTOGRAPH FOR ALL DATASETS. EXCEPT FOR MEMS DATASET, ALL OTHER DATASET HAVE SIGNIFICANT NUMBER OF PROTEINS WITH MEFF LESS THAN 2, WHICH DEMONSTRATES THE DIFFICULTY OF THIS PROBLEM.

# Chapter 3

# Unsupervised Structure Learning for Contact Prediction

## 3.1 Adaptive Clustering in Gaussian Graphical Models

Gaussian graphical models (GGMs) are widely used to describe real world data and have important applications in various fields such as computational biology, spectroscopy, climate studies, etc. Learning the structure of GGMs is a fundamental problem since it helps uncover the relationship between random variables and allows further inference. It is well known that the structure of a GGM, i.e., the conditional dependence of the underlying Gaussian vector, is encoded only by the zero pattern of its precision matrix. A straightforward method to estimate the precision matrix is to invert the empirical covariance matrix. In addition to the singularity issue when the dimension p is larger than the number of samples n, the precision matrix resulting from this method is usually not sparse and thus, the learned structure may greatly deviate from the real one. Graphical Lasso (GLASSO) is a popular approach for the estimation of the structure of a GGM. GLASSO maximizes the log-likelihood while penalizing the L1 norm of the precision matrix, which is used to favor a sparse graph. PSICOV used this version of GLASSO to predict contact map of a protein.

In many real-world applications the underlying graph or network that we want to estimate has block structure such that it can be divided into blocks where the inter-block dependence is much weaker than the intra-block dependence. For example, in protein-protein interaction networks, proteins with similar functions are more likely to form a pathway or a complex. Or in contact map, if amino acid at position i and j form a protein, it is very likely that (i±1, j) or (i, j±1) form a contact as well, because the Euclidian distance of (i, j)'s neighbor highly depends on the distance of (i, j). Therefore, it is of great interest to learn such a block-structured graph, which is also equivalent to clustering the variables into disjoint groups. The clustering would not be hard as long as we could estimate the graph accurately since we could simply use the connected components of the estimated graph as a clustering of variables. However, almost all the graph estimation methods such as GLASSO require some predefined parameters controlling the sparsity of the graph and different values of the parameters may lead to quite different clustering results. We may also apply those generic clustering algorithms such as k-means to the variables. However, these clustering algorithms are mainly designed for clustering observations rather than variables and they cannot differentiate direct couplings of variables from indirect couplings.

### 3.1.1 Related Approach

Suppose that $x = (x_1, x_2, \ldots, x_L)$ follows a L-dimensional multivariate Gaussian distribution. For simplicity we assume $x \sim N(0, \Sigma)$ because we can always pre-processing data by subtracting their means. And let $\Omega = [\Omega_{ij}]_{L \times L} = \Sigma^{-1}$ be its precision matrix. It is easy to prove that $x_i$ and $x_j$ are conditionally independent given all other random variables if and only if $\Omega_{ij} = 0$. Therefore, estimating the structure of a Gaussian graphical model is equivalent to estimating the zero pattern in $\Omega$.

Banerjee et al. [24] and Yuan and Lin [25] independently proposed a technique that can estimate the sparse precision matrix. They achieved this by maximizing the L1 penalized log-likelihood, i.e.

$$\widehat{\Omega} = \arg \max_{\Omega > 0} \log \det(\Omega) - tr(\Omega \widehat{\Sigma}) - \lambda ||\Omega||_1,$$

where λ is the tuning parameter that decides the sparsity level of the graph, $||\Omega||_1 = \sum_{ij} |\Omega_{ij}|$ and $\widehat{\Sigma} = \frac{1}{n} X^T X$ is the empirical covariance matrix. The problem can then be solved by a block coordinate decent algorithm called graphical lasso [22].

However, not much work has been done for learning the block structure in a GGM. When the block structure information is not known a priori, all the existing studies employ a Bayesian approach partially because it is hard to design a penalty term to enforce the block structure without leading to a computationally intractable problem. An example of such work is by Marlin and Murphy [26], who proposed a Bayesian model that use a stochastic block model as prior and then use variational Bayes to do inference. Further, they employ a heuristic method to determine the number of clusters. This method starts by putting all the variables in a single cluster, ant then split clusters iteratively to increase free energy. After computing the marginal MAP clustering information, they use group LASSO [27] to infer the precision matrix.

In another two similar approaches to learn a block-structured GGM, Marlin et al. [28] and Ambroise et al. [29] use latent variables to indicate group membership and Laplace distributions as priors for the precision matrix entries. The group membership in formation is used to choose the hyperparameters of the prior distributions. An Expectation-Maximization (EM) algorithm and a variational algorithm are then used, respectively, to learn the structure and estimate the graph.

Another relevant method is Dirichlet process variable clustering (DPVS) proposed by Palla et al. [30]. This work considers the variable clustering problem in a factor model setting and uses nonparametric Bayesian methods to cluster the variables. Specifically, they consider the model where the L variables can be estimated as follows.

$$x_j = g_j y_{z_j} + \epsilon_j, j = 1, \dots L,$$

where $z_j$ is the membership of the j-th variable, $y_z$ is a Gaussian distributed latent factor for group z, $g_j$ is the factor loadings, and $\epsilon_j$ is a Gaussian noise. In fact, x generated by this model forms a block-structured Gaussian graphical model and thus can be viewed as a special case of the model to be presented below.

### 3.1.2 The Nonparametric Bayesian Approach

We consider the problem of clustering the variables of a Gaussian graphical model. Suppose that $\Omega$, the precision matrix of $x = (x_1, x_2, \dots, x_L)$ is block diagonal after some permutation [31]. This is equivalent to assuming that the variables can be grouped into several clusters, and that the edges in the underlying graph only exist within each cluster. The clustering structure can be relaxed to a more general setting where a relatively small number of edges exist between clusters or the inter-cluster edges carry much smaller weight. We now propose a nonparametric Bayesian approach to model such settings

**Model** Suppose that $z = (z_1, z_2, \ldots, z_L)$ are hidden variables indicating the membership of $x_1, x_2, \ldots, x_L$, i.e., the $x_i$ and $x_j$ are in the same cluster if and only if $z_i = z_j$. In fact, z defines a partition over the set $\{1, 2, \ldots, L\}$. We assume that $z_1, z_2, \ldots, z_L$ are generated by a Chinese restaurant process CRP($\alpha$) [32], where $\alpha$ is the concentration parameter, controlling how diverse the clustering tends to be. The Chinese restaurant process defines a distribution over random partitions of positive integers, with the possible number of clusters being infinite. Specifically, $z_1, z_2, \ldots, z_L$ are exchangeable and can be sampled sequentially by the following conditional probability.

$$P(Z_i|Z_{1:i-1}, \alpha) = \begin{cases} \dfrac{\sum_{j<i} 1_{z_j=z_i}}{i-1+\alpha}, & \exists j < i: z_j = z_i \\ \dfrac{\alpha}{i-1+\alpha}, & \forall j < i: z_j \neq z_i \end{cases}$$

Further, when only considering the first L elements, a specific partition $\rho = (z_1, z_2, \ldots, z_L)$ is assigned with the following probability:

$$P(\rho) = \frac{\alpha^{\#clusters}\Gamma(\alpha)}{\Gamma(n+\alpha)} \prod_{cluster \in \rho} \Gamma(\#cluster).$$

For a given clustering z, we assume that the precision matrix is from a Wishart distribution defined over symmetric positive semidefinite matrices. As a prior distribution for the precision matrix, the Wishart distribution is conjugate to the multivariate Gaussian likelihood. The density function of $\Omega \sim Wishart_L(V, v)$ is

$$P(\Omega|V, v) = \frac{|\Omega|^{\frac{v-p-1}{2}} \exp\{-\frac{1}{2} tr(V^{-1}\Omega)\}}{2^{\frac{vp}{2}} |V|^{\frac{p}{2}} \Gamma_p(\frac{v}{2})},$$

where $\Gamma_p(\cdot)$ is the multivariate Gamma function, V is known as the scale matrix and $v$ as the degree of freedom. The expectation of above Wishart distribution is $vV$. Here, to reflect our knowledge about the clustering pattern based on z, we set the scale matrix V to have a block diagonal structure. In particular, we let

$$V = V(z, W) = \begin{cases} \dfrac{W_{ij}}{v}, & if\ z_j = z_i \\ 0, & if\ z_j \neq z_i, \end{cases}$$

where W is a prior guess of the precision matrix and we scale it by a factor of $\dfrac{1}{v}$ such that the expectation of remaining entries will be the same as in W.

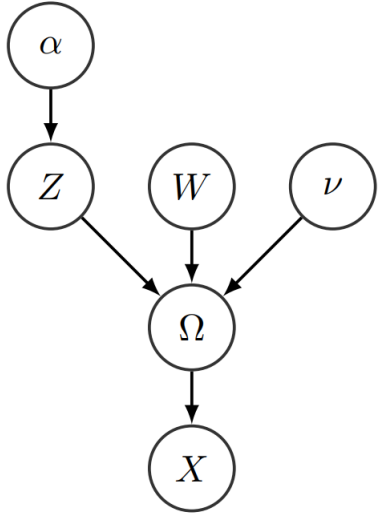

FIGURE 4 GRAPHICAL REPRESENTATION OF THE GENERATIVE MODE.

Thus, we have introduced a generative model to form a Gaussian graphical model with clustered variables. As shown in Fig. 4, our model can be summarized below.

$$Z|\alpha \sim CRP(\alpha)$$
$$\Omega|Z, W, v \sim Wishart_L(V(Z, W), v)$$
$$X|\Omega \sim N(0, \Omega^{-1}).$$

An alternative way to model block-structured GGMs is to assume that the precision matrix $\Omega$, given the clustering information Z, follows a block-wise Wishart distribution. Specifically, suppose that $z_1, z_2, \dots, z_L$ take values in $\{1, \dots, k\}$, and for z=1, …, k, let $I_z = \{i: z_i = z\}$ and $p_z = |I_z|$. Then we can assume the precision matrix $\Omega$ is from

$$\Omega_{I_z} \sim Wishart_{p_z}(V_z, v_z), for\ z = 1, \dots, k,$$
$$\Omega_{ij} = 0\ if\ z_i \neq z_j,$$

where $\Omega_{I_z}$ is the submatrix of $\Omega$ with indices $I_z$. In other words, the precision matrix is assumed to be block-structured, and each block is assumed to follow a Wishart distribution. Such an approach sets the off-block-diagonal entries of the precision matrix to exactly 0. In practice, this alternative approach also works for the case where weak dependence exists between clusters and performs similarly as the model we proposed above. Therefore, in this paper, we mainly discuss the model proposed first.

**Hyperparameter** There are three hyperparameters to be specified or tuned in the model, namely, $\alpha$, W and $\nu$. We discuss below our strategies of choosing them and the underlying reasons. The concentration parameter $\alpha$ of the Chinese restaurant process takes value in $(0, \infty)$. To improve the flexibility of the model, we can put a prior distribution on the hyperparameter $\alpha$, for which we use Gamma(1,1) throughout this paper. In fact, the inference results are similar with different choices of the priors as long as it has a support $(0, \infty)$.

The Wishart distribution of $\Omega$ is characterized by three parameters, z, W and $\nu$, where z is obtained from the Chinese restaurant process. Some methods such as empirical Bayes estimation [33] are proposed for the scale matrix without enforcing a block diagonal. We set W to the empirical precision matrix (i.e., $W = \widehat{\Omega}$), b which is a widely used method. For the case when p is smaller than n, we set W to be the GLASSO estimator with a small penalization parameter. From now on we will treat W as fixed, and denote V (z, W) as V (z).

For the degree of freedom $\nu$, a common choice, which is also the least informative one, would be to set $\nu = L$, the dimension of the matrix. In order to reflect our prior knowledge of the block structure, we set $\nu = \max\{L, n\}$ where n is the sample size. To see why this favors block diagonal structure of the precision matrix, consider the posterior distribution $P(\Omega|z, \nu, X)$ where X represents n i.i.d. samples. Because of the conjugacy, this is still a Wishart distribution, with expectation

$$\tilde{\Omega} = \left(\frac{\nu V(z)^{-1} + n\hat{\Omega}}{\nu + n}\right)^{-1},$$

where $\hat{\Sigma}$ is the sample covariance matrix. Notice that $V(z)^{-1}$ has a block diagonal structure, so the posterior mean somehow preserves the intra-cluster covariance structure while adding some shrinkage on the inter-cluster correlation. By choosing $\nu = \max\{L, n\}$, the shrinkage effect remains consistent for different L and n when n ≥ L. Besides, when L < n, such a choice will introduce more shrinkage on the off-block-diagonal entries, reflecting more strength from the prior knowledge of the block structure when we have insufficient data. Although there are some other sensible choices for the degree of freedom, such as putting a prior with a support on (L − 1, ∞), we choose $\nu = \max\{L, n\}$ throughout this paper, which turns out to work well for various settings regardless of L and n.

### 3.1.3 Inference

In this part, we describe the methods we have implemented to achieve variable clustering using the model introduced above. Specifically, given the data X, we would like to compute the posterior distribution of the latent variables, with special interest in the clustering information z. Note that for z this is a distribution over partitions of $\{1, 2, \ldots, L\}$. Although we can compute the posterior distribution P(z|X) with other variables integrated out analytically up to a normalization constant, the number of partitions on $\{1, 2, \ldots, L\}$ is known to be the Bell number, which grows faster than exponentially, hence making it computationally intractable to find the normalization constant and to directly sample from the posterior distribution.

**Gibbs Sampler** To explore the posterior distribution over the latent variables, we propose a Gibbs sampling method as follows. We update the elements of z one at a time. That is, we sample $z_i$ according to the conditional distribution P($z_i$ |X, Ω, $z_{-i}$, α) where $z_{-i}$ indicate all variables except $z_i$ . In particular,

$$P(z_i|X, \Omega, z_{-i}, \alpha) \propto P(X|\Omega)P(\Omega|z)p(z_i|z_{-i}, \alpha)$$

$$\propto \frac{|\Omega|^{\frac{v-L-1}{2}}}{|V(z)|^{\frac{v}{2}}} \exp(-\frac{1}{2}tr(V(z)^{-1}\Omega)P(z_i|z_{-i},\alpha), \quad (1)$$

where $p(z_i|z_{-i},\alpha)$ is given by

$$P(z_i = z|z_{-i},\alpha) = \begin{cases} \frac{p_{-i,z}}{L-1+\alpha}, & \exists j < i: z_j = z \\ \frac{\alpha}{L-1+\alpha}, & \forall j < i: z_j \neq z, \end{cases} \quad (2)$$

with $p_{-i,z}$ being the number of elements in cluster z excluding $z_i$, i.e., $p_{-i,z} = \sum_{j \neq i} 1_{z_j=z}$.

To update $\Omega$, we sample from $P(\Omega|X, z, \alpha)$ as follows:

$$\Omega|X, z, \alpha \sim Wishart_p((V(z)^{-1} + \sum_{i=1}^{n} X_i^T), \ n+v).$$

Alternatively, we can sample one element in z with $\Omega$ integrated out, i.e., using the following probability:

$$P(z_i|X, z_{-i}, \alpha) \propto P(X|z, \alpha)P(z_i|z_{-i}, \alpha)$$

$$= \int_\Omega P(X, \Omega|z, \alpha) d\Omega \cdot P(z_i|z_{-i}, \alpha)$$

$$\propto \int_\Omega P(X|\Omega)P(\Omega|z)d\Omega \cdot p(z_i|z_{-i}, \alpha)$$

$$\propto \frac{\Gamma_L(\frac{v+n}{2})}{\Gamma_L(\frac{v}{2})} \frac{|V(z)|^{\frac{n}{2}}}{|I_L + V\sum_{i=1}^{n} X_i X_i^T|^{\frac{v+n}{2}}}. \quad (3)$$

Since $z_i$ is discrete and the Wishart distribution is conjugate, it is easy to update z and $\Omega$ based on equation (1) and (2), or update z based on equation (3). We will use the latter one as our default Gibbs sampler.

To update the hyperparameter α, we compute

$$P(\alpha|X, z) \propto P(X|z)P(z|\alpha)P(\alpha)$$

$$\propto \frac{\alpha^{\#cluster(Z)}\Gamma(\alpha)}{\Gamma(L+\alpha)} P(\alpha).$$

This is a univariate distribution and we sample from it using slice sampling. [34].

With the conditional probability defined above, we have a Gibbs sampler for drawing samples from the posterior distribution of the latent variables z.

**Split-merge Metropolis-Hastings Updates** As mentioned in [35], the above-proposed Gibbs sampler may be inefficient. Because the Gibbs sampler updates the cluster membership incrementally, the Markov chain must pass through a series of low probability states to traverse between two isolated posterior modes. This leads to slow convergence and slow movement between two posterior modes. To tackle this limitation, we incorporate into our Gibbs sampler a split-merge Metropolis-Hastings procedure as proposed in [35] for the updating of the group membership Z. This split-merge Metropolis-Hastings procedure splits or merges the clusters using a restricted Gibbs sampling scan [35]. To exploit the major changes from the Metropolis-Hastings step, and the minor refinement from the Gibbs sampling step, we update z by alternating between the Gibbs sampler and the split-merge Metropolis-Hastings procedure. The whole procedure is summarized in Algorithm 2. See [35] for more details of the split-merge Metropolis-Hastings procedure.

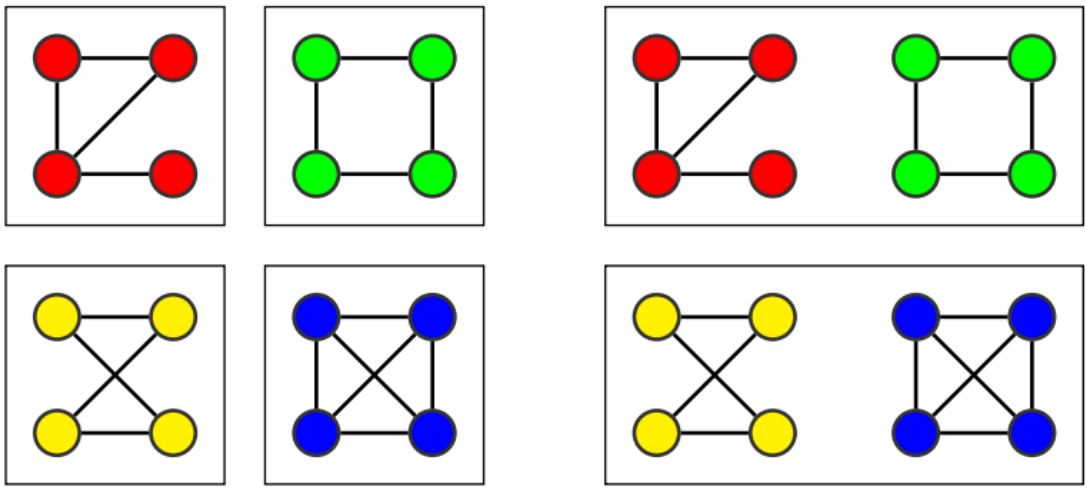

FIGURE 5 ILLUSTRATION OF DIFFERENT CLUSTERING RESULTS THAT BOTH MAKE SENSE.

When the data X is generated from a GGM with variables that can be clustered into disjoint groups, then the posterior distribution is very much likely to have multiple modes, corresponding to different clustering assignments. For example, the graphical model in figure 5 has 16 variables belonging to 4 groups, shown in 4 different colors. In this figure, the left part shows the most natural way of clustering the variables, while it also makes sense to cluster them in the way as shown on the right part of the figure. For this graphical model, there are 15 reasonable ways to cluster the 16 variables, which are expected to have much higher probabilities than all the others.

---

**Algorithm 1**

$\alpha^{(0)} \sim \text{Gamma}(1,1)$
$Z^{(0)} \sim \text{CRP}(\alpha^{(0)})$
**for** $m = 1$ to $M$ **do**
    **if** $m$ is odd **then**
        **for** $i = 1$ to $p$ **do**
            $Z_i^{(m)} \sim P(Z_i | X, Z_{-i}^{(m-1)}, \alpha^{(m-1)})$
        **end for**
    **else**
        Update $Z^{(m-1)}$ by split-merge MH procedure
    **end if**
    $\alpha^{(m)} \sim P(\alpha | X, Z^{(m)})$
**end for**

---

Most of the time, we are more interested in such reasonable clustering, especially the finest clustering, than in the posterior probability of one clustering. By the finest clustering, we mean that the one in which no cluster can be further divided into two disjoint sub-clusters (e.g. the clustering on the left in Fig. 5). This being said, rather than running the Markov chain for long enough until convergence, finding the posterior mode that corresponds to the finest clustering is good enough for our inference purpose. In practice, we start the Markov chain from a clustering that treats each variable as a single cluster and run the Algorithm 1 without split-merge procedure until it hits a local mode. We then report this state as our clustering of the variables. This method to some extent can be viewed as a greedy algorithm for finding the finest clusters, and we summarize it as Algorithm 2. Although greedy, as we shall see in the following section, it performs pretty well and efficiently on the synthetic data generated by both us and others as well as the real data.

---

**Algorithm 2**

$\alpha^{(0)} \sim \text{Gamma}(1,1)$
$Z^{(0)} = (1, 2, \ldots, p)$
**for** $m = 1$ to $M$ **do**
    **for** $i = 1$ to $p$ **do**
        $Z_i^{(m)} \sim P(Z_i | X, Z_{-i}^{(m-1)}, \alpha^{(m-1)})$
    **end for**
    **if** $\exists (z_1, \ldots, z_p)$ s.t. for $i = 1, \ldots, p$
$P(Z_i = z_i | X, Z_{-i}^{(m-1)}, \alpha^{(m-1)}) > 1 - \epsilon$ **then**
        **break**
    **end if**
    $\alpha^{(m)} \sim P(\alpha | X, Z^{(m)})$
**end for**
**output** $Z^{(m)}$

---

### 3.1.4 Experiments

**Synthetic Data** Here we present three experiments on synthetic data. The first experiment illustrates the relationship between posterior modes and clustering. The second one shows how well our method performs compared to some simple generic methods in a variety of settings. The third experiment evaluates our method using the synthetic data proposed in [30] and compares it with the method in [30].

**Modes and Clusterings** Suppose that our model consists of p variables of c clusters. To generate the data, we first assign each variable to one of the c clusters with probability 1/c. Then, we add an edge between two variables by probability Pin if they are in the same cluster or otherwise, by probability Pout. For each edge (i, j), we set $\Omega_{ij} = 0.3$.

Finally, to make sure that the precision matrix is positive definite, we set its diagonal element to the absolute value of the minimum eigenvalue of the current Ω plus 0.2.

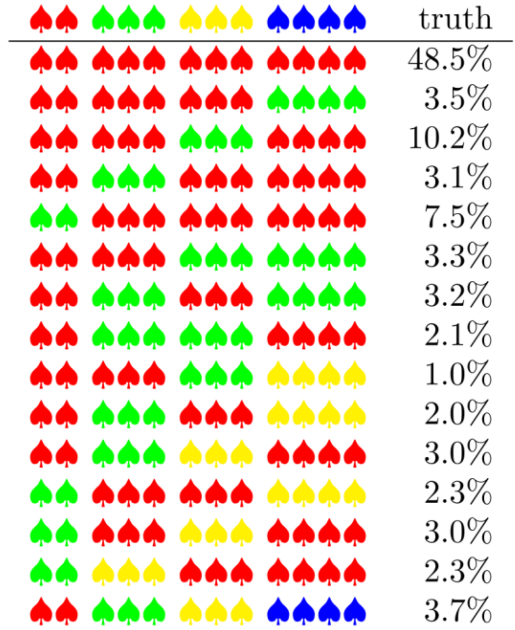

FIGURE 6  FREQUENCY OF GETTING TRAPPED AT THE POSTERIOR MODES. THE FIRST ROW REPRESENTS THE TRUE CLUSTERING ACCORDING TO WHICH WE GENERATE OUR DATA. DIFFERENT COLORS INDICATE DIFFERENT CLUSTERS.

We show a simple example to illustrate that the posterior modes correspond to all reasonable clusterings. Using the above-mentioned data generation method, we construct a Gaussian graphical model (GGM) with p = 12 variables and c = 4 clusters with sizes 2, 3, 3, and 4. We set $P_{in} = 1$ and $P_{out} = 0$, so the GGM has 4 fully connected components without any intercomponent edges. Then we generate n = 50 i.i.d. samples from this GGM. We run the Gibbs sampler for 1000 times starting from different starting points of (α, Z) drawn from their prior distributions. At each time we run the Gibbs sampler until it gets trapped at one mode of the posterior distribution, i.e., when the Markov chain has a very small chance (say, < 0.001) to traverse to another state. For all the 1000 simulations, the Markov chain always reaches one of the 15 partitions listed in figure 6, which also lists the frequency the Markov chain dwelling in each mode. The 15 modes are exactly all the possible combinations of the 4 true clusters, showing that the posterior modes and reasonable clusterings are closely related.

**Finding the Finest Clustering** Now we consider an example where we are interested in recovering the finest clustering. We generate the synthetic data using a GGM with p = 50, Pin = 1 and Pout = 0. We vary the experiment settings with different number of sample and number of clusters to test our method. For comparison, we have also implemented the spectral clustering [13] method. To use spectral clustering, we employ three different similarity measures to define the relationship between variables: the empirical covariance matrix calculated from the sample data, the empirical precision matrix and the precision matrix generated by GLASSO. Starting from the spectrum of these matrices, we perform dimensionality reduction and then use k-means to cluster the variables in the transformed space.

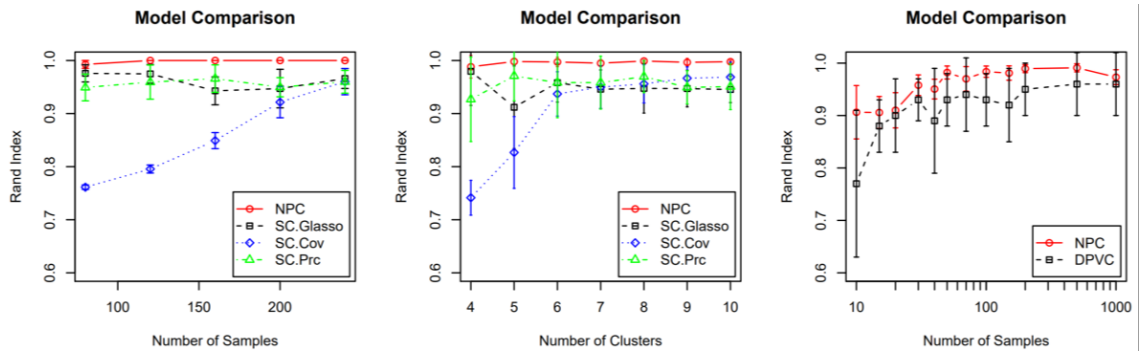

FIGURE 7 PERFORMANCE COMPARISON OF OUR METHOD NPC AND THE OTHERS IN TERMS OF THE AVERAGE RAND INDEX. FROM LEFT TO RIGHT, (A) P = 50, N = 100, AND THE NUMBER OF CLUSTERS RANGING FROM 4 TO 10; (B) P = 50, C = 6, AND THE NUMBER OF SAMPLES RANGING FROM 40 TO 240; (C) THE DATA IS GENERATED ACCORDING TO [30], WITH P = 20, C = 5, AND THE NUMBER OF SAMPLES RANGING FROM 10 TO 1000. SC.GLASSO, SC.COV AND SC.PRC STAND FOR SPECTRAL CLUSTERING WITH THREE DIFFERENT SIMILARITY MATRICES, AND DPVC FOR THE METHOD IN [30].

First, we set the number of clusters to 6, and then vary the number of samples from 80 to 240. For each set of samples, we conduct 10 independent simulations and compute the average Rand index, which is a widely used measure for clustering similarity. Rand index ranges from 0 to 1, with 1 indicating the perfect match. As shown in Fig. 3(a), our method outperforms spectral clustering regardless of the number of samples, while the accuracy for both methods improves as more samples are used. Note that spectral clustering requires a predefined value for the number of clusters, for which we use 4, the ground truth here. Second, we fix the number of samples to 100 and vary the number of clusters from 4 to 10. Spectral clustering is always fed with the true number of clusters as the parameter. As shown in Fig. 7(b), our method still has higher accuracy than spectral clustering in all the experiments, showing that our nonparametric Bayesian method can find the right number of clusters automatically

**Comparison with a Factor Model**  As mentioned before, Palla et al. [30] studies variable clustering in a different setting. Although their model is different from ours, the covariance structure is also a block diagonal one. Using the data generation method described in Palla et al's paper, we generate a set of synthetic data with p = 20 dimensions and c = 5 equally sized clusters (of 4 variables). For each cluster we sample $Y_{iz} \sim N(0, 1)$ for i = 1, . . . , n and z = 1, . . . , c, then $g_j \sim N(0, 1)$ for j = 1, . . . , p and finally sample $X_{ij} \sim N(g_j Y_{iz}, 0.1)$ for i and j where $z_j$ denotes the cluster of the j-th variable. We generate the test data sets with n, the number of samples, varying from 10 to 1000 and repeat 10 times for each n. As shown in Fig. 7(c), except for some small n, our method always has higher accuracy than the DPVC method proposed in [30].

**Stock Real Data** To test the performance of our method on a real data set, we apply our method to an equity dataset in the "huge" package [36], which consists of 1245 daily closing prices from January 1, 2003 to January 1, 2008 for 452 equities in the S&P 500 index. The stocks are divided into ten sectors including health care, utilities, energy, consumer staples, materials, telecommunications, industrials, consumer discretionary, and financials. Each sector has 6 to 70 stocks. Stocks in the same sector are expected to be more correlated with each other, and therefore tend to form a cluster. We run our method to cluster these stocks based upon their closing prices. We obtain 26 clusters with size larger than 2, in total covering 413 stocks. Compared to the crude manual 10-sector classification, our clustering is more fine-grained. As shown in Fig. 4, each stock is colored according to its true sector classification. Many clusters generated by our method consist of stocks sharing the same color. Our algorithm identifies 7 sectors with very little misclassification. Further examination shows that our clustering result is not only consistent with the true sector classification, but can also provide finer-grain classification. For example, our method divides the financial sector (in pink) into five small clusters, corresponding to five sub-sectors: property & casualty insurance, real estate investment trust, banks, diversified financial service, and other insurance companies. Our method also clusters some stocks of different sectors into the same group. For example, one of our clusters contains stock in both the materials and industrials sectors. This is not due to bad clustering. Instead it is because some stocks indeed belong to two different sectors. For example, many stocks in in the industry sector belong to industrial materials or industrial conglomerates.

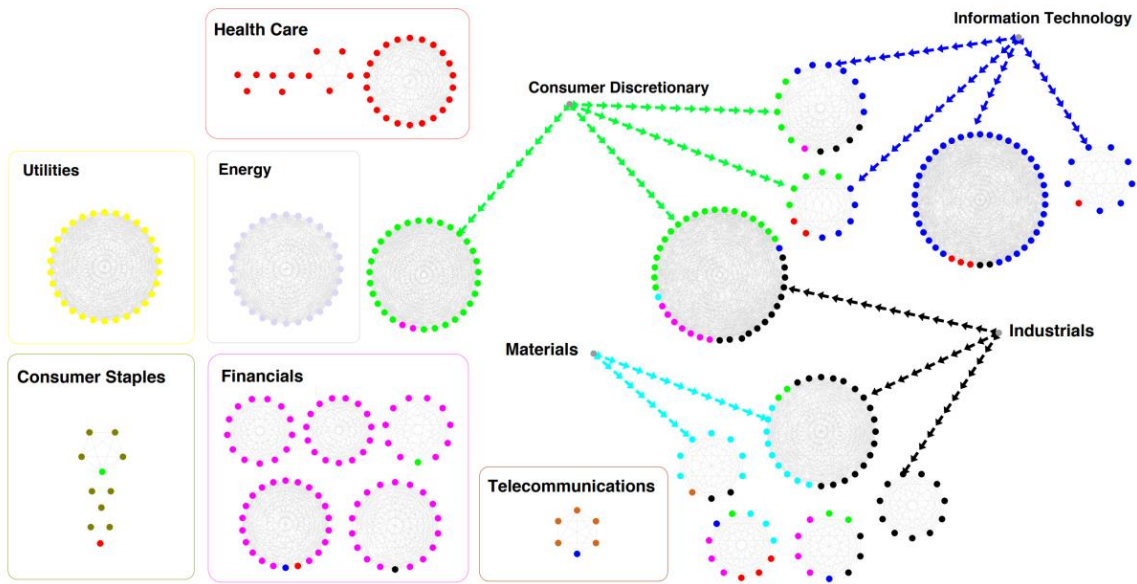

In addition, our clustering result is very stable and also accurate in terms of the Rand index. Running our method 100 times starting from different initial clustering, the mean and the standard deviation of the Rand Index are 0.89 and 0.007, respectively.

For comparison, we have also implemented the spectral clustering using the precision matrix estimated by GLASSO as the similarity measure. This reflects the basic idea of clustering the variables based on the estimated graph. This procedure requires specifying two parameters, namely, the number of clusters and the penalty parameter for GLASSO. Among numerous trials with the number of clusters ranging from 10 to 30 and different levels of sparsity of the estimated graph, the clustering results vary substantially. The Rand index ranges from 0.17 to 0.88, which are obtained with $K = 10$ and an estimated graph of 5074 edges, and $K = 29$ and a graph of 8600 edges, respectively. This comparison clearly shows the advantage of our method: parameter-free and self-adaptive to the data.

**Contact Data** Due to limitations of speed, we sampled proteins from CASP and CAMEO, which had a combined total of 182 proteins. After setting exclusion criteria such that the maximal length of the sequence was less than 100, there were about 40 proteins left for our subset. Since PISCOV assumes the multiple sequence alignment, it fits the Gaussian Graphical Model assumption well. We compared our model with PSICOV, which provides no clustering information. Please see Figure 9(a) for a detailed one-vs-one comparison. Overall, our method had a 0.088 top L/5 long-range prediction accuracy, while PSICOV had an accuracy of only 0.069. Neither method performed well with this dataset because there were only a few homologs for some of the proteins, failing the assumption of EC analysis.

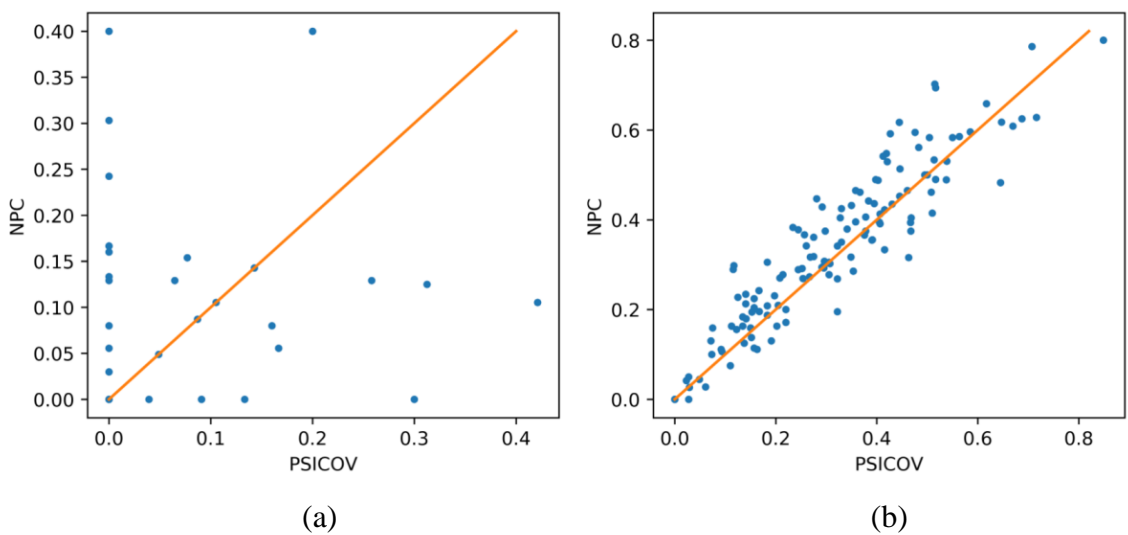

FIGURE 9 ONE VS ONE L/5 RESULT COMPARISON BETWEEN PSICOV AND ADAPTIVE CLUSTERING

To test the algorithm's performance more accurately, we further run both algorithms on a test set with more homologs. More detailed, MSA were generated for proteins with known structure and more than 1000 sequences from Pfam families. Similarly, only sequences with less than or equal to 100 AAs were kept, and it resulted in a set of 140 proteins. Overall, the proposed method achieves 0.469 top L/5 long-range prediction accuracy, which outperforms PSICOV's 0.441 accuracy. Please refer to figure 9(b) for detailed one-vs-one comparison, and note that both methods generated more reasonable results compare to CASP+CAMEO test set because ECA methods requires a large number of homologs.

We further investigate about why the proposed NPC approach outperforms corresponding PSICOV baseline by analyzing the results for the 10 proteins where the difference between NPC and PSICOV is maximum. We find that 6 out of 10 are mainly beta proteins[2], which is significantly higher than the proportion of mainly beta proteins in PDB (~ 24%). For a more detailed comparison, the predicted contact map of PSICOV and NPC, as well as the ground truth for 3PE9 are visualized in figure 10. By comparing the predictions within the green and yellow boxes in figure 10, it shows that NPC predicts a contact map more accurately by adding cluster constraints to the model if such clustering structures exist in the ground truth contact map. More specifically, NPC predicts 17 and 10 contacts correctly in green and yellow box, respectively, while PSICOV only predicts 11 and 6 correctly.

---

[2] The list of 10 proteins are **4AVR**, **3PE9**, 1QZM, **3LAG**, 2DYJ, **3VDJ**, **1B75**, 3WCQ, **4F2E** and 1VMG, where mainly beta proteins are in **bold**.

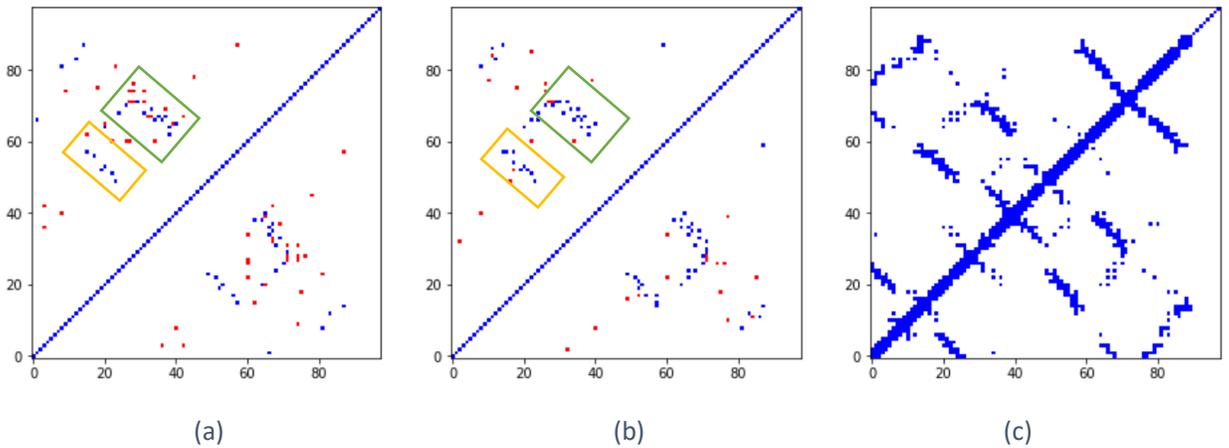

FIGURE 10 (A) PREDICTED CONTACT MAP OF PSICOV FOR 3PE9 (B) PREDICTED CONTACT MAP OF NPC FOR 3PE9. (C) 3PE9'S GROUND TRUTH CONTACT MAP.

## 3.2 Learning Structures by Infinite Dimensional Exponential Families

For parametric setting, a lot of work has been done other than GGMs [37-39]. In the context of exponential family graphical models, where the node conditional distribution given all the other nodes is a member of an exponential family, the structure is described by the non-zero coefficients [40]. Most existing approaches to learn the structure of a high-dimensional undirected graphical model are based on minimizing a penalized loss objective, where the loss is usually a log-likelihood or a composite likelihood and the penalty induces sparsity on the resulting parameter vector. See, for example [22], [41], [42], [43], [40] for more details. PlmDCA or CCMpred also use this exponential family assumption and similar objective functions. In this section, however, we focus on learning the structure of a pairwise graphical models without assuming a parametric class of models. The main challenge in estimating nonparametric graphical models is computation of the log normalizing constant. To get around this problem, we propose to use score matching [44, 45] as a divergence, instead of the usual KL divergence, as it does not require evaluation of the log partition function. The probability density function is estimated by minimizing the expected distance between the model score function and the data score function, where the score function is defined as gradient of the corresponding probability density functions. The advantage of this measure is that the normalization constant is canceled out when computing the distance. In order to learn the underlying graph structure, we assume that the logarithm of the density is additive in node-wise and edge-wise potentials and use a sparsity inducing penalty to select non-zero edge potentials. As we will prove later, our procedure will allow us to consistently estimate the underlying graph structure.

### 3.2.1 Notations

Let [n] denote the set $\{1, 2, \ldots, n\}$. For a vector $\theta = (\theta_1, \theta_2, \ldots, \theta_d)^T \in R^d$, let $||\theta||_p = \left(\sum_{i \in [d]} |\theta_i|^p\right)^{\frac{1}{p}}$ denote its $l_p$ norm. Let column vector vec(D) denote the vectorization of matrix D, cat(a, b) denote the concatenation of two vectors a and b, and $mat(a_1^T, \ldots, a_d^T)$ the matrix with rows give by $a_1^T, \ldots, a_d^T$. For $\chi \subseteq R^d$, let $L^p(\chi, p_0)$ denote the space of function for which the p-th power of absolute value is $p_0$ integrable ;and for $f \in L^p(\chi, p_0)$, let $||f||_{L^p(\chi, p_0)} = ||f||_p = \left(\int_\chi |f|^p dx\right)^{\frac{1}{p}}$ denote its $L^p$ norm. Throughout the paper, we denote H as Hilbert space and $<\cdot,\cdot>_H, ||\cdot||_H$ as corresponding inner product and norm.

For any operator $C: H_1 \to H_2$, we use ||C|| denote the usual operator norm, which is define as

$$||C|| = \inf\left\{a \geq 0: ||Cf||_{H_2} \leq a||f||_{H_1} \text{ for all } f \in H_1\right\};$$

and $||C||_{HS}$ to denote its Hilbert-Schmidt norm, which is defined as

$$||C||_{HS}^2 = \sum_{i \in I} ||Ce_i||_{H_2}^2,$$

where $e_i$ is an orthonormal basis of H for an index set I. Also, we use R(C) to denote operator C's range space. For any $f \in H_1$ and $g \in H_2$, let $f \otimes g$ denote their tensor product.

### 3.2.2 Background

Besides GGMs, one way to estimate structure is the pseudo-likelihood method. It estimates the neighborhood of a node a by the non-zeros of the solution to a regularized linear model,

$$\hat{\theta}_s = \arg\min_\theta \frac{1}{n}||x_s - x_{-s}\theta||_2^2 + \lambda||\theta||_1 \quad (4).$$

The estimated neighborhood is then $\widehat{N}(s) = \{a: \theta_{sa} \neq 0\}$.

Another way to specify a parametric graphical model is by assuming that each node-conditional distributions is a part of the exponential family [40]. Specifically, the conditional distribution of $x_s$ given $x_{-s}$ is assumed to be

$$P(x_s|x_{-s}) = \exp\left(\sum_{t \in N(s)} \theta_{st} x_s x_t + C(x_s) - D(x_{-s}, \theta)\right) \quad (5),$$

where C is the base measure, D is the e log-normalization constant and N(s) is the neighborhood a the node s. Similar to (4), the neighborhood of each node can be estimated by minimizing the negative log-likelihood with $l_1$ penalty on θ. The optimization is tractable when the normalization constant D can be easily computed based on the model assumption. For example, under Poisson graphical model assumptions for count data, the normalization constant is $-\exp(\sum_{t \in N(s)} \theta_{st} x\_t)$. When using the neighborhood estimation, the graph can be estimated as the union of the neighborhoods of each node, which leads to consistent graph estimation [40, 46].

**Generalized Exponential Family and RKHS** We say H is a reproducing kernel Hilbert space (RKHS) associate with kernel $k: \chi \times \chi \to R_+$ if and only if for each $x \in \chi$, the following two conditions are satisfied:
  (1) $k(\cdot, x) \in H$ and

(2) it has reproducing properties such that $f(x) = <f, k(\cdot, x)>_H$ for all $f(\cdot) \in H$, where k is a symmetric and positive semidefinite function.

Denote the RKHS H with kernel k as $H(k)$. For any $f \in H(k)$, there exists a set of $x_i$ and $\alpha_i$, such that $f(\cdot) = \sum_{i=1}^{\infty} \alpha_i k(\cdot, x_i)$. Similarly for any $g \in H(k), g(\cdot) = \sum_{j=1}^{\infty} \beta_j k(\cdot, y_j)$, the inner product of f and g is defined as $<f, g>_H = \sum_{i,j=1}^{\infty} \alpha_i \beta_k k(x_i, y_j)$. Therefore the norm of f simply is $||f||_H = \sqrt{\sum_{i,j} \alpha_i \alpha_j k(x_i, x_j)}$. The summation is guaranteed to be larger than or equal to zero because the kernel k is positive semidefinite.

We consider the exponential family in infinite dimensions[47], where
$$P = \{p_f(x) = e^{f(x) - A(f)} q_0(x), x \in \chi; f \in F\}$$
and the function space F is defined as
$$F = \{f \in H(k): A(f) = \log \int_\chi e^{f(x)} q_0(x) dx < \infty\},$$
where $q_0(x)$ is the base measure, A(f) is a generalized normalization constant such that $p_f(x)$ is a valid probability density function. To see it as a generalization of the exponential family, we show some examples that can generate useful finite dimension exponential families:

- Normal: $\chi = R, k(x, y) = xy + x^2 y^2$
- Poisson: $\chi = N, k(x, y) = xy$
- Exponential: $\chi = R_+, k(x, y) = xy$.

For more details, please refer to [47].

When learning structure of a graphical model, we will further impose structural conditions on H(k) in order ensure that F consists of additive functions.

**Score Matching** Score matching is a convenient procedure that allows for estimating a probability density without computing the normalizing constant [44, 45] . It is based on minimizing Fisher divergence

$$J(p||p_0) = \frac{1}{2} \int p(x) || \frac{\partial \log p(x)}{\partial x} - \frac{\partial \log p_0(x)}{\partial x} ||_2^2 \, dx,$$

where $\frac{\partial \log p(x)}{\partial x} = (\frac{\partial \log p(x)}{\partial x_1}, ..., \frac{\partial \log p(x)}{\partial x_d})$ is the score function. Observe that for $p(x, \theta) = \frac{1}{Z(\theta)} q(x, \theta)$ the normalization constant $Z(\theta)$ cancels out in the gradient computation, which makes the divergence independent of Z(θ). Since the score matching objective involves the unknown oracle probability density function $p_0$, it is typically not computable. However, under some mild conditions which we will discuss in method section, the above score matching definition can be rewritten as

$$J(p||p_0) = \int p_0(x) \sum_{i \in [d]} \frac{1}{2} \left( \frac{\partial \log p(x)}{\partial x_i} \right)^2 + \frac{\partial^2 \log p(x)}{\partial x_i^2} \, dx.$$

After substituting the expectation with an empirical average, we get

$$\hat{J}(p||p_0) = \frac{1}{n} \sum_{a \in [n]} \sum_{i \in [d]} \frac{1}{2} \left( \frac{\partial \log p(X_a)}{\partial x_i} \right)^2 + \frac{\partial^2 \log p(X_a)}{\partial x_i^2} \quad (6),$$

Compared to maximum likelihood estimation, minimizing $\hat{J}(p||p_0)$ is computationally tractable. While we will be able to estimate $p_0$ only up to a scale factor, this will be sufficient for the purpose of graph structure estimation.

### 3.2.3 Methods

**Model Formulation and Assumptions** We assume that the true probability density function $p_0$ in P. Furthermore, for simplicity we assume that

$$\log p_0(x) = f(x) = \sum_{i \leq j, (i,j) \in S} f_{0,ij}(x_i, x_j),$$

where $f_{0,ii}(x_i, x_i)$ is a node potential and $f_{0,ij}(x_i, x_j)$ is an edge potential. The set S denotes the edge set of the graph. Extensions to models where potentials are defined over larger cliques are possible. We further assume that $f_{0,ij} \in H_{ij}(k_{ij})$, where $H_{ij}$ is a RKHS with kernel $k_{ij}$. To simplify the notation, we use $f_{0,ij}(x)$ or $k_{ij}(\cdot, x)$ to denote $f_{0,ij}(x_i, x_j)$ and $k_{ij}(\cdot, (x_i, x_j))$. If the context is clear, we drop the subscript for norm or inner product. Define

$$H(S) = \{f = \sum_{i,j \in S} f_{ij} | f_{ij} \in H_{ij}\}$$

as a set of functions that decompose as sum of bivariate functions on edge set S. Note that H(S) is also (a subset of) a RKHS with the norm $||f||^2_{H(S)} = \sum_{i,j \in S} ||f_{ij}||^2_{H_{ij}}$ and kernel $k = \sum_{i,j \in S} k_{ij}$.

Let $\Omega(f) = ||f||_{H,1} = \sum_{i \leq j} ||f_{ij}||_{H_{ij}}$. For any edge set S (not necessarily the true edge set), we denote $\Omega_S(f_S) = \sum_{s \in S} ||f_s||_{H_s}$ as the norm $\Omega$ reduced to S. Similarly, denote its dual norm as $\Omega_S^* = \max_{\Omega_S(g_S) \leq 1} <f_S, g_S>$.

Under the assumption that the unknown $f_0$ is additive, the loss function becomes

$$J(f) = \frac{1}{2} \int p_0(x) \sum_{i \in [d]} \left(\frac{\partial f(x)}{\partial x} - \frac{\partial f_{0(x)}}{\partial x_0}\right)^2 dx$$

$$= \frac{1}{2} \sum_{i \in [d]} \sum_{j,j' \in [d]} <f_{ij}, -f_{0,ij} \int p_0(x) \frac{\partial k_{ij}(\cdot, (x_i, x_j))}{\partial x_i} \otimes \frac{\partial k_{ij'}(\cdot, (x_i, x_{j'}))}{\partial x_i} dx (f_{ij'} - f_{0,ij'})>$$

$$= \frac{1}{2} \sum_{i \in [d]} \sum_{j,j' \in [d]} <f_{ij} - f_{0,ij}, C_{ijij'}(f_{ij'} - f_{0,ij'})>.$$

Intuitively, C can be viewed as a $d^2$ matrix, and the operator at position (ij, ij') is $C_{ij,ij'}$. For general (ij, i'j'), $i \neq i'$, the corresponding operator simply is 0. Define $C_{SS'}$ as

$$\int p_{0(x)} \sum_{ij \in S, (i',j') \in S'} \frac{\partial k_{ij}(\cdot, (x_i, x_j))}{\partial x_i} \otimes \frac{\partial k_{i'j'}(\cdot, (x_{i'}, x_{j'}))}{\partial x_i} dx,$$

which intuitively can be treated as a sub matrix of C with rows S and columns S'. We will use this notation intensively in the main theorem and its proof.

Following [48], we make the following assumptions.

A1. Each $k_{ij}$ is twice differentiable on $\chi \times \chi$

A2. For any i and $\tilde{x} \in \chi_j = [a_j, b_j]$, we assume that

$$\lim_{x_i \to a_i^+ \text{ or } b_i^-} \frac{\partial^2 k_{ij}(x,y)}{\partial x_i \partial y_i}\Big|_{y=x} p_0^2(x) = 0,$$

where $x = (x_i, \tilde{x}_j)$ and $a_i, b_i$ could be $-\infty$ or $+\infty$.

A3. This condition ensures that $J(p||p_0) < \infty$ for any $p \in P$ [for more details see [48]]:

$$\left\| \frac{\partial k_{ij}(\cdot, x)}{\partial x_i} \right\|_{H_{ij}} \in L^2(\chi, p_0), \left\| \frac{\partial^2 k_{ij}(\cdot, x)}{\partial x_i^2} \right\|_{H_{ij}} \in L^2(\chi, p_0).$$

A4. The operator $C_{SS}$ is compact and the smallest eigenvalue $\omega_{min} = \lambda_{min}(C_{SS}) > 0$.

A5. $\Omega_{S^c}^*[C_{S^cS}C_{SS}^{-1}] \leq 1 - \eta$, where $\eta > 0$

A6. $f_0 \in R(C)$, which means there exists $\gamma \in H$, such that $f_0 = C\gamma$, where $f_0$ is the oracle function.

We will discuss the definition of operator C and $\Omega$ in next section. Compared with [40], A4 can be interpreted as the dependency condition and the A5 is the incoherence condition, which is a standard condition for structure learning in high dimensional statistical estimators.

**Estimation Procedure** We estimate f by minimizing the following penalized score matching objective

$$\min_f \hat{L}_\mu(f) = \hat{J}(f) + \frac{\mu}{2} ||f||_{H,1}$$

$$s.t. f_{ij} \in H_{ij},$$

where $\hat{J}(f)$ is given in (6). The norm $||f||_{H,1} = \sum_{i \leq j} ||f_{ij}||_{H_{ij}}$ is used as a sparsity inducing penalty. A simplified form of $\hat{J}(f)$ is given below that will lead to efficient algorithm for solving above optimization problem.

The following theorem states that the score matching objective can be written as a penalized quadratic function on f.

**Theorem 3.1.**

(i) The score matching objective can be represented as

$$L_\mu(f) = \frac{1}{2}<f-f_0, C(f-f_0)> + \frac{\mu}{2}||f||_{H,1},$$

Where $C = \int p_0(x) \sum_{i\in[d]} \frac{\partial k(\cdot,x)}{\partial x_i} \otimes \frac{\partial k(\cdot,x)}{\partial x_i} dx$ is a trace operator.

(ii) Give observed data $X_{n\times d}$, the empirical estimator of $L_\mu$ is

$$\hat{L}_\mu(f) = \frac{1}{2}<f, \hat{C}f> + \sum_{i\leq j} <f_{ij}, -\hat{\xi}_{ij}> + \frac{\mu}{2}||f||_{H,1} + const \quad (7),$$

where $\hat{C} = \frac{1}{n}\sum_{a\in[n]}\sum_{i\in[d]} \frac{\partial k(\cdot,X_a)}{\partial x_i} \otimes \frac{\partial k(\cdot,X_a)}{\partial x_i}$, and $\hat{\xi}_{ij} = \frac{1}{n}\sum_{a\in[n]} \frac{\partial^2 k_{ij}(\cdot,(X_{ai},X_{aj}))}{\partial x_i^2} + \frac{\partial^2 k_{ij}(\cdot,(X_{ai},X_{aj}))}{\partial x_j^2}$.

Theorem 3.2 (i) The solution to (7) can be represented as

$$\widehat{f_{ij}} = \sum_{b\in[n]} \beta_{bij} \frac{\partial k_{ij}(\cdot,(X_{bi},X_{bj}))}{\partial x_i} + \beta_{bji} \frac{\partial k_{ij}(\cdot,(X_{bi},X_{bj}))}{\partial x_j} + \alpha_{ij}\hat{\xi}_{\_ij}$$

where $i \leq j$.

(ii) Minimizing (7) is equivalent to minimizing the following quadratic function

$$\frac{1}{2n}\sum_{ai}\left(\sum_{bj}(\beta_{bij}G^{ab}_{ij11} + \beta_{bji}G^{ab}_{ij12}) + \sum_j \alpha_{ij}h^{1a}_{ij}\right)^2$$

$$+ \sum_{i\leq j}(\sum_b (\beta_{bij}h^{1b}_{ij} + \beta_{bji}h^{2b}_{ij}) + \alpha_{ij}\left|\left|\hat{\xi}\right|\right|^2_{ij}) + \frac{\mu}{2}||f||^1_H$$

$$= \frac{1}{2n}\sum_{ai}(D^T_{ai}\theta)^2 + E^T\theta + \frac{\mu}{2}\sum_{i\leq j}\sqrt{\theta^T_{ij}F_{ij}\theta_{ij}}$$

where $G^{ab}_{ijrs} = \frac{\partial^2 k_{ij}(X_a,X_b)}{\partial x_r \partial y_s}, h^{rb}_{ij} =< \frac{\partial k_{ij}(\cdot,X_b)}{\partial x_r}, \hat{\xi}_{ij} >$ are constant that only depends on X. $\theta = cat(vec(\alpha), vec(\beta))$ is the vector parameter and $\theta_{ij} = cat(\alpha_{ij}, vec(\beta_{\cdot ij}))$ is a group of parameters. $D_{ai}, E$ and $F$ are corresponding constant vectors and matrices based on G, h and the order of parameters. Then the above problem can be solved by group lasso.

Let $\hat{f}^\mu = \arg\min_{f \in H} \hat{L}_\mu(f)$ denote the solution, then we can estimate the graph as follows:

$$\hat{S}_\mu = \{(i,j): ||\hat{f}_{ij}^\mu|| \neq 0\}.$$

That is, the graph is encoded in the sparsity pattern of $\hat{f}_\mu$.

Next, we study statistical properties of the proposed estimator. Let S denote the true edge set and $S^C$ its complement. We prove that $\hat{S}_\mu$ recovers S with high probability when the sample size n is sufficiently large. Denote $D = mat(D_{11}^T, \ldots, D_{ai}^T, \ldots, D_{nd}^T)$. We will need the following result on the estimated operator $\hat{C}$,

Proposition 3.1 (Lemma 5 in [48])

1. $||\hat{C} - C||_{HS} = O_{p_0}(n^{-\frac{1}{2}})$
2. $||(C + \mu L)^{-1}|| \leq \frac{1}{\mu \min diag(L)}, ||C(C + \mu L)^{-1}|| \leq 1, where\ \mu > 0$ and L is diagonal with positive constants.

With these preliminary results, we have the following main theorems.

Theorem 3.3 Assume that conditions A1 to A7 are satisfied. The regularization parameter $\mu$ is selected at the order of $n^{-\frac{1}{4}}$ and satisfied $\mu \leq \frac{\eta \kappa_{min} \omega_{min}}{4(1-\eta)\kappa_{max}\sqrt{|S|}+\frac{\eta}{5}}$, where $\kappa_{min} = \min_{s \in S} ||f_s^*|| > 0$, and $\kappa_{max} = \max_{s \in S} ||f_s^*|| > 0$. Then $P(\hat{S}_\mu = S) \to 1$.

### 3.2.4 Results

**Synthetic Data** We illustrate performance of our method on two simulations. In our experiments, we use the same kernel defined as follows:

$$k(x,y) = \exp\left(-\frac{||x-y||_2^2}{2\sigma^2}\right) + r(x^T y + c)^2,$$

that is, the summation of a Gaussian kernel and a polynomial kernel. We set $\sigma^2 = 1.5$, r = 0.1 and c = 0.5 for all the simulations.

We report the true positive rate vs false positive rate (ROC) curve to measure the performance of different procedures. Let S be the true edge set, and let $\hat{S}_\mu$ be the estimated graph. The true positive rate is defined as $TPR_\mu = \frac{|S=1 \text{ and } \hat{S}_\mu=1|}{|S=1|}$, and false positive rate is $FPR_\mu = \frac{|\hat{S}^\mu=1 \text{ and } S=0|}{|S=0|}$, where $|\cdot|$ is the cardinality of the set. The curve is then plotted based on 100 uniformly sampled regularization parameters and based on 20 independent runs.

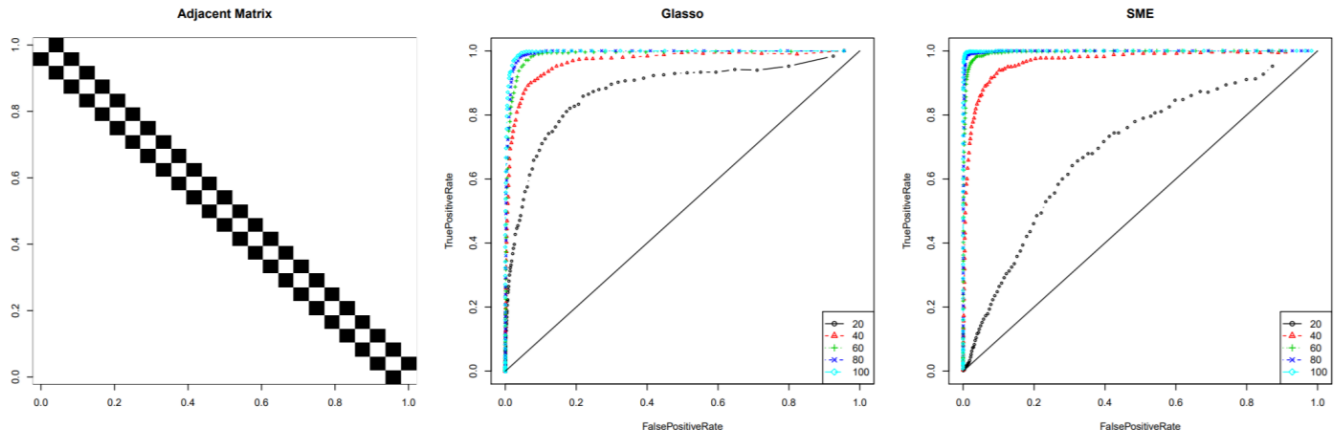

FIGURE 11 THE ESTIMATION RESULTS FOR GAUSSIAN GRAPHICAL MODELS. LEFT: THE ADJACENT MATRIX OF TRUE GRAPH. CENTER: THE ROC CURVE OF GLASSO. RIGHT: THE ROC CURVE OF SCORE MATCHING ESTIMATOR (SME)

In the first simulation, we apply our algorithm to data sampled from a simple chain graph-based Gaussian model (see figure 10 for details), and compare its performance with GLASSO. We use the same sampling method as in [36] to generate the data: we set $\Omega s = 0.4$ for s ∈ S and its diagonal to a constant such that $\Omega$ is positive definite. We set the dimension d to 25 and change the sample size n ∈ {20, 40, 60, 80, 100} data points.

Except for the low sample size case (n = 20), the performance of our method is comparable with GLASSO, without utilizing the fact that the underlying distribution is of a particular parametric form. Intuitively, to capture the graph structure, the proposed nonparametric method requires more data because of much weaker assumptions.

To further show the strength of our algorithm, we test it on a nonparanormal (NPN) distribution [42]. A random vector $x = (x_1, \ldots, x_d)$ has a nonparanormal distribution if there exist functions $(f_1, \ldots, f_d)$ such that $(f_1(x_1), \ldots, f_d(x_d)) \sim N(\mu, \Sigma)$. When f is monotone and differentiable, the probability density function is given by

$$P(x) = \frac{1}{(2\pi)^{\frac{p}{2}}|\Sigma|^{\frac{1}{2}}} \exp\left(-\frac{1}{2}(f(x) - \mu)^T \Sigma^{-1}(f(x) - \mu)\right) \prod_j |f_j'|.$$

Here the graph structure is still encoded in the sparsity pattern of $\Omega = \Sigma^{-1}$, that is, $x_i \perp x_j | x_{-i,j}$ if and if $\Omega_{ij} = 0$ [42].

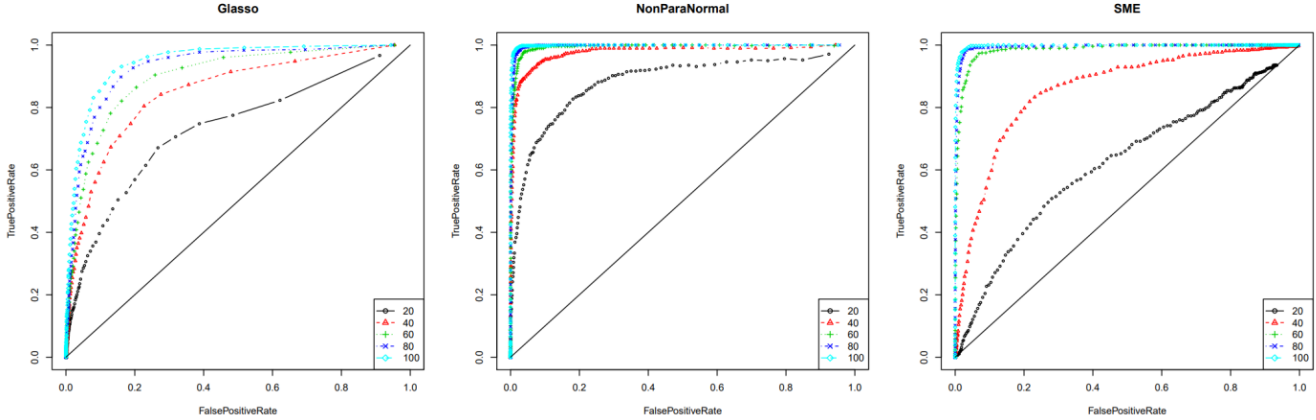

FIGURE 12 THE ESTIMATED ROC CURVES OF NONPARANORMAL GRAPHICAL MODELS FOR GLASSO (LEFT), NPN (CENTER) AND SME (RIGHT).

In our experiments we use the "Symmetric Power Transformation" [42], that is

$$f_j(z_j) = \sigma_j \left(\frac{g_0(z_j - \mu_j)}{\sqrt{\int g_0^2(t - \mu_j)\phi\left(\frac{t - \mu_j}{\sigma_j}\right) dt}}\right) + \mu_j,$$

where $g_0(t) = sign(t)|t|^\alpha$, to transform data. For comparison with GLASSO, we first use a truncation method to Gaussianize the data, and then apply graphical lasso to the transformed data. See [42] for details. From figure 11, without knowing the underlying data distribution, the score matching estimator outperforms GLASSO, and show similar results to nonparanormal when the sample size is large.

**Contact Data** We used the same protein subset as detailed in the prior section, except we remove the MSAs with more than 2000 homologs due to high computation cost. CCMpred handled the data quite well because the assumption of categorical data fits the data assumption exactly, therefore it outperformed PSICOV by a large margin (0.139 vs 0.069) on CASP+CAMEO test set. While our method assumes the data is in a much broader assumption, it still performs relatively well. It has a 0.136 top L/5 long-range contact prediction accuracy, which is roughly the same as CCMpred's accuracy. For Pfam subset, CCMpred achieves 0.512 accuracy for top L/5 long-range prediction, which slightly outperforms the proposed method with (0.485). These results suggest that our algorithm learns the underlying data distribution without knowing it in advance.

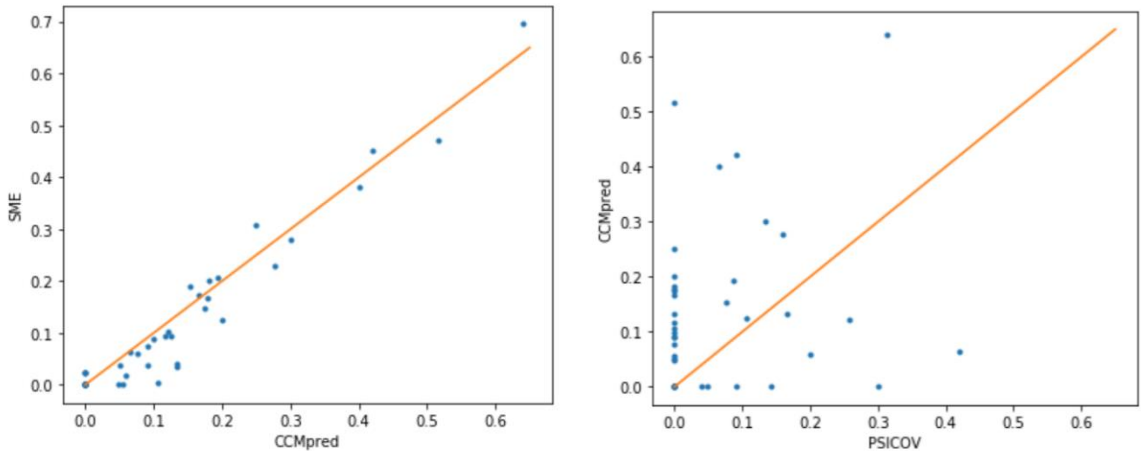

FIGURE 13 (LEFT) ONE VS ONE L/5 RESULT COMPARISON BETWEEN CCMPRED AND SME (RIGHT) ONE VS ONE L/5 RESULT COMPARISON BETWEEN PSICOV AND CCMPRED

We also want to point out that the results above indicate that the proposed approach is not quite useful for contact prediction as it performs similar to CCMpred, yet much slower.

On the other hand, the contributions are mainly from two folds, (1) the proposed algorithm could learn the graphical model structure under a much broader assumption since this assumption covers basically all exponential families; (2) the proposed algorithm is able to recover the graph structure with probability 1 even under this broader assumption.

# Chapter 4

# Contact Prediction by Deep Learning

In this chapter, we will explore how to use supervised machine learning methods to predict contact. Typically, the supervised machine learning approach outperforms the evolutionary coupling analysis approach because it (1) predicts contact using more varied information from many more features; (2) directly takes the output of evolutionary coupling analysis as an input feature; and (3) uses the protein's 3-dimensional structural information for training. Existing methods include SVMSEQ [15], PconC2 [16], MetaPSICOV [17], PhyCMAP [19], and CoinDCA-NN [18]. Among those methods, MetaPSICOV performs best [17], but the prediction quality of all of those methods is insufficient for accurate contact-assistant protein folding. Consequently, we propose a better contact prediction method, especially for proteins without large numbers of sequence homologs, and compare it extensively against MetaPSICOV.

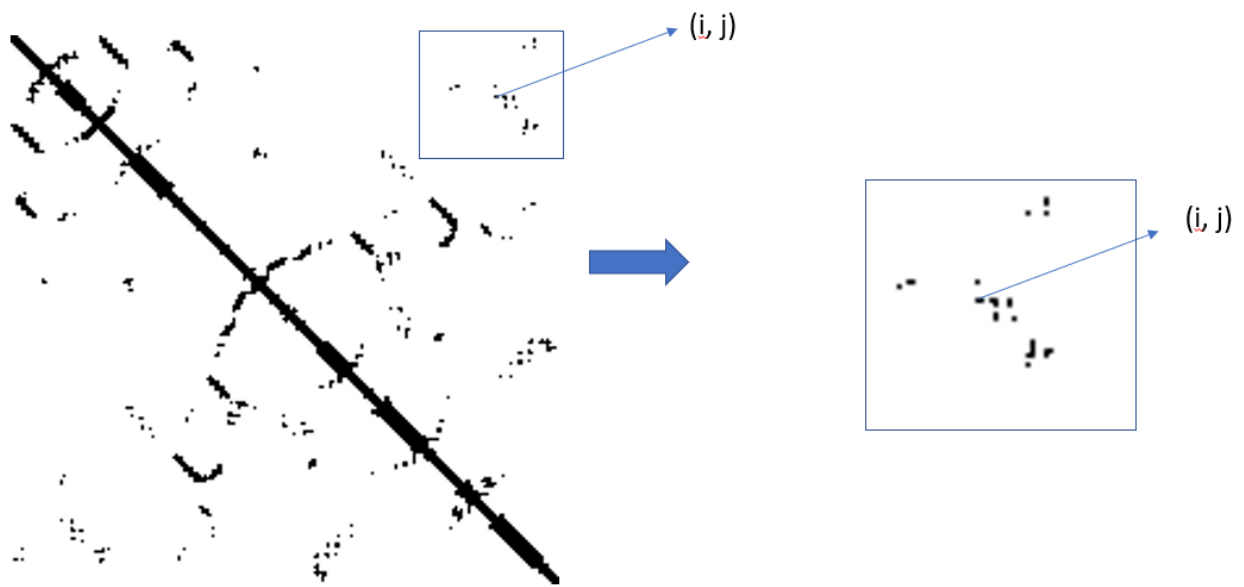

FIGURE 14 WE EXTRACT ALL FEATURES AROUND POSITION (I,J) WITHIN A FIX WINDOW SIZE W. THIS WINDOW WILL GENERATE A SET OF FEATURE WITH SIZE 2W+12W+1NFEATURES, AND THE LABEL FOR THIS FEATURE IS THE CONTACT LABEL AT (I,J).

We first tried to improve the contact prediction method based on MetaPSICOV, which uses a shallow, fully connected neural network. To predict a contact between amino acids i and j, instead of using features from a fixed window around i and j, we used the features within a fixed window (size w) on the **image** level, i.e., all features at (i', j') such that max(i-i', j-j') ≤ w; see Figure 13 for a visualized example. Thus, not only did we use more information by including more features, but we also maintained the spatial relationships of all features. The resulting feature size is (2w+1)×(2w+1)×(number of features), which makes it quite similar to an image with more than 3 channels and thus more convenient for convolution neural networks (CNN) [49, 50].

Denoting the sequence length as L, we used the following features to build our model:
(1) Protein sequence profile, i.e., position-specific scoring matrix with dimensions Lx20;
(2) Predicted probability of secondary structures for the target sequence by using convolutional neural fields [51]. The resulting feature dimension is Lx3;
(3) Predicted 3-state solvent accessibility by using RaptorX-Property[52]. The resulting feature dimension is also Lx3;
(4) CCMpred score. The dimension for this feature is LxL;
    a. Unlike MetaPSICOV, we chose not to additionally use the PSICOV score because compared to CCMpred, it is too time consuming and offers too marginal of an improvement.
(5) Mutual information and pairwise potential [53, 54]. The dimensions for both are LxL.

We refer to the first three sets of features—sequence profile, predicted probability of secondary structures, and predicted 3-state solvent accessibility—as 1D features because they correspond to each single amino acid in sequence. After concatenating them, we have a Lx26 matrix as our final 1D feature. The last two sets of features, CCMpred score and mutual information and pairwise potential, are 2D features because they correspond to each pair of amino acids. After concatenating them, we have a LxLx3 matrix as our final 2D feature.

To use CNN conveniently, we convert those 1D features into 2D features by using an operation such as outer product. More specifically, we use the 1D features for each residue $\{f_1, f_2, \ldots, f_L\}$, where f is a k-dimensional vector that stores the feature information. Then, for each pair of residues i and j, we concatenate $f_i, f_{(i+j)/2}$ and $f_j$ into a single vector and use it as one input feature for this pair. The dimensions of the resulting converted features are then $L \times L \times 3k$.

Next, we introduce the architecture of CNN and each of its components and usage.

## 4.1 Introduction to CNN

Deep learning, especially convolutional neural network, has been a very popular tool in computer vision [49, 55], natural language processing [56, 57], and computational biology [58, 59]. Yann Lecun initially proposed LeNet for handwritten characters with only 5 layers in 1989 [50]. With the development of hardware and datasets, large scale training of neural networks became possible. Later in 2012, Alexnet was proposed, which is almost the same as LeNet in terms of architecture. It has 8 layers, but a much greater number of parameters (60 million). CNN then gained much attention from the community when it won the 2012 ImageNet Large Scale Visual Recognition Challenge [49].

A typical CNN has five components, which includes a convolutional layer (conv layer); an activation function, which usually is rectified linear unit (RELU); batch normalization (BN) [60]; pooling; and fully connected (FC) layers. We will go through each of them in this section, and then explain our proposed architecture.

**Convolutional Layer**

Denote the input as $L \times L \times n_f$, where L is the sequence length and $n_f$ is the number of features. The conv layer is a set of learnable filters, each with size $s \times s \times n_f$, where s is typically a small odd number, like 3, 5 or 7. The convolutional layer starts sliding each filter across the width and height of input features and computes the dot product between the filter and features at any position. Note that after the convolution operation, the size of output is typically a little bit smaller than L—more precisely, with new length L+1-s. In this thesis, we always pad the input with zeros, such that the output and input have the same length after conv layer's operation.

**Batch Normalization Layer**

To increase the stability of the training process, batch normalization (BN) is introduced. It first normalizes the output of the previous layer by subtracting the batch mean and dividing by its standard deviation. More specifically, given a batch of output of the previous layer $y_B$, for feature k, we first obtain

$$\widehat{y_k} = \frac{y_k - mean_B(y_{\cdot,k})}{std_B(y_{\cdot,k})},$$

where $mean_B$ and $std_B$ denote the mean and standard deviation for feature k in batch B, respectively. Then, BN makes the optimization method (e.g., ADAM [61]) execute "denormalizaton" by introducing two parameters for scaling and shifting the activation outputs. More specifically, set the output as

$$y_k = \alpha \hat{y}_k + \beta,$$

where $\alpha\ and\ \beta$ are two trainable parameters.

**RELU Layer**

The RELU is an activation layer that takes the output of the previous layer and clamps all the negative values in it to zero. In other words,

$$RELU(Y) = \max(0, Y).$$

It does not bring in any trainable parameters, but it is very helpful because it introduces non-linearity and allows the model to fit a large set of possible functions.

**Pooling Layer**

The function of the pooling layer is to progressively reduce the size of the previous layer by using max, average, or other operations. Not only does it reduce the computational cost and number of parameters in the model, but it also helps increase the receptive field of later representation. The most common pooling layer is a down sampled layer with a filter size of 2x2 and samples the max out of a 2x2 region. We use this common max pooling layer as our pooling layer.

With the increasing depth of CNN, vanishing gradient [62] is a more and more severe problem in optimization, as the gradient in the earlier layers is extremely small. To solve this problem, Karen Simonyan and Andrew Zisserman proposed VGG [63] with 16 and 19 layers, which uses a much smaller filter size[3] (3x3) to ease optimization. Later, residual network [55, 64] was proposed as a more efficient way to tackle the problem by adding a short path from one layer to the next. The gradient can then pass through the path to help train earlier layers. Another short path method was proposed in dense network [65], and it concatenates features from all of the previous layers to pass the gradient rather than adding them. Instead of making CNN deeper, wide residual network [66] chooses to make the network wider by adding more feature maps.

## 4.2 Convolutional Neural Network Architectures

In this section, we introduce the aforementioned influential and popular CNN architectures, including (1) LeNet [50]; (2) Very Deep Convolutional networks (VGG) [63]; (3) Residual Network (ResNet) [55, 64]; (4) Dense Network (DenseNet) [65]; and (5) Wide Residual Network (Wide ResNet) [66].

### 4.2.1 LeNet

---

[3] Please refer LetNet in next section for the definition of filter size

We first introduce the terms we will use in this thesis, and then the simplest LeNet architecture. The input features have the shape LxLx1, where L is the length and width of the input image, and 1 is the number of input features. It is followed by a convolutional layer that has an output called the "feature map." A feature map is generated by applying a non-linear function (often called an "activation function") on the convolution of a sub-region of input features with a linear filter.[4] In LeNet, there are 6 such filters, each with a size of 5x5. Next, LeNet applies a max-pooling layer on the output of the convolutional layer. A max-pooling layer divides the input feature into a set of rectangles, and the output is the maximum value for each such rectangle. This is followed by another convolutional and max-pooling stack. Finally, the output of the last layer (max-pooling, in this case) is flattened and connected to two fully connected layers with hidden neurons 120 and 84, respectively.

### 4.2.2 VGG

The architectures before VGG often use a much larger filter size, such as 9x9 and 11x11 in AlexNet [67]. VGG, on the other hand, uses a much smaller 3x3 filter. The authors show that multiple 3x3 filters can be as effective as a larger filter size, yet with a much fewer number of parameters, which allows for training of a deeper network (16 or 19 weighted layers). The architecture of VGG often has 5 convolutional blocks, and each block has 2 or 3 convolutional layers. Three fully connected (FC) layers are connected to the output of the convolutional layers to conduct the classification. Since the rise of VGG, a 3x3 or 5x5 filter size has been preferred to a larger filter size. The hyperparameters for the VGG are number of layers per block, number of blocks, and number of hidden neurons in an FC layer.

### 4.2.3 Residual Network

When we try to train a deeper neural network, exploding and vanishing gradients make the optimization difficult for the first few layers [55, 68, 69]. For example, with the same training and test dataset, a deeper neural network results in not only higher test errors, but also higher training errors, which suggests that the greater difficulty arises in optimization, rather than in overfitting due to the deeper network.

---

[4] For visualization of convolution operation, please refer to https://github.com/vdumoulin/conv_arithmetic

Residual network was the first one (concurrently with highway networks[68]) that formally introduced the idea of 'short-cut' connections in CNN to tackle this problem. The 'short-cut' connections in ResNet refer to the output of the previous layer plus the output of the current layer. Those small units can be stacked further to form a much deeper network. Since the gradient is able to pass through the short-cut connections without vanishing or exploding, it is possible to train a much deeper network within a reasonable amount of time. The authors of the ResNet even trained a CNN with more than 100 layers without optimization difficulties [55].

Later on, they explored further by comparing different ways to combine the "activation layer," "convolutional layer," and "batch normalization layer" [60], and showed that "batch normalization → activation → convolutional layer" yields better performance [64]. In our experiments, we use this refined version of ResNet. The hyperparameters for the ResNet are the number of small units per block, number of blocks, and number of hidden neurons in an FC layer.

### 4.2.4 DenseNet

To further improve the information flow, instead of summing the output from the **previous** layer, DenseNet redefines the "short-cut" connection by **concatenating all** layers with each other. The information then can be passed through easily since each layer has all the feature maps of the previous layer. Suppose each layer produces n feature maps, the l-th layer will have $l*(n-1) + n_0$ feature maps, where $n_0$ is the number of features in the input. The author proposed using a small n to prevent the network from growing too wide to be optimized and showed that a small n is sufficient to obtain state-of-the-art results. We use n=12 in all our experiments. The hyperparameters for DenseNet are the number of blocks in the network and number of hidden neurons in an FC layer.

### 4.2.5 Wide Residual Network

Instead of making a network deeper, the authors of Wide ResNet explored the trade-offs between depth (number of layers) and width (number of feature maps) for ResNet. To keep the number of parameters under control with more than 100 layers, the author of ResNet made the network as thin as possible. As [68, 70] suggested, a very deep network may not be necessary because many layers contribute very little to the objective function, and only a few layers can learn the useful representation. To solve this problem, they built a ResNet with k times more width, yet much fewer layers, and their model can achieve higher performance on several benchmarks. The hyperparameters for the model are width parameter k, number of units in each block, number of blocks, and number of hidden neurons in an FC layer.

## 4.3 Image Level Convolutional Neural Network

To predict the likelihood of residue i and residue j forming a contact, a pipeline based on traditional supervised machine learning approaches was developed. First, it extracts features from a window centered at residue i and j and concatenates them. Next, a binary classifier, such as a shallow neural network or SVM, is applied. Similarly, to predict the contact between residue i and j by using CNNs, features from a two-dimensional (2D) window ranging from i-window to i+window and j-window to j+window are extracted from raw features. Note that unlike traditional approaches that flatten all input features into one dimension, the 2D spatial structure of the input features is maintained in CNNs. We refer to this approach as an image-level CNN.

### 4.3.1 Sampling Procedure

In general, we cannot use all available pairs for training because theoretically, there are $\frac{L(L-1)}{2}$ possible pairs of (i, j) for a sequence with length L, where i, j indicates the position of two amino acids in the target protein. With 6,000 sequences with average lengths of

100, it amounts to about 30 million points of training data that are extremely unbalanced because there are too many negative examples. Hence, we used a sampling strategy to (1) make the dataset smaller and easier to train and (2) balance the dataset.

First, because there are only a few positive examples, which are very important, we keep *all* of them for training. For negative examples, we discard them randomly with a probability of 80% for medium-range pairs and 96% for long-range pairs. Note that we don't differentiate between medium- and long-range models, and we train and predict them together using the same model. The final training data has about 2 million training examples, and the proportion of positive examples to negative examples is roughly 1:7.

## 4.4 Pixel Level Convolutional Neural Network

We propose another end-to-end deep learning architecture that takes the whole sequence as the input, and predicts the whole contact map directly, which can boost the speed and accuracy even further. Our proposed deep learning model consists of two deep convolutional neural networks. The first component is a one-dimensional (1D) CNN that can reduce the sequential feature (i.e., sequence profile, predicted secondary structure, and solvent accessibility) dimension and learn its feature representation simultaneously. After mapping the sequential features to a lower dimensional space, we convert the learned 1D features to a two-dimensional matrix through outer concatenation. The output is then fed into a 2D convolutional neural network with pairwise features (i.e., co-evolution information, pairwise contact, distance potential, and CCMpred's output). Finally, the resulting contact map is obtained by a fully connected layer on the output of the 2D CNN [58, 71].

Next, we will take ResNet as our example and explain the details of our proposed architecture. We change the ResNet block to VGG, DenseNet, and Wide ResNet to further explore the effects of different architectures.

### 4.4.1 ResNet Model Details

The network consists of two residual networks, each consisting of residual blocks that are concatenated together. Figure 14 shows an example of a residual block that consists of two convolution layers and two activation layers. In this figure, $X_i$ and $X_{i+1}$ are the input and output of this block, respectively. The activation layer is RELU, a nonlinear transformation of input without any parameters. Let $f(X_i)$ denote the result of $X_i$ going through two activation layers and two convolutional layers. Then $X_{i+1}$ is equal to $X_i + f(X_i)$. Since $f(X_i)$ is the difference between $X_{i+1}$ and $X_i$, $f$ is also called a residual function, and this neural network with multiple residual blocks is called residual network. To speed up training and convergence, we also add a batch normalization layer [60] before each activation layer, which normalizes its input to have a mean of 0 and a standard deviation of 1.

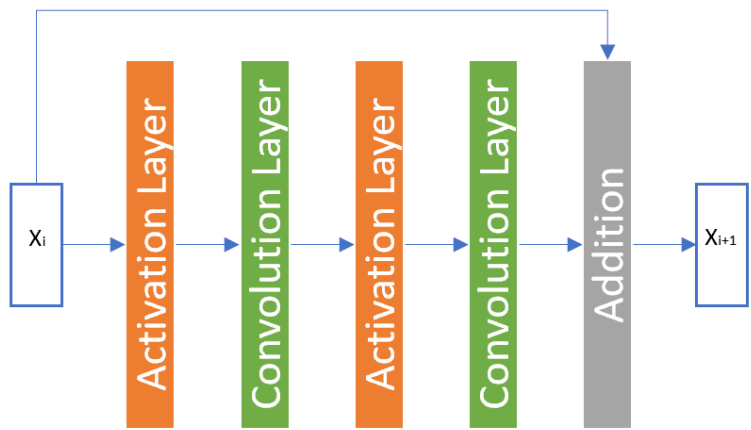

FIGURE 15 A BLOCK OF RESIDUAL NETWORK WITH X$_I$ AND X$_{I+1}$ BEING INPUT AND OUTPUT, RESPECTIVELY. EACH BLOCK CONSISTS OF TWO CONVOLUTIONAL LAYERS AND TWO ACTIVATION LAYERS.

Next, we introduce some of the hyperparameter settings. The filter size (i.e., window size) used by a 1D convolution layer is 17 while the filter size in 2D convolution layer is 3×3. By stacking many residual blocks together, even if we use a small window size at each convolution layer, our network can still model long-range dependency between two different positions. We fix the depth (i.e., the number of convolution layers) of the 1D residual network to 6, but we vary the depth of the 2D residual network for tuning. Our experimental results show that with about 60 hidden neurons and about 60 convolution layers for the 2D residual network, our model can obtain good performance. Note that it has been shown that for image classification, a convolutional neural network with a smaller window size but many more layers usually outperforms a network with a larger window size but fewer layers. Furthermore, a 2D convolutional neural network with a smaller window size also has a fewer parameters than a network with a larger window size. Typically, ResNet assumes that inputs have fixed dimension, while our network needs to take variable-length proteins as input. Additionally, there is no pooling layer in the whole model, so many layers are necessary to increase the size of the receptive field and thus model the long-range correlation.

**1D convolution** Roughly, a 1D convolution operation is matrix-vector multiplication. Let X and Y (with dimensions L×m and L×n, respectively) be the input and output of a 1D convolutional layer, respectively. Let the window size be 2w+1 and s = (2w+1)×m. The convolutional operator that transforms X to Y can be represented as a 2D matrix with dimension n×s, denoted as C. Note that the shape of C does not depend on protein length and each convolutional layer may have a different C. Let $X_i$ be a submatrix of X centered at residue i ($1 \leq i \leq L$) with dimension (2w+1)×m, and $Y_i$ be the i-th row of Y. We may calculate $Y_i$ by first flattening $X_i$ to a vector of length s and then multiplying C and the flattened $X_i$.

**Training with proteins of different lengths** Our network can take variable-length proteins as input. We train our deep network in a minibatch mode, which is used regularly in deep learning. That is, for each training iteration, we use a minibatch of proteins to calculate the gradient and update the model parameters. A minibatch may have one or several proteins. We sort all training proteins by length and group proteins of similar lengths into minibatches. Considering that most proteins have a length of up to 600 residues, proteins in a minibatch often have the same length. In the case that they do not, we add zero padding to shorter proteins. Our model is a fully convolutional network; therefore it is protein-length independent, and two different minibatches are allowed to have different protein lengths. At inference stage, since our network can take variable-length input, we do not need to cut a long protein into segments in predicting contact maps. Instead, we predict all contacts of a protein simultaneously, and there is no need to use zero padding since only a single protein is predicted in a batch.

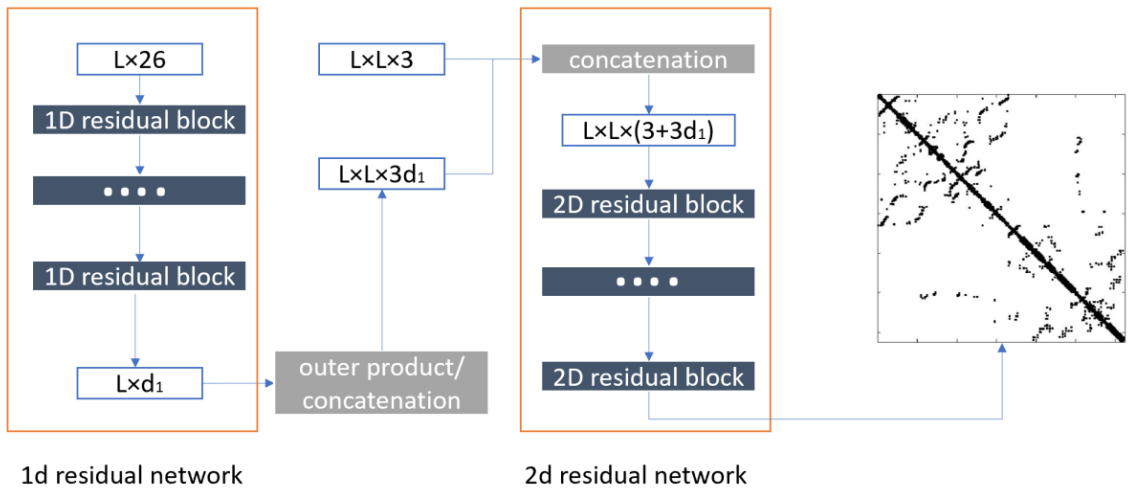

FIGURE 16 THE OVERALL NETWORK ARCHITECTURE OF THE DEEP LEARNING MODEL. MEANWHILE, L IS PROTEIN SEQUENCE LENGTH AND N IS THE NUMBER OF HIDDEN NEURONS IN THE LAST 1D CONVOLUTIONAL LAYER. FIGURE IS FROM [58]

When we explore other CNN architectures, the 1D CNN model is *fixed* like ResNet's since it only has six layers, whose outputs are further transformed into 2D features, and we *vary* the architecture of 2D CNN model only. As introduced above, we investigate what occurs when we change the architecture to VGG, DenseNet, and Wide ResNet.

**4.4.2 Training Procedures**

For all the architectures, we fix the following hyperparameters for training:
1. The number of epochs is fixed as 36. Typically, it will converge at around epoch 25.

2. The $l_2$ regularizer is set as 0.0001.
3. A stochastic gradient descent method and a momentum of 0.9 with initialized step size 0.01 and 0.001 and ADAM [72] with step size 0.001 and 0.0001 are applied for optimization; the step size is also reduced by 10 times at epochs 18 and 27. The epoch and initial step size are selected by the loss function in the valid dataset.
4. The unit for each block (ResNet and Wide ResNet) is fixed as 4. We vary the depth of each network (number of blocks) until it can fully utilize the GPU memory on a single card (12 GB).
5. For the final output layer, the number of hidden layers is 2 and the number of hidden units is fixed at 100.
6. Each architecture is run three times with different initialization; the one with the best valid loss is then evaluated by the test data.
7. The window size for an image-level CNN is 20.

We use a maximum-likelihood approach to train model parameters. That is, we maximize the occurrence probability of the native contacts (and non-contacts) of the training proteins. Therefore, the loss function is defined as the negative log-likelihood averaged over all the residue pairs of the training proteins. Since the ratio of contacts among all the residue pairs is very small, to make the training algorithm converge quickly, we assign a larger weight to the residue pairs forming a contact. The weight is assigned such that the total weight assigned to contacts is approximately 1/8 of the number of non-contacts in the training set.

## 4.5 Results

We evaluate the performance of previously mentioned state-of-the-art techniques MetaPSICOV, the image level ResNet (ResNet-IL), and pixel level VGG, ResNet, DenseNet, and Wide ResNet by the accuracy of top L/k (k=1, 2, 5, 10) predicted contacts. Note that the medium- and long-range contacts are defined as a pair of residues with sequence distance falling within [12, 24) and >= 24, respectively.

As shown in tables 3(a) to 3(c), ResNet-IL outperforms MetaPSICOV by 10% to 14% on long-range L/10 contact prediction and 7% to 11% on medium-range L/10 contact by simply replacing a 3-layer neural network or SVM with a 50-layer residual network. For long-range L/5 and L/2, the improvements are 10% to 15% and 11% to 14%, and for medium-range L/5 and L/2, the improvements are 7% to 18% and 4% to 6%. In summary, ResNet-IL significantly improves long-range and medium-range contact prediction accuracy on all test datasets over the previous state-of-the-art predictor, metaPSICOV.

| Method | Medium | | | | Long | | | |
|---|---|---|---|---|---|---|---|---|
| | L/10 | L/5 | L/2 | L | L/10 | L/5 | L/2 | L |
| CCMpred | 0.33 | 0.27 | 0.19 | 0.13 | 0.37 | 0.33 | 0.25 | 0.19 |
| MetaPSICOV | 0.69 | 0.59 | 0.42 | 0.28 | 0.60 | 0.54 | 0.45 | 0.35 |
| ResNet-IL | 0.76 | 0.67 | 0.48 | 0.32 | 0.74 | 0.69 | 0.59 | 0.47 |
| VGG | 0.82 | 0.73 | 0.53 | 0.34 | 0.79 | 0.75 | 0.65 | 0.52 |
| ResNet | **0.84** | **0.74** | **0.54** | **0.36** | **0.83** | **0.79** | **0.70** | **0.55** |
| DenseNet | 0.82 | 0.73 | **0.54** | 0.35 | 0.81 | 0.77 | 0.66 | 0.54 |
| WideResNet | 0.82 | **0.74** | **0.54** | 0.35 | 0.81 | 0.76 | 0.68 | 0.54 |

(a)

| Method | Medium | | | | Long | | | |
|---|---|---|---|---|---|---|---|---|
| | L/10 | L/5 | L/2 | L | L/10 | L/5 | L/2 | L |
| CCMpred | 0.27 | 0.22 | 0.14 | 0.10 | 0.30 | 0.26 | .20 | 0.15 |
| MetaPSICOV | 0.53 | 0.45 | 0.32 | 0.22 | 0.47 | 0.42 | 0.33 | 0.25 |
| ResNet-IL | 0.62 | 0.53 | 0.38 | 0.26 | 0.61 | 0.56 | 0.45 | 0.34 |
| VGG | 0.65 | 0.58 | 0.41 | 0.27 | 0.65 | 0.62 | 0.50 | 0.38 |
| ResNet | 0.68 | 0.59 | **0.42** | **0.28** | **0.69** | **0.65** | **0.54** | **0.41** |
| DenseNet | 0.68 | 0.58 | **0.42** | 0.27 | 0.68 | 0.64 | 0.53 | **0.41** |
| WideResNet | **0.70** | **0.61** | **0.42** | **0.28** | 0.66 | 0.64 | **0.54** | **0.41** |

(b)

| Method | Medium | | | | Long | | | |
|---|---|---|---|---|---|---|---|---|
| | L/10 | L/5 | L/2 | L | L/10 | L/5 | L/2 | L |
| CCMpred | 0.36 | 0.26 | 0.15 | 0.10 | 0.52 | 0.45 | 0.31 | 0.21 |
| MetaPSICOV | 0.49 | 0.40 | 0.27 | 0.18 | 0.61 | 0.55 | 0.42 | 0.30 |
| ResNet-IL | 0.59 | 0.47 | 0.31 | 0.20 | 0.71 | 0.65 | 0.53 | 0.39 |
| VGG | 0.64 | 0.52 | 0.33 | 0.21 | 0.76 | 0.71 | 0.59 | 0.44 |
| ResNet | **0.66** | **0.53** | **0.34** | **0.22** | **0.79** | **0.74** | **0.63** | **0.47** |
| DenseNet | 0.65 | **0.53** | **0.34** | 0.21 | 0.77 | 0.72 | 0.60 | 0.46 |
| WideResNet | 0.65 | **0.53** | **0.34** | 0.21 | 0.77 | 0.73 | 0.61 | 0.46 |

(c)

TABLE 2 CONTACT PREDICTION ACCURACY FOR RESNET WITH DIFFERENT FEATURES AND CCMPRED ON (A) 105 CASP PROTEINS, (B) 76 CAMEO PROTEINS AND (C) 398 MEMBRANE PROTEINS

Furthermore, the results can be boosted by using our proposed pixel-level CNNs. Pixel-level ResNet outperforms the corresponding image-level ResNet on long-range contact prediction by almost 10% on all three test datasets. To be more specific, 9%, 8%, and 8% on L/10, 10%, 9%, and 9% on L/5, and 11%, 9%, and 10% on L/2. For medium-range L/10 and L/5 contact predictions, the improvements are also significant. Figure 16 (a) indicates that ResNet can improve almost on all targets over ResNet-IL, which justifies the benefit of the proposed pixel-level architecture.

For pixel-level deep learning models only, more advanced CNN architectures (e.g., ResNet, Wide ResNet, and DenseNet) generate very similar results, only slightly outperforming the old-fashioned VGG. See Figures 16(b) to 16(d) for detailed one-to-one accuracy comparisons between VGG, WideResNet, DenseNet, and ResNet on top L/5 predictions.

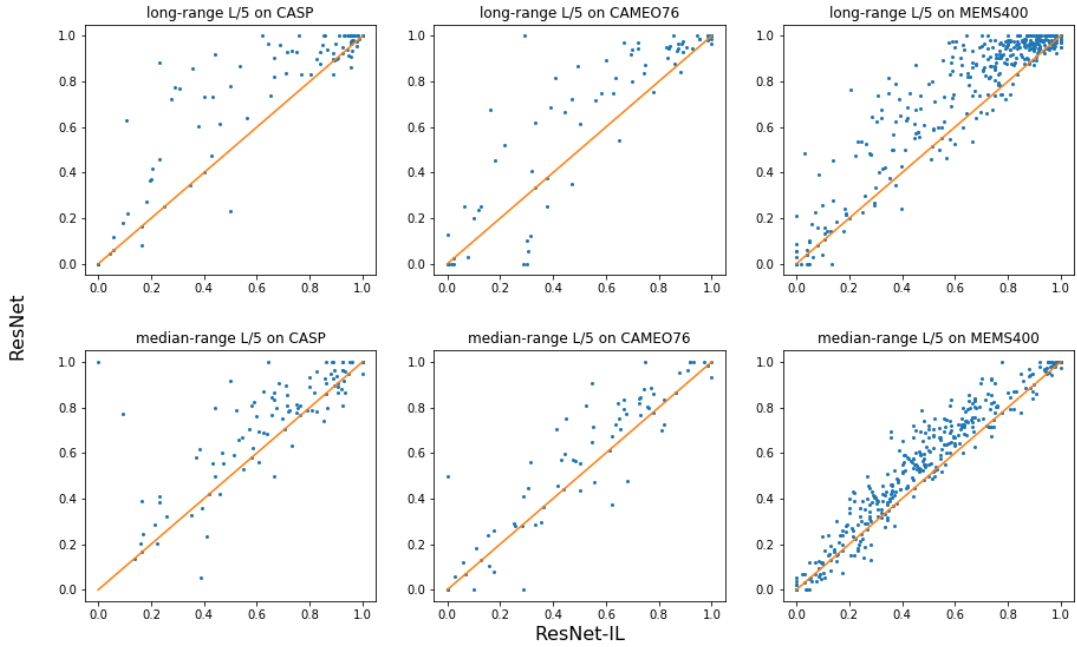

(a)

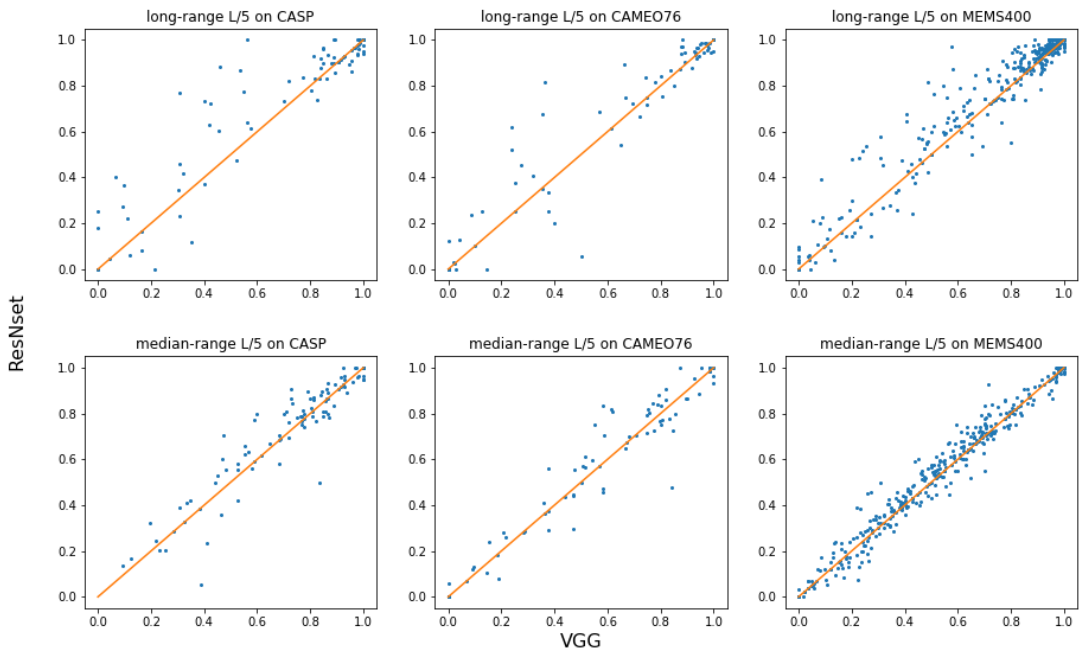

(b)

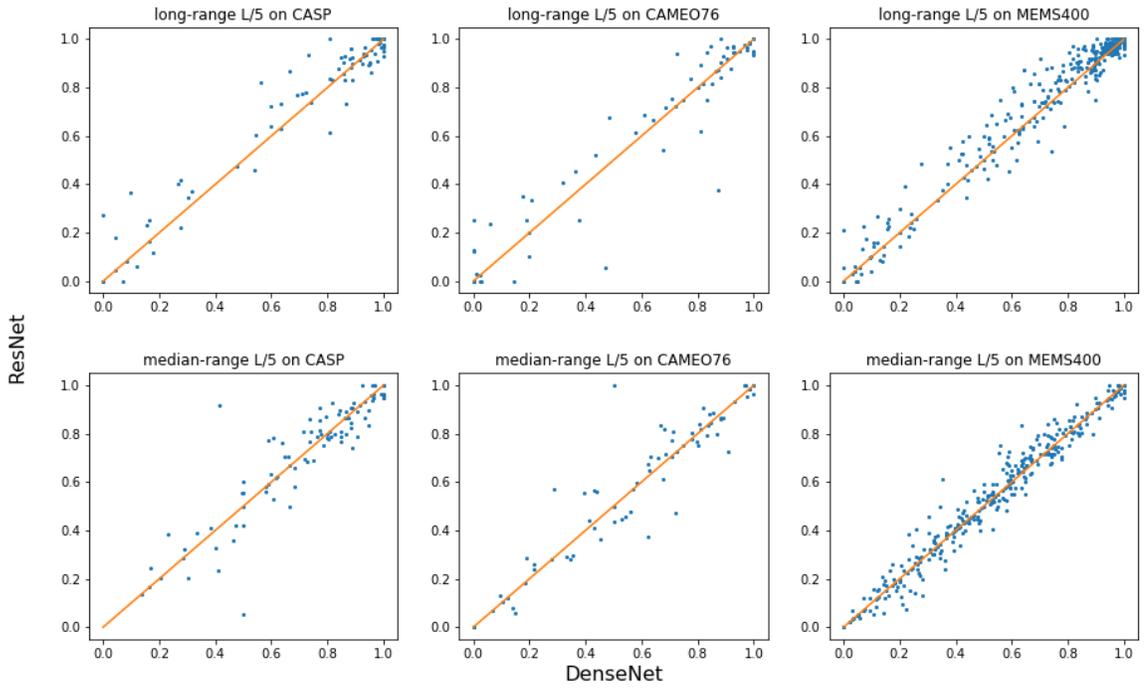

(c)

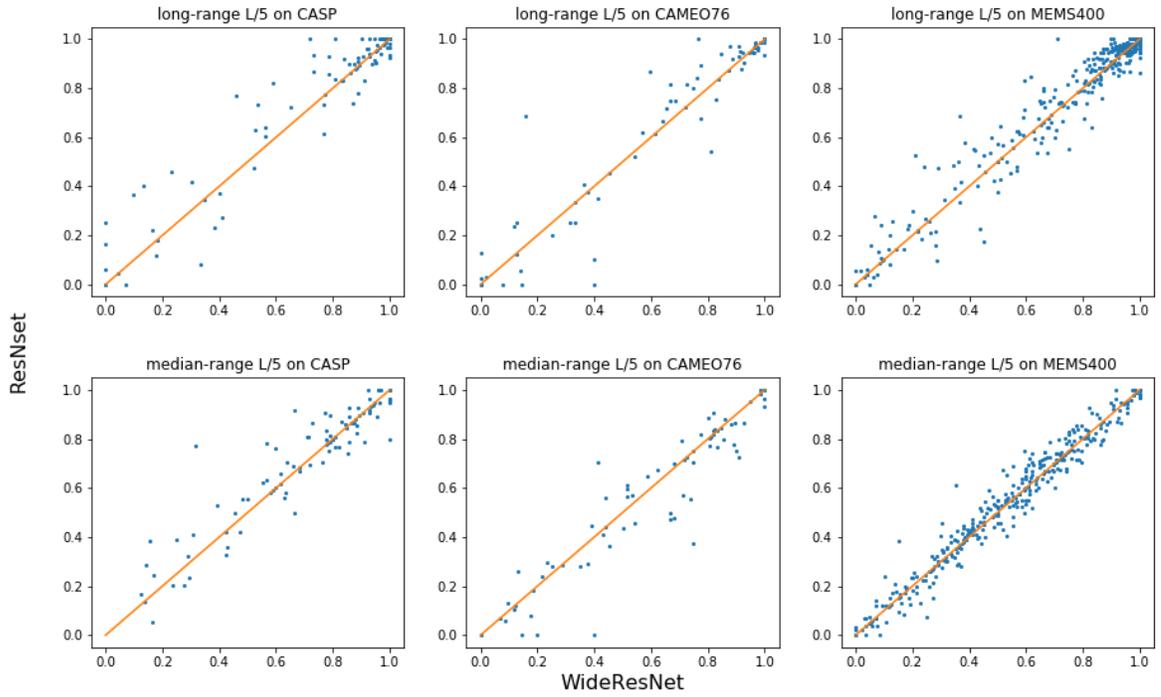

(d)

FIGURE 17 ONE-TO-ONE PERFORMANCE COMPARISON BETWEEN (A) RESNET-IL AND RESNET, (B) VGG AND RESNET, (C) DENSENET AND RESNET AND (D) WIDE RESNET AND RESNET, WHERE EACH DOT INDICATES THE CONTACT PREDICTION ACCURACY.

To test the methods' performance with respect to the number of homologs, we use Meff to measure the number of effective sequence homologs in multiple sequence alignment. Note that Meff can be seen as the number of non-redundant sequence homologs when 70% sequence identity is used as the cutoff to remove redundancy. We group the test proteins in CASP/CAMEO and MEMS into 10 bins based on their ln(Meff) and compute the average prediction accuracy in each bin. The first three bins are merged for membrane proteins because they have a small number of proteins.

Figure 17 indicates that for top L/5 long-range contact predictions, almost all pixel-level models significantly outperform image-level ResNet for all ln(Meff), especially when ln(Meff) is less than 5. More advanced architectures with short paths have very similar performances, yet they all outperform VGG.

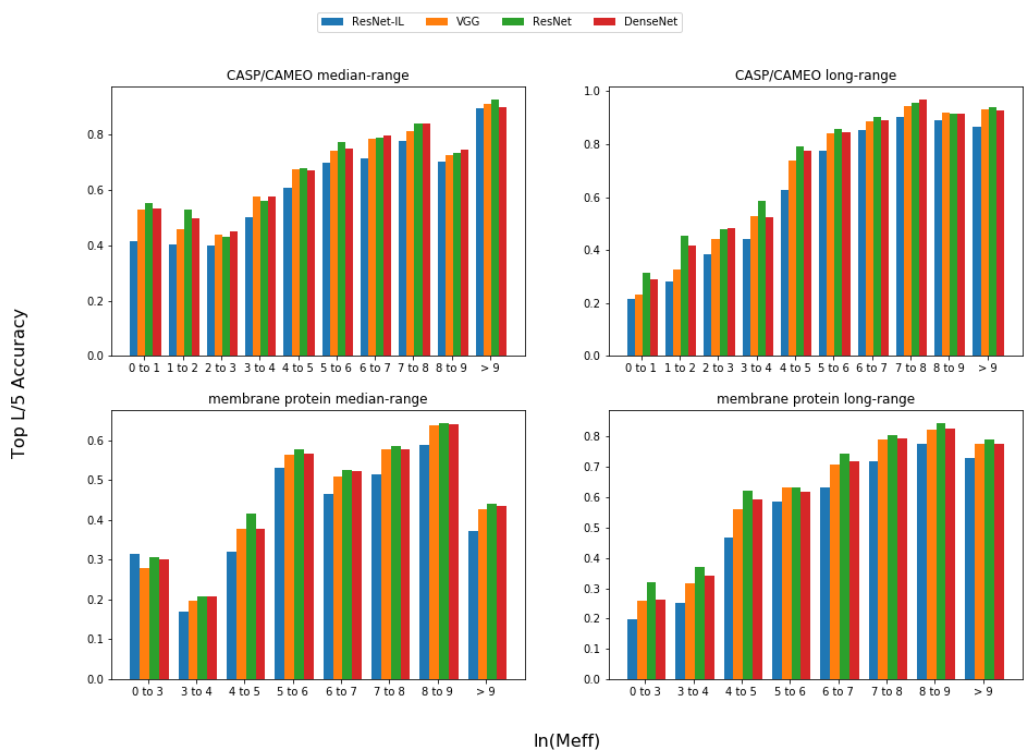

FIGURE 18 TOP L/5 MEDIUM- (LEFT) AND LONG-RANGE(RIGHT) CONTACT PREDICTION ACCURACY FOR RESNET-IL(BLUE), VGG(ORANGE), RESNET(GREEN) AND DENSENET(RED) WITH RESPECT TO HOMOLOGOUS INFORMATION MEASURED BY LN(MEFF). THE 105 CASP AND 76 CAMEO RESULTS ARE DISPLAYED IN TOP PANELS, AND 398 MEMBRANE RESULTS ARE DISPLAYED IN BOTTOM PANELS.

## 4.6 Importance of Features

To test the efficiency of our 2D features, we remove all of the sequence-level features and 1D embedding part of pixel-level models and run experiments on our best-performing ResNet architecture. We started with ResNet with only the CCMpred feature (ResNet[CCMpred]). Even with only one feature, CNN can capture the relationships between amino acids much better than raw CCMpred, with a 10% to 20% accuracy gain on both medium- and long-range predictions based on Tables 4(a) to 4(c). Figure 18(a) shows there are significant improvements on almost all targets. As shown in Table 3, if we add all other 2D features (ResNet[2D]) as well, the accuracy improves by another 10%, which is also significant; see Figure 18(b) for more details. ResNet with all features (ResNet [All]), however, can only improve significantly on the CASP 105 dataset, while on both CAMEO and MEMS, the accuracy gain is only around 2% to 5%. Those results prove that the main contribution of the model is from the 2D features, even though there are only four of them. Figure 18(c) also indicates a similar one-to-one performance.

| Method | Medium | | | | Long | | | |
|---|---|---|---|---|---|---|---|---|
| | L/10 | L/5 | L/2 | L | L/10 | L/5 | L/2 | L |
| ResNet [All] | 0.84 | 0.74 | 0.54 | 0.36 | 0.83 | 0.79 | 0.70 | 0.55 |
| ResNet [2D] | 0.77 | 0.68 | 0.50 | 0.33 | 0.79 | 0.73 | 0.64 | 0.51 |
| ResNet [CCMpred] | 0.63 | 0.55 | 0.40 | 0.26 | 0.65 | 0.60 | 0.52 | 0.41 |
| CCMpred | 0.40 | 0.32 | 0.21 | 0.14 | 0.43 | 0.39 | 0.31 | 0.23 |

(a)

| Method | Medium | | | | Long | | | |
|---|---|---|---|---|---|---|---|---|
| | L/10 | L/5 | L/2 | L | L/10 | L/5 | L/2 | L |
| ResNet [All] | 0.68 | 0.59 | 0.42 | 0.28 | 0.69 | 0.65 | 0.54 | 0.41 |
| ResNet [2D] | 0.63 | 0.56 | 0.40 | 0.26 | 0.65 | 0.61 | 0.50 | 0.38 |
| ResNet [CCMpred] | 0.46 | 0.41 | 0.30 | 0.20 | 0.53 | 0.49 | 0.40 | 0.30 |
| CCMpred | 0.27 | 0.22 | 0.14 | 0.10 | 0.30 | 0.26 | 0.20 | 0.15 |

(b)

| Method | Medium | | | | Long | | | |
|---|---|---|---|---|---|---|---|---|
| | L/10 | L/5 | L/2 | L | L/10 | L/5 | L/2 | L |
| ResNet [All] | 0.66 | 0.53 | 0.34 | 0.22 | 0.79 | 0.74 | 0.63 | 0.47 |
| ResNet [2D] | 0.63 | 0.51 | 0.32 | 0.21 | 0.77 | 0.72 | 0.60 | 0.45 |
| ResNet [CCMpred] | 0.59 | 0.47 | 0.31 | 0.20 | 0.73 | 0.68 | 0.55 | 0.41 |
| CCMpred | 0.36 | 0.26 | 0.15 | 0.10 | 0.52 | 0.45 | 0.31 | 0.21 |

(c)

TABLE 3 CONTACT PREDICTION ACCURACY FOR RESNET WITH DIFFERENT FEATURES AND CCMPRED ON (A) 105 CASP PROTEINS, (B) 76 CAMEO PROTEINS AND (C) 398 MEMBRANE PROTEINS.

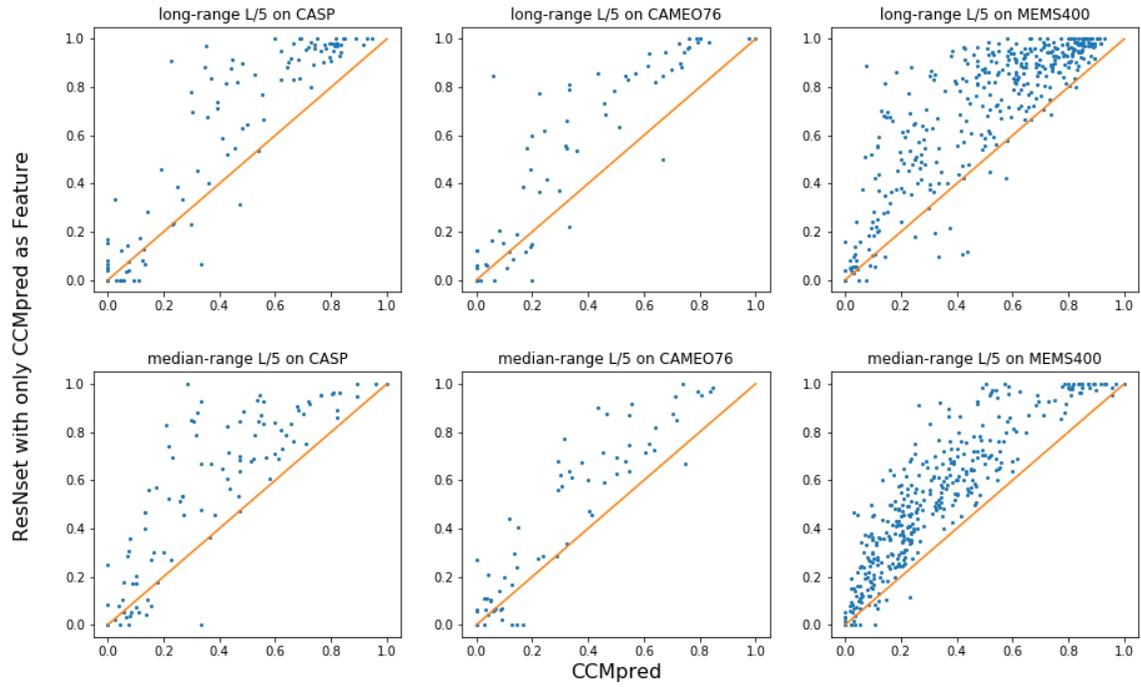

(a)

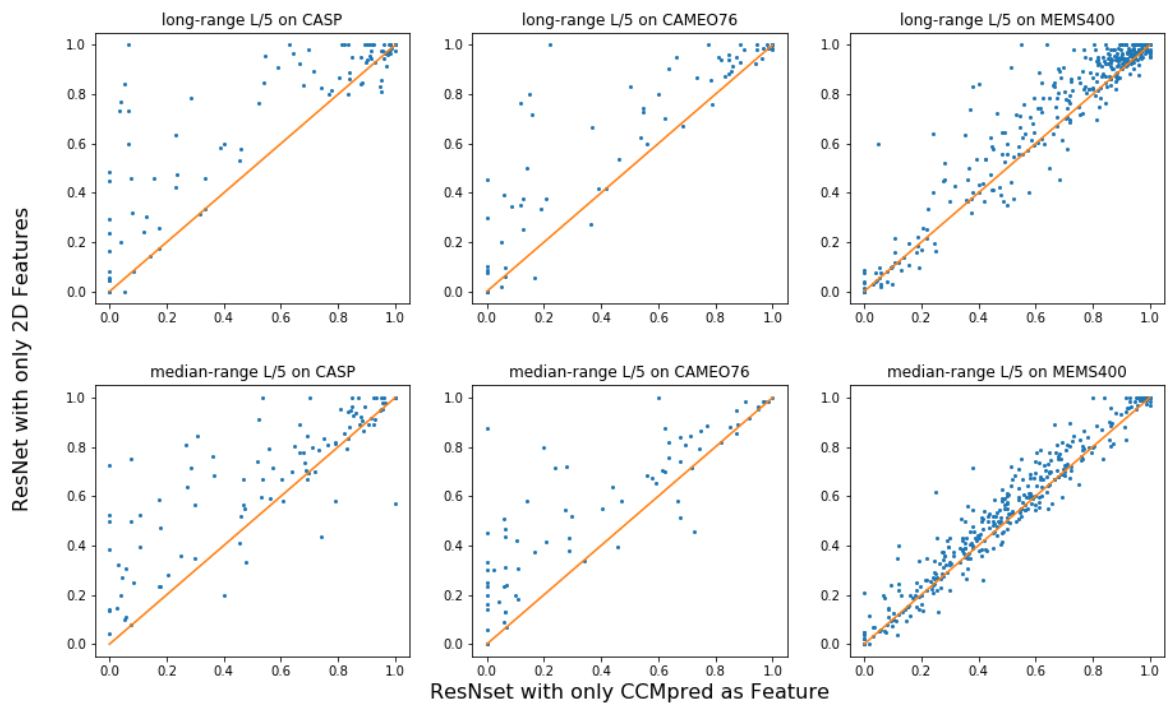

(b)

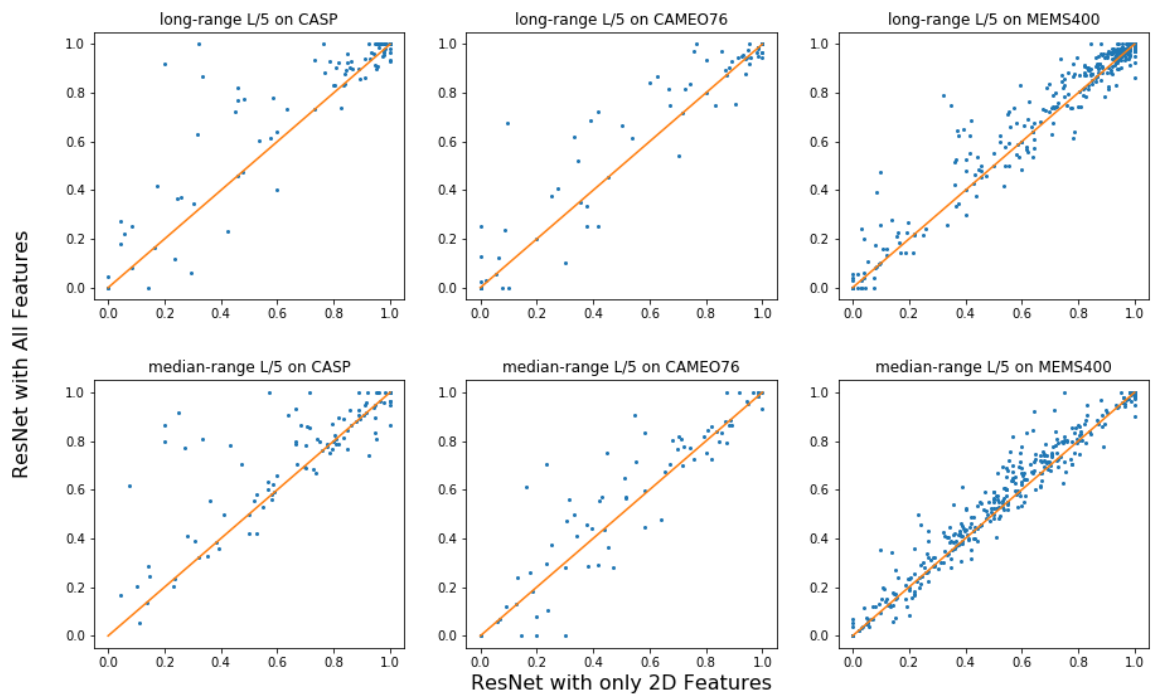

(c)

FIGURE 19 ONE-TO-ONE COMPARISON BETWEEN (A) CCMPRED ALONE AND RESNET BASED ON CCMPRED AS FEATURE ONLY, (B) RESNET WITH 2D FEATURES AND RESNET WITH CCMPRED AS FEATURE ONLY AND (C) RESNET WITH 2D FEATURES AND RESNET WITH ALL FEATURES.

Figure 19 illustrates the power of 2D features. We can see that no matter what the ln(Meff) is, ResNet[CCMpred] always performs better than the baseline method CCMpred by a large margin, and ResNet with 2D features (ResNet[2D]) has similar results to ResNet[All]. However, when the number of homologs is large (log(Meff) >=7), the top L/5 medium- and long-range contact prediction accuracy for ResNet with CCMpred as feature, ResNet with 2D features, and ResNet with all features have very similar results. On the other hand, when the number of homologs is small, ResNet[2D] significantly outperforms ResNet[CCMpred], except for membrane protein top L/5 medium-range contact prediction, for which the two methods obtain similar results.

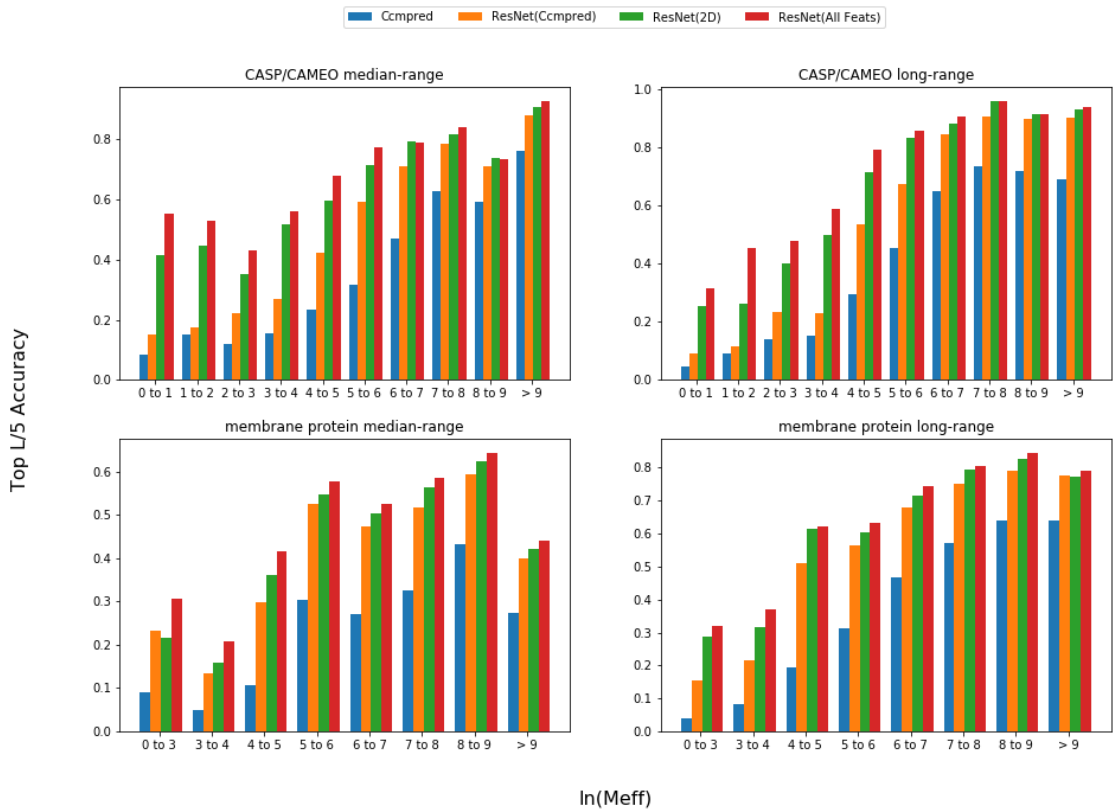

FIGURE 20 TOP L/5 MEDIUM- AND LONG-RANGE CONTACT PREDICTION ACCURACY FOR CCMPRED (BLUE), RESNET WITH ONLY CCMPRED AS FEATURE(ORANGE), RESNET WITH ALL 2D FEATURES (GREEN) AND RESNET WITH ALL FEATURES(RED) WITH RESPECT TO HOMOLOGOUS INFORMATION MEASURED BY LN(MEFF). THE 105 CASP AND 76 CAMEO RESULTS ARE DISPLAYED IN TOP PANELS, AND 398 MEMBRANE RESULTS ARE DISPLAYED IN BOTTOM PANELS.

## 4.7 Distance-based Contact Prediction

In this section, we explore the possibility of building a regression model that predicts continuous distances and binarized the predicted distances ($\leq 8$Å) to obtain the predicted contacts. Note that [80] also explores this direction by treating it as a classification problem and introducing more bins to discretize interatom distance. Specifically, we use the model architecture presented in figure 15, and the cross-entropy loss function is changed to mean square error ($L_{MSE}(y, \hat{y}) = ||y - \hat{y}||^2$), and a relative mean square error ($L_{R-MSE}(y, \hat{y}) = \left|\left|\frac{y-\hat{y}}{y}\right|\right|^2$). To remove the outlier effect, all the interatom distances that are larger than or equal to 15 Å are normalized to 15 Å.

| Method | Medium | | | | Long | | | |
|---|---|---|---|---|---|---|---|---|
| | L/10 | L/5 | L/2 | L | L/10 | L/5 | L/2 | L |
| CCMpred | 0.33 | 0.27 | 0.19 | 0.13 | 0.37 | 0.33 | 0.25 | 0.19 |
| MetaPSICOV | 0.69 | 0.59 | 0.42 | 0.28 | 0.60 | 0.54 | 0.45 | 0.35 |
| ResNet | **0.84** | **0.74** | **0.54** | **0.36** | **0.83** | **0.79** | **0.70** | **0.55** |
| DB-MSE | 0.76 | 0.68 | 0.47 | 0.32 | 0.74 | 0.68 | 0.57 | 0.44 |
| DB-R-MSE | 0.67 | 0.57 | 0.43 | 0.30 | 0.65 | 0.56 | 0.44 | 0.34 |

TABLE 4 CONTACT PREDICTION ACCURACY FOR DISTANCE-BASED (DB) METHOD BY USING MSE LOSS AND R-MSE LOSS ON 105 CASP PROTEINS.

The results are presented in table 4. This approach (DB-MSE) is able to outperform the traditional method like CCMpred or MetaPSICOV but still achieves significantly lower performance than the deep learning-based classification approach we introduced in the previously section. Besides, the MSE loss is clearly superior to relative MSE loss in this scenario as the spread between their performances is quite significant.

We further investigate with the underlying reason for this performance discrepancy considering we are using the same model architecture for both approaches. To this end, we visualize the distribution of predicted distance given the ground-truth distances that are from 2~3 Å, 3~4 Å, …, and 7~8 Å respectively, i.e., the corresponding pair of AAs are contacts. From figure 21, we find out that the model performs quite well when ground-truth distances are less than 5 Å, with the model classify most of pairs as contacts. However, it mis-classifies a significant amount of pairs with ground-truth distance ranging from 6 to 8 Å.

One possible reason is that the regression loss function does not treat 8 Å as a special boundary and this introduces more errors for contact prediction; for example, if the ground-truth distance is 7.5 Å, the MSE loss for predicted value as 6.5 Å and 8.5 Å are the same (both are 1 Å$^2$), but the latter prediction is wrong when it is used to predict contact.

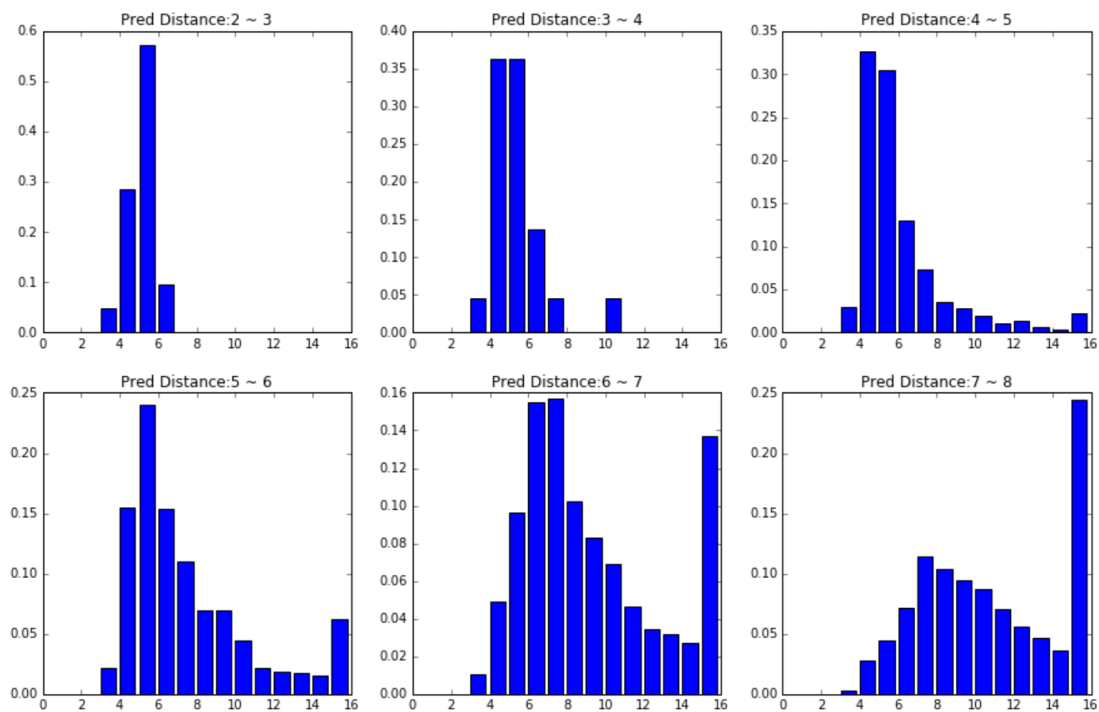

FIGURE 21 GROUND TRUTH DISTANCE DISTRIBUTION WHEN PREDICTED DISTANCES ARE RANGING FROM 2 TO 8.

## 4.8 Computing the Diversity Score of Contact Maps

High contact prediction accuracy does not necessarily lead to better folding, and predictions' diversity may also play an important role. For example, as shown in Figure 20, ResNet[CCMpred] detects 87 true contacts out of 98 predictions with an accuracy of 0.8878, while CCMpred itself only detects 70 true contacts with an accuracy of 0.7143. Thus, in terms of accuracy, there is a more than 17% gain for ResNet[CCMpred]. On the other hand, CCMpred is much more diverse than ResNet; it not only detects all blocks of contacts in ResNet[CCMpred], but also detects several novel diverse contacts that don't appear in ResNet[CCMpred] at all.

### 4.8.1 Definition of Diversity Score

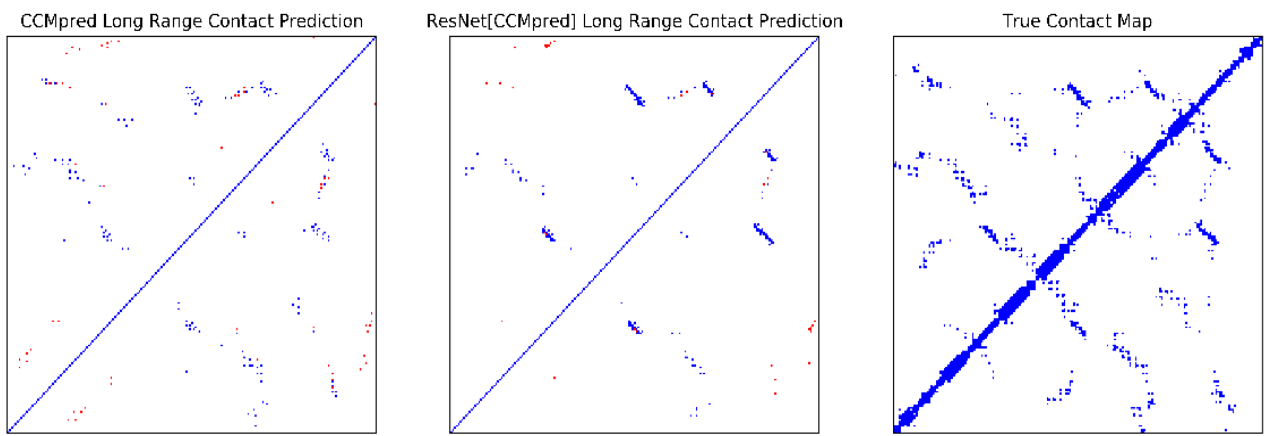

FIGURE 22 THE VISUALIZATION FOR TOP L/2 LONG RANGE CONTACT MAP FOR CCMPRED'S (LEFT), RESNET[CCMPRED] (MIDDLE) AND TRUE CONTACT MAP (RIGHT) ON TARGET T0805-D1 FROM CASP 11, WHERE BLUE COLORS INDICATES CORRECTLY PREDICTION CONTACTS, AND RED COLOR INDICATES WRONGLY PREDICTED CONTACTS. DIAGONAL OF PREDICTIONS ARE FILLED BY BLUE COLOR FOR VISUALIZATION PURPOSE ONLY.

To fairly compare two methods, say method A and method B, we design a novel diversity score that not only includes accuracy, but also takes diversity into consideration. Denote the **correctly** predicted contact map of method A and B as $\{A_1, A_2 \ldots A_M\}$, and $\{B_1, B_2 \ldots B_N\}$ out of L predictions, where each $A_i$ or $B_j$ is a two-dimensional vector indicating the position of two amino acids. To quantify the number of novel contacts that A detects while B fails to do so ($|A-B|$) or vice versa ($|B-A|$), we search for predictions in B that are close to $A_i$ for each i. More specifically, for each $A_i$, we search all predictions in B and compute their distance by $|A_i(0)-B_j(0)| + |A_i(1)-B_j(1)|$ for all $B_j$. If the minimum distance over B's predictions is larger than a predefined distance threshold ($\Delta$), say 2, we call $A_i$ a novel diversity contact that only A detects. Otherwise, it suggests that there is a contact in B that can provide similar information and does not count as a novel diversity contact.

To compute $|A-B|$, assume, for example, $A = \{A_1 = (1, 27), A_2 = (10, 40), A_3 = (2, 28)\}$, $B = \{B_1= (1, 28), B_2= (7, 48)\}$ and the distance threshold $\Delta$ as 2. For $A_1$, its distance to $B_1$ and $B_2$ are 1 and 27, respectively, and the minimum of those distances is 1, which is less than 2. Therefore, $A_1$ is not a novel diversity contact to B. Meanwhile, $A_2$ is a novel contact because its minimum distance to B is 11, and $A_3$ is not a novel contact because its minimum distance to B is also 1. Overall, only one contact in A is a novel contact over B, so |A-B| = 1. Similarly, we have |B-A|=1 because $B_2$ is a novel contact to A. Although A has one more correct prediction, the novelty of the two methods is the same based on the proposed diversity score. Finally, the final novel score is defined as |A-B| (or |B-A|) divided by the total number of predictions. Its value ranges from 0 to 1, which indicates the percentage of novel contacts in A over B (or B over A).

### 4.8.2 Diversity-Inducing Algorithm

We then compare CCMpred with ResNet[CCMpred] through diversity scores (with Δ=2) over three test datasets. For long-range top L/2 contact predictions, although ResNet[CCMpred] has much higher accuracy than CCMpred itself (as shown in Figure 21), its diversity scores are roughly the same as those of CCMpred in CASP105 and CAMEO76 test datasets, with p-values equal to 0.7703 and 0.0010, respectively; thus, neither of them are very significant. However, on membrane protein test datasets, ResNet[CCMpred] significantly outperforms CCMpred, with a p-value of 1.10x10-57.

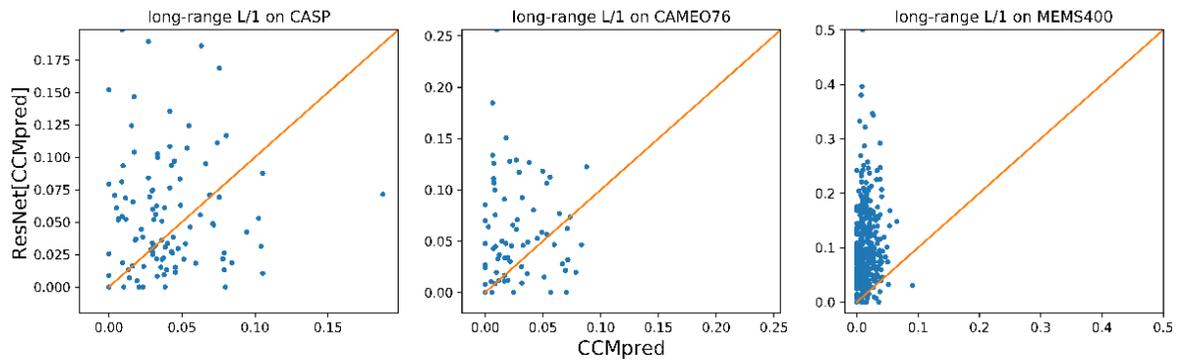

(a)

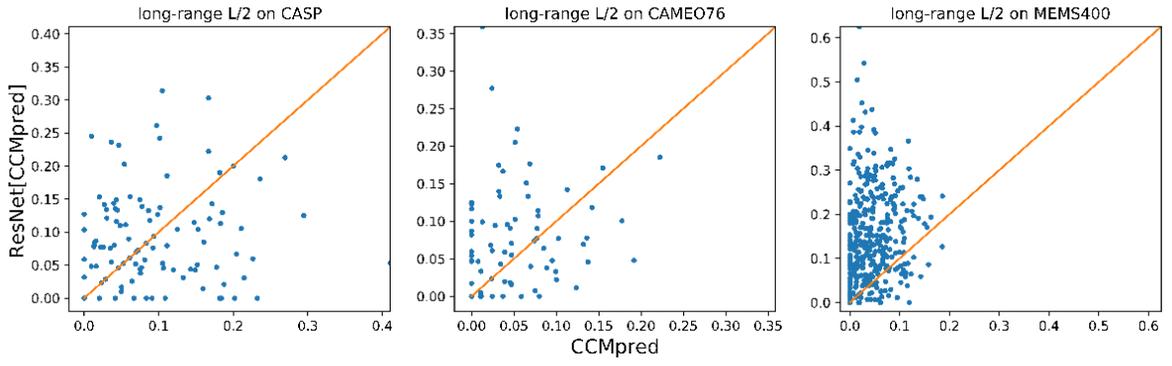

(b)

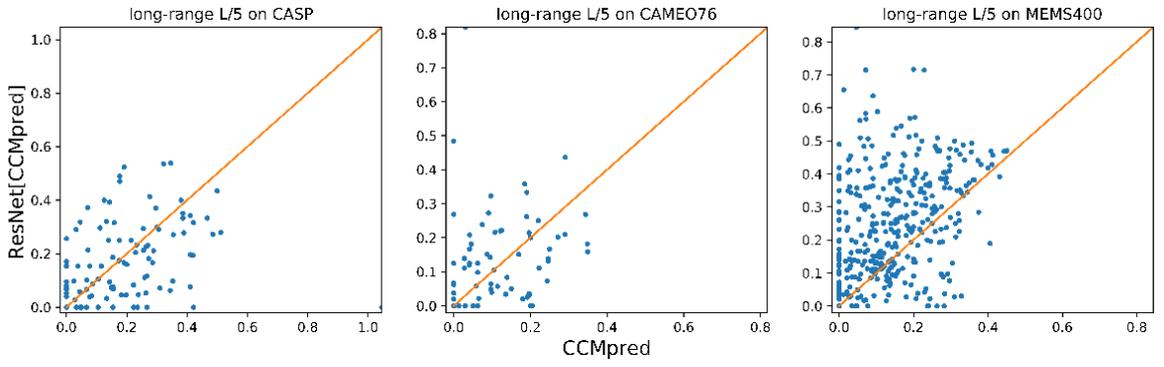

(c)

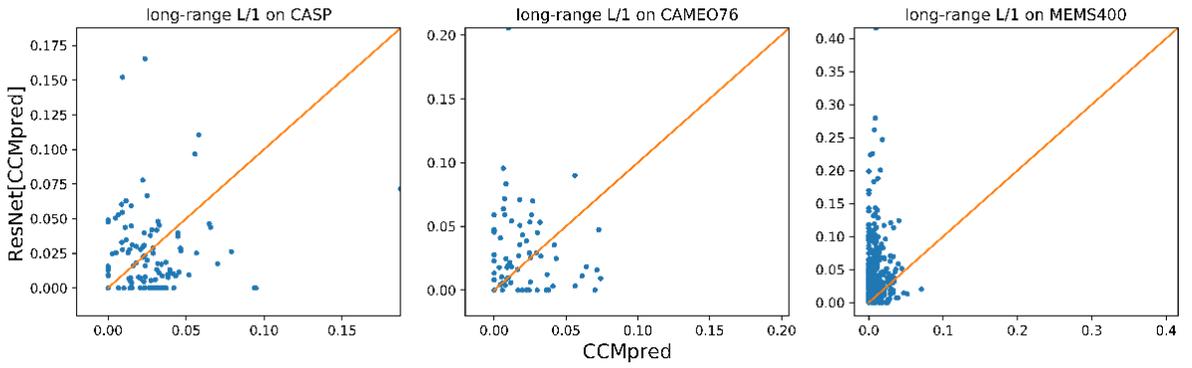

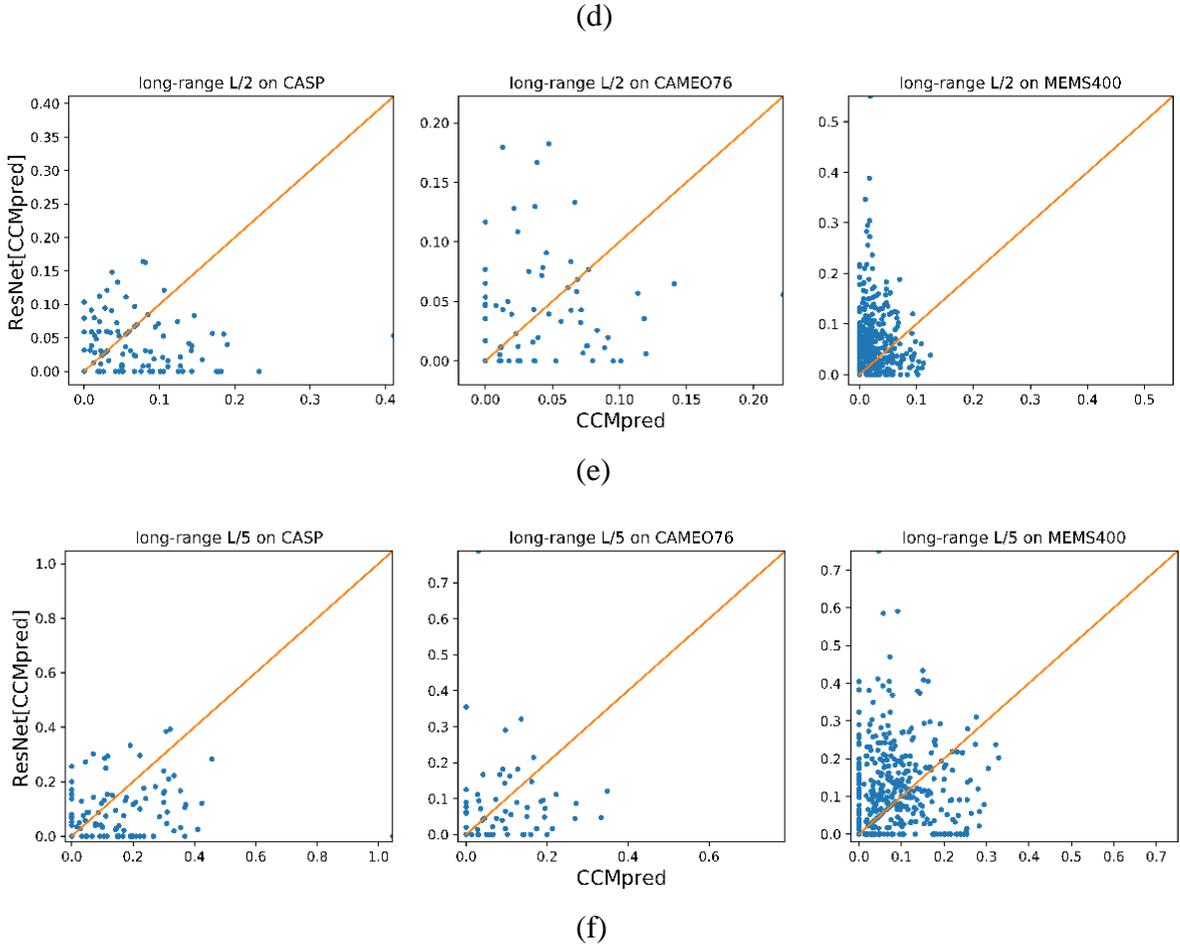

FIGURE 23 THE NOVEL SCORE COMPARISON BETWEEN CCMPRED (A) AND RESNET[CCMPRED](B) ON LONG-RANGE TOP L, L/5 AND L/10 CONTACT PREDICTION FOR THRESHOLD AS 2 (A,B,C) AND 4(D,E,F). THE X AXIS IS PERCENTILE OF |A-B|, AND Y AXIS IS PERCENTILE OF |B-A|.

The reason behind this observation is that ResNet (or any other CNN architecture) tends to infer the whole block of contacts, since i89t models the relationship between neighbors by convolution operations. We take T0805-D1 from CASP 11 as an example. As shown in Figure 22, there is a clear cluster structure in the heat map of ResNet[CCMpred]'s top predictions. In contrast, the heat map for CCMpred's raw score is much more diverse. Therefore, predicting contacts by simply ranking all the probabilities in ResNet[CCMpred] will have a lower diversity score than in CCMpred.

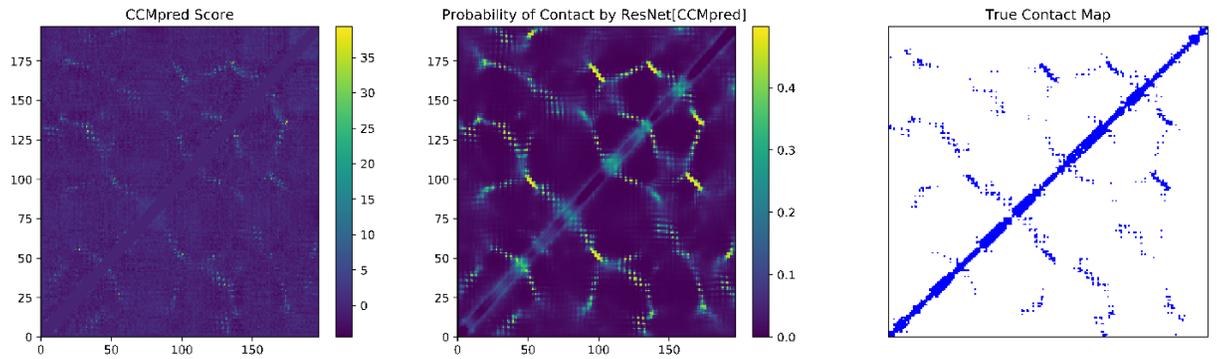

FIGURE 24 THE VISUALIZATION FOR CCMPRED'S Z-SCORE(LEFT), RESNET[CCMPRED]'S PROBABILITY (MIDDLE) AND TRUE CONTACT MAP(RIGHT) ON TARGET T0805-D1 FROM CASP 11.

Hence, we propose a new ranking method that can induce diversity to make predictions. More specifically, we first rank the predictions by probability, denoting them as $P_1 \geq P_2 \geq \ldots \geq P_N$, where N is the number of valid pairs. We then initialize the prediction set A as $\{P_1\}$. For $P_2$, we compute its minimum distance to prediction set A, and if it is smaller than a pre-defined threshold ($\Lambda$), say 2, then we skip $P_2$ because there exists a similar contact in set A. Otherwise, we add $P_2$ to A. We run this process continually for $P_3$, $P_4$, and so forth, until the size of A equals to the number of desired predictions. The resulting prediction set A will be very diverse because the minimum distance between any two predictions is at least $\Lambda$. We then compare the performance of ResNet[CCMpred] with that of CCMpred when using our proposed algorithm. As shown in Figure 23, after the diversity ranking algorithm is introduced, ResNet[CCMpred] performs significantly better than CCMpred on all three test datasets, with p-values of $9.49 \times 10^{-7}$, $4.56 \times 10^{-7}$, and $3.57 \times 10^{-84}$, respectively. For a detailed comparison, please refer to Table 5.

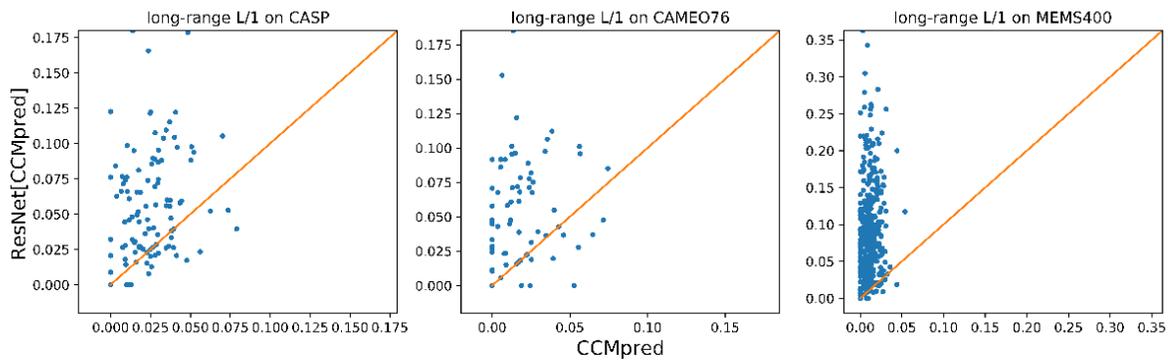

(a)

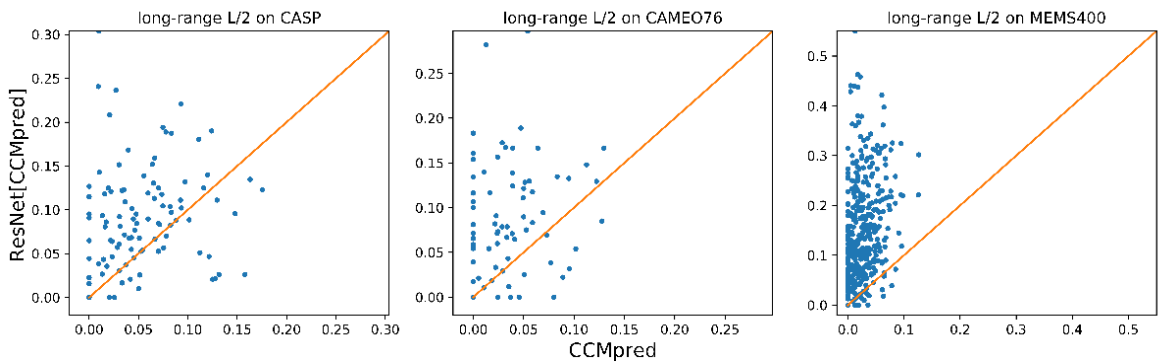

(b)

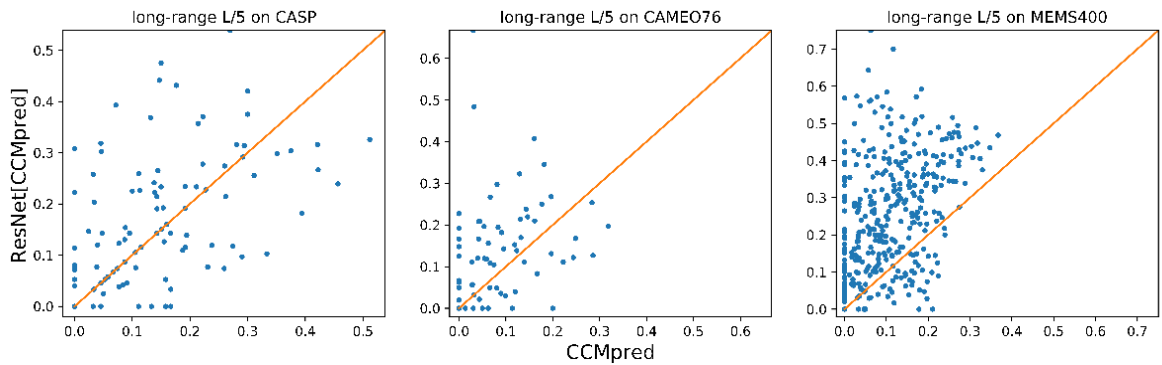

(c)

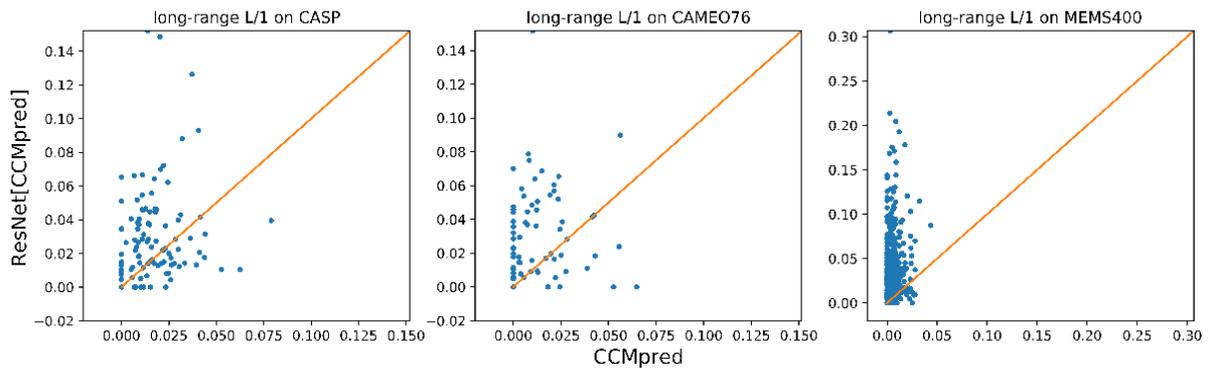

(d)

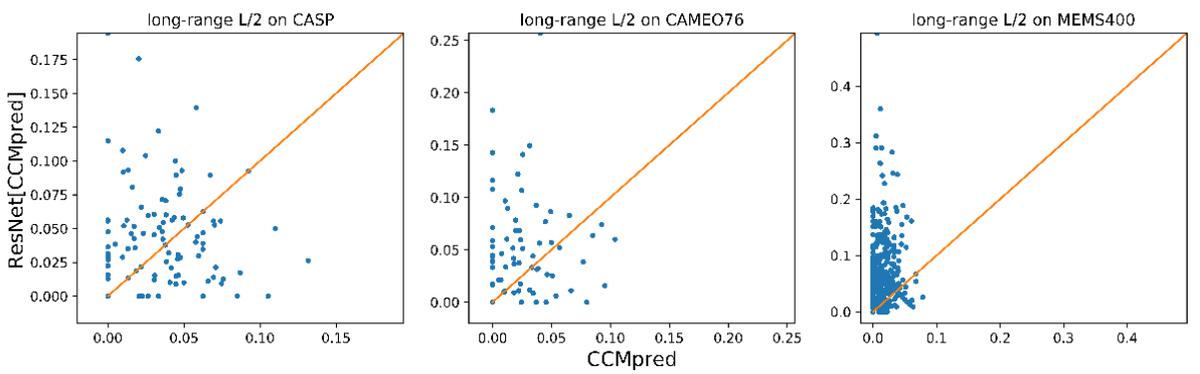

(e)

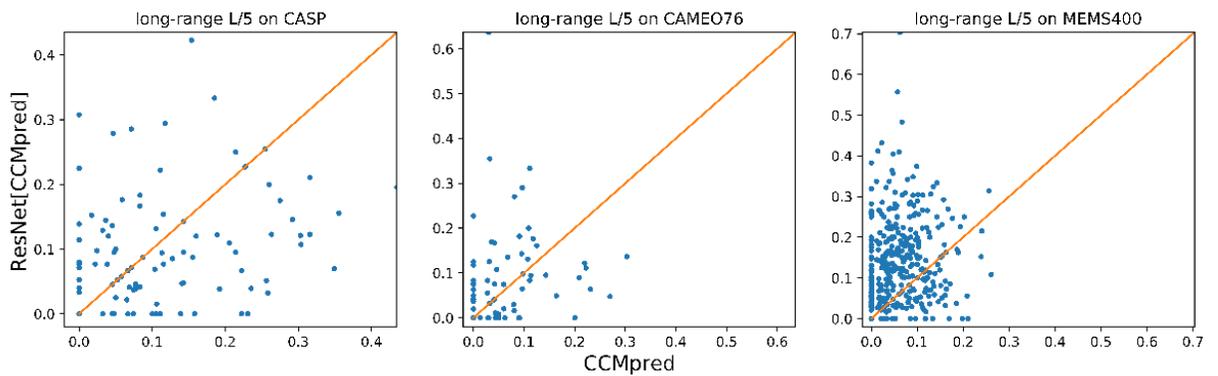

(f)

Figure 25 The novel score comparison between CCMpred (A) and ResNet[CCMpred](B) on long-range top L/5, L/2 and L contact prediction by proposed ranking algorithm for threshold as 2(a, b, c) and 4(d, e, f). The x axis is percentile of |A-B|, and y axis is percentile of |A-B|, and y axis is percentile of |B-A|.

|  | Simple Ranking Novel Score | | | Diverse Induce Ranking Novel Score | | |
| --- | --- | --- | --- | --- | --- | --- |
|  | CCMpred | ResNet[CCMpred] | p-value | CCMpred | ResNet[CCMpred] | p-value |
| L/5 | 0.1902 | 0.1678 | 0.2960 | 0.1422 | 0.1643 | 0.1940 |
| L/2 | 0.0939 | 0.0909 | 0.7702 | 0.0525 | 0.0886 | $9.5 \times 10^{-7}$ |
| L | 0.0391 | 0.0537 | 0.0058 | 0.0250 | 0.0564 | $5.5 \times 10^{-13}$ |

(a)

|  | Simple Ranking Novel Score | | | Diverse Induce Ranking Novel Score | | |
| --- | --- | --- | --- | --- | --- | --- |
|  | CCMpred | ResNet[CCMpred] | p-value | CCMpred | ResNet[CCMpred] | p-value |
| L/5 | 0.1064 | 0.1245 | 0.3589 | 0.0790 | 0.1253 | 0.0077 |
| L/2 | 0.0514 | 0.0774 | 0.0095 | 0.0357 | 0.0843 | $4.6 \times 10^{-8}$ |
| L | 0.0265 | 0.0554 | 0.0005 | 0.0185 | 0.0511 | $5.9 \times 10^{-10}$ |

(b)

|  | Simple Ranking Novel Score | | | Diverse Induce Ranking Novel Score | | |
| --- | --- | --- | --- | --- | --- | --- |
|  | CCMpred | ResNet[CCMpred] | p-value | CCMpred | ResNet[CCMpred] | p-value |
| L/5 | 0.1388 | 0.2346 | $1.5 \times 10^{-20}$ | 0.1052 | 0.2405 | $5.8 \times 10^{-44}$ |
| L/2 | 0.0437 | 0.1423 | $1.1 \times 10^{-57}$ | 0.0260 | 0.1430 | $3.6 \times 10^{-84}$ |
| L | 0.0141 | 0.0950 | $1.8 \times 10^{-78}$ | 0.0107 | 0.0844 | $1.6 \times 10^{-84}$ |

(c)

|  | Simple Ranking Novel Score | | | Diverse Induce Ranking Novel Score | | |
| --- | --- | --- | --- | --- | --- | --- |
|  | CCMpred | ResNet[CCMpred] | p-value | CCMpred | ResNet[CCMpred] | p-value |
| L/5 | 0.1558 | 0.0893 | 0.0002 | 0.1119 | 0.0920 | 0.1259 |
| L/2 | 0.0714 | 0.0389 | $1.2 \times 10^{-5}$ | 0.0344 | 0.0431 | 0.0569 |
| L | 0.0274 | 0.0250 | 0.5297 | 0.0167 | 0.0299 | $2.7 \times 10^{-5}$ |

(d)

|  | Simple Ranking | | | Diverse Induce Ranking | | |
| --- | --- | --- | --- | --- | --- | --- |
|  | CCMpred | ResNet[CCMpred] | p-value | CCMpred | ResNet[CCMpred] | p-value |
| L/5 | 0.0833 | 0.0710 | 0.4531 | 0.0604 | 0.0800 | 0.1795 |
| L/2 | 0.0398 | 0.0393 | 0.9390 | 0.0251 | 0.0490 | 0.0001 |
| L | 0.0195 | 0.0273 | 0.0731 | 0.0125 | 0.0285 | $1.8 \times 10^{-5}$ |

(e)

|  | Simple Ranking | | | Diverse Induce Ranking | | |
| --- | --- | --- | --- | --- | --- | --- |
|  | CCMpred | ResNet[CCMpred] | p-value | CCMpred | ResNet[CCMpred] | p-value |
| L/5 | 0.0810 | 0.1189 | $2.4 \times 10^{-8}$ | 0.0575 | 0.1271 | $4.1 \times 10^{-29}$ |
| L/2 | 0.0236 | 0.0617 | $7.9 \times 10^{-25}$ | 0.0134 | 0.0678 | $6.4 \times 10^{-53}$ |
| L | 0.0084 | 0.0416 | $1.4 \times 10^{-37}$ | 0.0057 | 0.0401 | $3.1 \times 10^{-53}$ |

(f)

TABLE 5 CONTACT PREDICTION NOVEL SCORE AND P-VALUE COMPARISON SIMPLE RANKING AND PROPOSED DIVERSE INDUCE RANKING ON (A) 105 CASP PROTEINS, (B) 76 CAMEO PROTEINS AND (C) 398 MEMBRANE PROTEINS FOR THRESHOLD EQUALS TO 2 (A, B, C) AND 4 (D, E, F).

### 4.8.3 Comparison with Entropy Score

To tackle the same dispersion problem discussed above, CASP12 introduced an Entropy Score (ES) to measure the diversity of the contact map. The score is computed based on the relative drop of entropy due to geometric constraints on the protein shape with respect to the entropy of an extended state without any constraints [74]. It is defined as

$$ES_{ext} = \frac{E|0 - E|C}{E|0} \cdot 100\%,$$

where $E|0$, $E|C$ are entropy values calculated for the protein without any constraints, and with a set of contacts (C), respectively. E | (0, C) are computed as the expectation of Shannon's information entropy computed for residue-residue distance under the uniform probability distribution assumption, i.e.,

$$E|x = \frac{1}{\text{\# all pairs}} \sum_{i \neq j} \ln(U_{ij} - L_{ij}),$$

where $U_{ij}, L_{ij}$ are the lower and upper bounds of residue-residue distances. Based on [74], for contacts, $U_{ij}$ is set as 8.0 Angstroms. For non-contact, the upper limit is set as $U_{ij} = 3.8 \cdot |i - j|$, while the lower bound is always set as 3.2. [74] computed the above score on correctly predicted contacts (true positives) only.

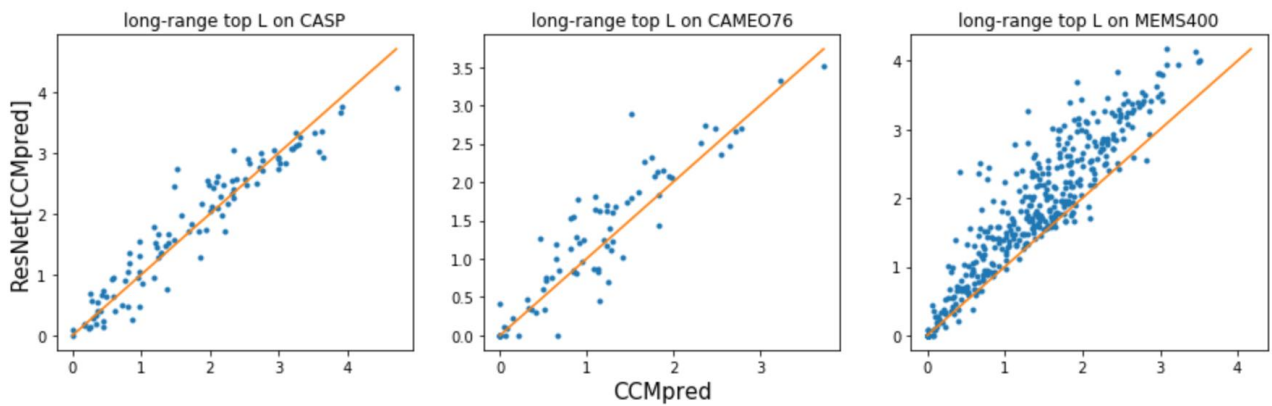

FIGURE 26 THE ES SCORE COMPARISON BETWEEN CCMPRED AND RESNET[CCMPRED]. THE ES SCORES ARE COMPUTED BASED ON TOP L LONG-RANGE CONTACT PREDICTIONS.

Since ResNet[CCMpred] tends to predict contacts from same region, from Figure 24 we can see that it performs similarly to CCMpred on long-range contact predictions in terms of ES for both CASP and CAMEO datasets, which is consistent with our novelty score results. This also shows the proposed novelty score can measure the diversity of predicted contacts.

### 4.8.4 Comparison with Coverage Ratio Score

We also compare novelty score with the coverage ratio score proposed in [80], which is defined as the coverage ratio of ground truth contacts by the top-n correctly predicted ones. More detailed, denote $dist_{cb}(A, B) = |i - k| + |j - l|$, where $dist_{cb}$ is the city-block distance between contact $A = (i, j)$ and contact $B = (k, l)$. We say a native contact A is covered by top-n predictions by $\tau$ if $\min_{B \in CP_n} dist_{cb}(A, B) \leq \tau$, where $CP_n$ is the set of correctly predicted contacts, and $\tau$ is a non-negative integer. Intuitively, if we could find a correctly predicted contact that is closed to the native contact A, we treat A as covered. When $\tau = 0$, the coverage ratio score simply reduced to recall [80].

|  | 0 | 1 | 2 | 3 | 4 | 5 |
|---|---|---|---|---|---|---|
| CCMpred | 22.06 | 36.87 | 45.66 | **51.94** | **57.05** | **60.58** |
| ResNet[CCMpred] | **32.56** | **41.63** | **46.63** | 51.18 | 54.86 | 57.34 |

TABLE 6 THE AVERAGE COVERAGE RATIO SCORE OF TOP L LONG RANGE PREDICTIONS FOR CASP12 DATASET. THE DISTANCE THRESHOLD IS RANGING FROM 0 TO 5.

The results are presented in table 6. Similar to diversity score or entropy score, ResNet[CCMpred] is better in terms of accuracy since it has higher coverage ratio score for smaller distance thresholds ($\tau = 0, 1$), it fails to predict more diverse contacts as its coverage ratio score is lower than CCMpred for larger distance thresholds ($\tau = 4, 5$).

# Chapter 5

# Conclusions

In this thesis, we first tried to improve unsupervised structure learning in graphical models in ECA. Since there is a cluster structure in contacts, we presented a nonparametric model that can cluster variables in a GGM into correlated groups, by exploiting block structure in a GGM and making use of an efficient MCMC algorithm. Our method performs well on both synthetic and real data and can successfully identify the underlying block structure. In particular, our method does not need a predefined value for the number of clusters. Instead, it can automatically determine it based on the data, thanks to our nonparametric approach.

Next, considering that plmDCA used the Potts model and pseudo-likelihood to tackle the computation of the partition function, we proposed a new procedure for learning the structure of a nonparametric graphical model. Our procedure is based on minimizing a penalized score-matching objective, which can be performed efficiently by using existing group lasso solvers. A particularly appealing aspect of our approach is that it does not require computing the normalization constant. Therefore, our procedure can be applied to a very broad family of infinite dimensional exponential families, including the Potts model. We have established that the procedure recovers the true underlying graphical structure with high probability under mild conditions. However, due to speed limitations of the algorithm, we could only investigate its performance on very short proteins.

In the future, we plan to investigate more efficient algorithms to solve the representer theorem for longer proteins, since it is often the case that $C$ is well structured and can be efficiently approximated.

Finally, we proposed two deep learning methods that could improve the accuracy of the current state-of-the-art predictor significantly. The first deep learning model treated the contact prediction as an image classification problem, and the architecture is simply a shallow convolutional neural network. This simple model already improved upon metaPSICOV on all of our test sets and generated more accurate contact predictions. By utilizing the latest breakthrough from computer vision community, we then proposed a new architecture with an ultra-deep residual network that could take the whole protein into consideration and predicts all contacts of a protein simultaneously. This new architecture further improved the performance of metaPSICOV by a much larger margin, which highlights the power of deep learning.

We also did an ablation study to investigate the impact of different features, demonstrating that the 2D features are very essential to the algorithm. With only one feature generated from CCMpred, we already improved the long-range top L/10 prediction accuracy of the original CCMpred by almost 20% on all three test sets. In addition, we also proved that the 1D features are less important because the long-range top L/10 accuracy only drops 3 to 5% even when we remove all of them. We then explore the possibility to predicting contact map by treating it as a regression problem, and find it underperforms previous classification approach. Lastly, we proposed a diversity-inducing score that could directly evaluate the predictor's dispersion. In the experiments, we showed that ResNet[CCMpred] has a very similar diversity score as CCMpred, even though there is a large margin in the accuracy of their predictions, which reveals that methods based on CNN tend to generate predictions that are close on sequence level and do not encourage diversity. We then proposed an algorithm that could generate diverse predictions and showed that it works well on CNN based methods.

With rapid progress of advanced language model such as BERT [81], researchers started to model protein sequence representations by using transformers [82], and it could be potentially beneficial to contact prediction as currently those 1D features do not contribute much to the final performance. For future work, we expect it could be fruitful to investigate more accurate sequence representation using large-scale protein sequences, and incorporate them into contact prediction algorithms to further boost the performance.

# References

1. Kim, D.E., et al., *One contact for every twelve residues allows robust and accurate topology-level protein structure modeling.* Proteins: Structure, Function, and Bioinformatics, 2014. **82**(S2): p. 208-218.
2. Adhikari, B., et al., *CONFOLD: residue-residue contact-guided ab initio protein folding.* Proteins: Structure, Function, and Bioinformatics, 2015. **83**(8): p. 1436-1449.
3. Wang, S., et al., *CoinFold: a web server for protein contact prediction and contact-assisted protein folding.* Nucleic acids research, 2016. **44**(W1): p. W361-W366.
4. Di Lena, P., K. Nagata, and P. Baldi, *Deep architectures for protein contact map prediction.* Bioinformatics, 2012. **28**(19): p. 2449-2457.
5. Bateman, A., et al., *The Pfam protein families database.* 2004. **32**(suppl_1): p. D138-D141.
6. Cocco, S., R. Monasson, and M.J.P.c.b. Weigt, *From principal component to direct coupling analysis of coevolution in proteins: low-eigenvalue modes are needed for structure prediction.* 2013. **9**(8): p. e1003176.
7. Lapedes, A., B. Giraud, and C.J.a.p.a. Jarzynski, *Using sequence alignments to predict protein structure and stability with high accuracy.* 2012.
8. Weigt, M., et al., *Identification of direct residue contacts in protein–protein interaction by message passing.* 2009. **106**(1): p. 67-72.
9. Marks, D.S., et al., *Protein 3D structure computed from evolutionary sequence variation.* 2011. **6**(12): p. e28766.
10. Jones, D.T., et al., *PSICOV: precise structural contact prediction using sparse inverse covariance estimation on large multiple sequence alignments.* Bioinformatics, 2011. **28**(2): p. 184-190.
11. Balakrishnan, S., et al., *Learning generative models for protein fold families.* 2011. **79**(4): p. 1061-1078.

12. Seemayer, S., M. Gruber, and J. Söding, *CCMpred—fast and precise prediction of protein residue–residue contacts from correlated mutations.* Bioinformatics, 2014. **30**(21): p. 3128-3130.

13. Ekeberg, M., et al., *Improved contact prediction in proteins: using pseudolikelihoods to infer Potts models.* Physical Review E, 2013. **87**(1): p. 012707.

14. De Juan, D., F. Pazos, and A.J.N.R.G. Valencia, *Emerging methods in protein co-evolution.* 2013. **14**(4): p. 249.

15. Wu, S. and Y. Zhang, *A comprehensive assessment of sequence-based and template-based methods for protein contact prediction.* Bioinformatics, 2008. **24**(7): p. 924-931.

16. Skwark, M.J., et al., *Improved contact predictions using the recognition of protein like contact patterns.* PLoS computational biology, 2014. **10**(11): p. e1003889.

17. Jones, D.T., et al., *MetaPSICOV: combining coevolution methods for accurate prediction of contacts and long range hydrogen bonding in proteins.* Bioinformatics, 2014. **31**(7): p. 999-1006.

18. Ma, J., et al., *Protein contact prediction by integrating joint evolutionary coupling analysis and supervised learning.* Bioinformatics, 2015. **31**(21): p. 3506-3513.

19. Wang, Z. and J. Xu, *Predicting protein contact map using evolutionary and physical constraints by integer programming.* Bioinformatics, 2013. **29**(13): p. i266-i273.

20. Dunn, S.D., L.M. Wahl, and G.B. Gloor, *Mutual information without the influence of phylogeny or entropy dramatically improves residue contact prediction.* Bioinformatics, 2007. **24**(3): p. 333-340.

21. Itoh, K. and M.J.P.o.t.N.A.o.S. Sasai, *Flexibly varying folding mechanism of a nearly symmetrical protein: B domain of protein A.* 2006. **103**(19): p. 7298-7303.

22. Friedman, J., T. Hastie, and R. Tibshirani, *Sparse inverse covariance estimation with the graphical lasso.* Biostatistics, 2008. **9**(3): p. 432-441.

23. Abola, E.E., F.C. Bernstein, and T.F. Koetzle, *The protein data bank*, in *Neutrons in Biology*. 1984, Springer. p. 441-441.

24. Banerjee, O., et al. *Convex optimization techniques for fitting sparse Gaussian graphical models*. in *Proceedings of the 23rd international conference on Machine learning*. 2006. ACM.
25. Yuan, M. and Y.J.B. Lin, *Model selection and estimation in the Gaussian graphical model.* 2007. **94**(1): p. 19-35.
26. Marlin, B.M. and K.P. Murphy. *Sparse Gaussian graphical models with unknown block structure*. in *Proceedings of the 26th Annual International Conference on Machine Learning*. 2009. ACM.
27. Yuan, M. and Y.J.J.o.t.R.S.S.S.B. Lin, *Model selection and estimation in regression with grouped variables.* 2006. **68**(1): p. 49-67.
28. Marlin, B.M., M. Schmidt, and K.P. Murphy. *Group sparse priors for covariance estimation*. in *Proceedings of the Twenty-Fifth Conference on Uncertainty in Artificial Intelligence*. 2009. AUAI Press.
29. Ambroise, C., J. Chiquet, and C.J.E.J.o.S. Matias, *Inferring sparse Gaussian graphical models with latent structure.* 2009. **3**: p. 205-238.
30. Palla, K., Z. Ghahramani, and D.A. Knowles. *A nonparametric variable clustering model*. in *Advances in Neural Information Processing Systems*. 2012.
31. Sun, S., Y. Zhu, and J. Xu. *Adaptive variable clustering in gaussian graphical models*. in *Artificial Intelligence and Statistics*. 2014.
32. Pitman, J., *Combinatorial stochastic processes*. 2002, Technical Report 621, Dept. Statistics, UC Berkeley, 2002. Lecture notes for ….
33. Haff, L.J.T.A.o.S., *Empirical Bayes estimation of the multivariate normal covariance matrix.* 1980: p. 586-597.
34. Neal, R.M.J.A.o.s., *Slice sampling.* 2003: p. 705-741.
35. Jain, S., R.M.J.J.o.c. Neal, and G. Statistics, *A split-merge Markov chain Monte Carlo procedure for the Dirichlet process mixture model.* 2004. **13**(1): p. 158-182.
36. Zhao, T., et al., *The huge package for high-dimensional undirected graph estimation in R.* 2012. **13**(Apr): p. 1059-1062.

37. Tang, Q., S. Sun, and J. Xu. *Learning scale-free networks by dynamic node specific degree prior*. in *International Conference on Machine Learning*. 2015.
38. Sun, S., H. Wang, and J. Xu. *Inferring block structure of graphical models in exponential families*. in *Artificial Intelligence and Statistics*. 2015.
39. Sun, S., M. Kolar, and J. Xu. *Learning structured densities via infinite dimensional exponential families*. in *Advances in Neural Information Processing Systems*. 2015.
40. Yang, E., et al. *Graphical models via generalized linear models*. in *Advances in Neural Information Processing Systems*. 2012.
41. Jeon, Y. and Y.J.S.S. Lin, *An effective method for high-dimensional log-density ANOVA estimation, with application to nonparametric graphical model building.* 2006: p. 353-374.
42. Liu, H., J. Lafferty, and L.J.J.o.M.L.R. Wasserman, *The nonparanormal: Semiparametric estimation of high dimensional undirected graphs.* 2009. **10**(Oct): p. 2295-2328.
43. Ravikumar, P., M.J. Wainwright, and J.D.J.T.A.o.S. Lafferty, *High-dimensional Ising model selection using ℓ1-regularized logistic regression.* 2010. **38**(3): p. 1287-1319.
44. Hyvärinen, A.J.J.o.M.L.R., *Estimation of non-normalized statistical models by score matching.* 2005. **6**(Apr): p. 695-709.
45. Hyvärinen, A.J.C.s. and d. analysis, *Some extensions of score matching.* 2007. **51**(5): p. 2499-2512.
46. Meinshausen, N. and P.J.T.a.o.s. Bühlmann, *High-dimensional graphs and variable selection with the lasso.* 2006. **34**(3): p. 1436-1462.
47. Canu, S. and A.J.N. Smola, *Kernel methods and the exponential family.* 2006. **69**(7-9): p. 714-720.
48. Sriperumbudur, B., et al., *Density estimation in infinite dimensional exponential families.* 2017. **18**(1): p. 1830-1888.

49. Krizhevsky, A., I. Sutskever, and G.E. Hinton. *Imagenet classification with deep convolutional neural networks*. in *Advances in neural information processing systems*. 2012.

50. LeCun, Y., et al., *Backpropagation applied to handwritten zip code recognition.* 1989. **1**(4): p. 541-551.

51. Wang, S., et al., *Protein secondary structure prediction using deep convolutional neural fields.* 2016. **6**: p. 18962.

52. Wang, S., et al., *RaptorX-Property: a web server for protein structure property prediction.* 2016. **44**(W1): p. W430-W435.

53. Miyazawa, S. and R.L.J.M. Jernigan, *Estimation of effective interresidue contact energies from protein crystal structures: quasi-chemical approximation.* 1985. **18**(3): p. 534-552.

54. Betancourt, M.R. and D.J.P.s. Thirumalai, *Pair potentials for protein folding: choice of reference states and sensitivity of predicted native states to variations in the interaction schemes.* 1999. **8**(2): p. 361-369.

55. He, K., et al. *Deep residual learning for image recognition*. in *Proceedings of the IEEE conference on computer vision and pattern recognition*. 2016.

56. dos Santos, C. and M. Gatti. *Deep convolutional neural networks for sentiment analysis of short texts*. in *Proceedings of COLING 2014, the 25th International Conference on Computational Linguistics: Technical Papers*. 2014.

57. Kim, Y.J.a.p.a., *Convolutional neural networks for sentence classification.* 2014.

58. Wang, S., et al., *Accurate de novo prediction of protein contact map by ultra-deep learning model.* 2017. **13**(1): p. e1005324.

59. Angermueller, C., et al., *Deep learning for computational biology.* 2016. **12**(7): p. 878.

60. Ioffe, S. and C.J.a.p.a. Szegedy, *Batch normalization: Accelerating deep network training by reducing internal covariate shift.* 2015.

61. Kingma, D.P. and J.J.a.p.a. Ba, *Adam: A method for stochastic optimization.* 2014.

62. Pascanu, R., T. Mikolov, and Y. Bengio. *On the difficulty of training recurrent neural networks*. in *International Conference on Machine Learning*. 2013.

63. Simonyan, K. and A.J.a.p.a. Zisserman, *Very deep convolutional networks for large-scale image recognition.* 2014.

64. He, K., et al. *Identity mappings in deep residual networks*. in *European conference on computer vision*. 2016. Springer.

65. Huang, G., et al. *Densely connected convolutional networks*. in *CVPR*. 2017.

66. Zagoruyko, S. and N.J.a.p.a. Komodakis, *Wide residual networks.* 2016.

67. Sermanet, P., et al., *Overfeat: Integrated recognition, localization and detection using convolutional networks.* 2013.

68. Srivastava, R.K., K. Greff, and J. Schmidhuber, *Highway networks.* arXiv preprint arXiv:1505.00387, 2015.

69. He, K. and J. Sun. *Convolutional neural networks at constrained time cost*. in *Proceedings of the IEEE Conference on Computer Vision and Pattern Recognition*. 2015.

70. Huang, G., et al. *Deep networks with stochastic depth*. in *European Conference on Computer Vision*. 2016. Springer.

71. Wang, S., et al., *Analysis of deep learning methods for blind protein contact prediction in CASP12.* 2018. **86**: p. 67-77.

72. Kingma, D. and J. Ba, *Adam: A method for stochastic optimization.* arXiv preprint arXiv:1412.6980, 2014.

73. Pietro Di Lena, et al., Deep architectures for protein contact map prediction, BIOINFORMATICS, 2012

74. Joery Schaarschmidt, et al., Assessment of contact predictions in CASP12: Co-evolution and deep learning coming of age, Proteins, 2017

75. Lapedes A, et al., Using Sequence Alignments to Predict Protein Structure and Stability with High Accuracy. arXiv:12072484. 2002

76. Morcos, Faruck, et al. "Direct-coupling analysis of residue coevolution captures native contacts across many protein families." Proceedings of the National Academy of Sciences 108.49 (2011): E1293-E1301.

77. Atchley, William R., et al. "Correlations among amino acid sites in bHLH protein domains: an information theoretic analysis." Molecular biology and evolution 17.1 (2000): 164-178.
78. Burger, Lukas, and Erik Van Nimwegen. "Disentangling direct from indirect co-evolution of residues in protein alignments." PLoS computational biology 6.1 (2010): e1000633.
79. Lapedes, Alan S., et al. "Correlated mutations in models of protein sequences: phylogenetic and structural effects." *Lecture Notes-Monograph Series* (1999): 236-256.
80. Xu, Jinbo. "Distance-based protein folding powered by deep learning." *Proceedings of the National Academy of Sciences* 116.34 (2019): 16856-16865.
81. Devlin, Jacob, et al. "Bert: Pre-training of deep bidirectional transformers for language understanding." *arXiv preprint arXiv:1810.04805* (2018).
82. Rives, Alexander, et al. "Biological structure and function emerge from scaling unsupervised learning to 250 million protein sequences." *bioRxiv* (2019): 622803.